\documentclass[twocolumn,twocolappendix,trackchanges]{aastex701}
\usepackage{graphicx}
\usepackage{txfonts}
\usepackage{hyperref}
\hypersetup{
  colorlinks   = true, 
  urlcolor     = blue, 
  linkcolor    = blue, 
  citecolor   = blue 
}

\usepackage{ulem}
\usepackage{float}
\usepackage{ragged2e}
\usepackage{threeparttable}
\usepackage{soul}
\usepackage{setspace}
\usepackage{lipsum}
\usepackage{afterpage}
\usepackage{longtable}
\usepackage{tabularx}
\usepackage{subcaption}   

\begin{document}

\title{An SRG/eROSITA Study of Be X-ray binaries in Quiescence}

\author[orcid=0000-0001-7221-935X,gname=Ajith,sname='Balu']{Ajith Balu}
\affiliation{Raman Research Institute, Sadashivanagar, Bengaluru, Karnataka - 560080, India}
\email[show]{ajithb@rrimail.rri.res.in}  

\author[orcid=0000-0002-7391-5776,gname=Kinjal, sname='Roy']{Kinjal Roy} 
\affiliation{Raman Research Institute, Sadashivanagar, Bengaluru, Karnataka - 560080, India}
\email{kinjal@rrimail.rri.res.in}

\author[0000-0002-8775-5945,gname=Biswajit,sname='Paul']{Biswajit Paul}
\affiliation{Raman Research Institute, Sadashivanagar, Bengaluru, Karnataka - 560080, India}
\email{bpaul@rri.res.in}

\begin{abstract}

Be X-ray binaries (BeXRBs) are mostly transient X-ray binaries with usually a neutron star as the compact object and a Be star as their optical companion. We performed a comprehensive analysis of BeXRBs in quiescence with the extended ROentgen Survey with an Imaging Telescope Array (eROSITA), the soft X-ray instrument on board the Spectrum-Roentgen-Gamma satellite, using its first all-sky survey (eRASS1) data. The transient BeXRBs spend majority of time in their quiescent state, but are often studied during their outbursts. A quiescence study is useful to probe the emission mechanism of BeXRBs at low mass accretion rates. The spectra of BeXRBs detected in the eRASS1 main catalogue could be well fitted with a powerlaw as well as a blackbody model. In the absence of detection of a BeXRB, we report the luminosity upper limit for its position in the sky. The X-ray luminosity for the quiescent BeXRBs detected with eROSITA have $L_{X}\sim10^{33-36}$ erg s$^{-1}$ for both absorbed powerlaw and blackbody models in most cases. For the absorbed blackbody model, the emission seems to come from a fraction of the neutron star's surface. The quiescence luminosity of BeXRBs are used to distinguish between different accretion models at low mass accretion rates, like accretion in the propeller regime or from a cold accretion disc. The detection space of BeXRBs in quiescence for the eRASS1 data is compared against previously reported luminosities across multiple X-ray observatories.

\end{abstract}


\section{Introduction} 

High-mass X-ray binaries (HMXBs) have a subset of systems called the Be X-ray binaries (BeXRBs), which are X-ray binaries (XRBs) with Be stars as the optical companion, accompanied by a neutron star (NS) in most cases. Be stars are fast-rotating early-type stars that are known to have emission lines in their optical spectra along with an excess infrared emission \citep{ Struve_1931ApJ_Be_stars, Slettebak_1979SSR_Be_stars}. Such emission features originate from an extended circumstellar disc at the equator of the main-sequence companion. H$\alpha$ is one such emission line in the Balmer series of hydrogen observed from the disc envelope structure; sometimes lines in emission are also observed for He or Fe \citep{Hanuschik1996AnA}.

BeXRBs are mostly transient in nature \citep{PabloReig2011Ap}, and are usually studied during their outbursts. The outbursts in BeXRBs are classified into two kinds, i) type I outbursts and ii) type II outbursts. The type I outbursts are periodic in nature and occur during the periastron passage of the compact object at each orbit, covering a fraction of the orbit with the peak \text{X-ray} luminosities $L_x\lesssim10^{37}\text{ erg s}^{-1}$. Meanwhile, type II outbursts are rare and irregular in nature and may last for several orbits of the binary system with their peak luminosity reaching around $10^{37}-10^{38}\text{ erg s}^{-1}$, close to the Eddington luminosity of a NS. Some BeXRBs are persistent and show less X-ray variation with no large transient outbursts, low X-ray luminosity $L_X\lesssim10^{35}$ erg s$^{-1}$ with slowly rotating NS with $P_{spin}\gtrsim100$ s \citep{ PabloReig2011Ap, Pradhan2013MNRAS_SW_J20006p3210, La_Palombara_2025AN_persistent_BeXRB}.

Most transient BeXRBs are studied in details during their outburst phase when they are bright, which allows investigation of changes in accretion mechanism across a wide range of X-ray luminosity and mass accretion rate. The magnetic field strength of NSs has also been estimated for many BeXRBs from their cyclotron resonant scattering feature (CRSF), an absorption line in X-ray spectra which was first detected in Her X$-$1 at $\sim$40 keV \citep{Truemper_1978ApJ_herxm1_cyclo, Maitra_2017JApA_crsf, Staubert_2019AnA} followed by a detection in the BeXRB \text{4U 0115$+$63} at \text{$\sim$20 keV} \citep{Wheaton_1979_4U0115p63_cyclo}. A cyclotron line measurement is useful in calculating the magnetospheric radius and to put luminosity limits for different emission mechanisms \citep{ Illarionov_1975AnA_propeller_regime, Tsygankov_2016AnA_prop_eff_4U_0115p63_N_V_0332p53} discussed later in this paper.

Pulsations in BeXRBs during quiescence have been detected down to luminosity of $L_{X}\sim10^{33}-10^{34}$ erg s$^{-1}$, and with a pulse fraction of $50-70\%$ \citep{Mukherjee2005aanda_3a0535p262,Tsygankov2017MNRAS_pulsars_in_quiescence,Reig2018aanda_IGR_J06074p2205}. Observation of BeXRBs in quiescence may help understand the accretion and emission mechanism of the sources at low luminosity. Spectral study and possible pulse detections at low luminosity may help further develop the accretion models used to describe BeXRB in quiescence. In the present work, we analyse data of the BeXRB sources detected by the extended ROentgen Survey with an Imaging Telescope Array (eROSITA) X-ray telescope. The eROSITA sky survey data allows for study of many BeXRB sources outside their outburst.

In this study, we explore the discovery space of BeXRBs for data from the first all-sky survey by eROSITA (eRASS1) on-board Spectrum-Roentgen-Gamma (SRG) \citep{Merloni2024AnA}. We compare these results against the traditional form of studying BeXRBs with long dedicated pointed observations made by instruments such as \textit{BeppoSAX} \citep{Boella_1997AnAS_BeppoSAX}, \textit{Chandra} \citep{Weisskopf_2000SPIE_chandrapaper}, or \textit{XMM-Newton} \citep{Jansen_2001AnA_XMM_newton}. eROSITA as a survey mission, allows investigations of nearby sources in one hemisphere of the sky. Other long-term survey instruments, like all-sky monitors, such as the MAXI mission \citep{Matsuoka2009PASJ_MAXI}, the Swift/BAT \citep{ Barthelmy2005SSRv_SWIFT_BAT, Krimm2013ApJS_SWIFT_BAT}, and RXTE$/$ASM \citep{Levine_1996ApJ_RXTE_ASM} do not have the sensitivity to detect BeXRBs in quiescence.

In Section \ref{sec:observation}, we discuss briefly about the eROSITA X-ray observatory and its data products used for the X-ray sources. In Section \ref{sec:results}, we present our spectral analysis of BeXRB sources that were observed in their quiescence state and search for any relation between spectral parameters for different models. We provide upper limits where sources are not detected. This is followed by a survey of previous reports on BeXRBs in quiescence from different X-ray observatories to compare them against the eRASS1 data. In \text{Section \ref{sec:discussion}}, we discuss the spectral models used for fitting and their physical significance. We then compare the detection space of BeXRBs for the eRASS1 survey to previous studies, and also possible low-luminosity accretion models for different \text{X-ray} sources.

\begin{table*}
\caption{List of the BeXRB sources and their properties. Best-fit spectral parameters for the \text{0.5$-$8 keV} eRASS1 data fitted for the two spectral models, i.e., i) \texttt{tbabs} $\times$ \texttt{powerlaw}, and ii)  \texttt{tbabs} $\times$ \texttt{bbodyrad}. The upper-limits on flux has a one-sided confidence level of 95$\%$.    }
\label{tab:spectral_model_fit}
\renewcommand{\arraystretch}{2.5}
\centering
\tiny
\setlength\tabcolsep{3pt}
\begin{tabular}{l|l|cccc|cccccccc}
\hline
   &                        &                       & {\scriptsize Source Properties}  &            &               &               &                               &                           &   & {\scriptsize Spectral Fitting}             \\  \hline
   & Source                 & Distance              & Orbital               & Spin Period           & Cyclotron     & Exposure      & N$_H$                         & $\Gamma$                  & kT$_{bb}$                 & R$_{BB}$                  & Observed Flux                         & Luminosity                    & c-stat        \\
   &                        & (pc)                  & Period                & (s)                   & Line          & (seconds)     & ($10^{22}$ cm$^{-2}$)         &                           & (KeV)                     & (km)                      & $(0.5-10\text{ keV})$                  & $(0.5-10\text{ keV})$         & (dof)         \\
   &                        &                       & (d)                   &                       & (keV)         &               &                               &                           &                           &                           & ($10^{-12}$ erg s$^{-1}$ cm$^{-2}$)   & ($10^{34}$ erg s$^{-1}$)    & \\  \hline
1  & 1A 0535$+$262          & $1790^{+76}_{-71}$    & $110.3\pm0.3$ [1]     & 103.41 [2]            & 45.5 [3]      & 213.3         & $0.71\pm0.08$                 & $0.8\pm0.1$               & $-$                       & $-$                       & $293^{+24}_{-25}$                     & $11.3^{+2.0}_{-1.8}$          & 408(462)      \\
   &                        &                       &                       &                       &               &               & $0.27\pm0.05$                 & $-$                       & $1.25^{+0.07}_{-0.06}$    & $0.51^{+0.06}_{-0.05}$    & $198^{+14}_{-17}$                     & $7.6^{+1.3}_{-1.2}$           & 438.89(462)   \\  \hline
2  & 2S 1417$-$624          & $7400^{+3200}_{-1800}$ & $42.12\pm0.03$ [4]   & 17.64 [5]         & 100 [6]  & 278.9         & $3.5^{+3.1}_{-2.2}$           & $1.2^{+1.7}_{-1.4}$       & $-$                       & $-$                       & $<9.9$                                & $<13.1$                       & 21.94(23)     \\
   &                        &                       &                       &                       &               &               & $2.3^{+2.0}_{-1.4}$           & $-$                       & $1.2^{+1.4}_{-0.4}$       & $<1.2$                    & $<5.9$                                & $<7.8$                        & 22.11(23)     \\  \hline
3  & 4U 0728$-$25           & $7900^{+1200}_{-1100}$ & $34.55\pm0.01$ [7]  & 103.145 [8]          & $-$           & 219.2         & $0.7^{+0.2}_{-0.1}$           & $1.3^{+0.2}_{-0.2}$       & $-$                       & $-$                       & $49.5^{+9.3}_{-8.5}$                  & $37^{+21}_{-14}$        & 184.66(238)   \\
   &                        &                       &                       &                       &               &               & $0.23\pm0.08$                 & $-$                       & $0.94^{+0.09}_{-0.08}$    & $1.6^{+0.5}_{-0.4}$       & $32.2^{+3.6}_{-4.9}$                  & $24^{+11}_{-9}$         & 203.19(238)   \\  \hline
4  & 4U 1036$-$56           & $4990^{+400}_{-320}$  & $61.0\pm0.2$ [9]     & 854.75 [10]           & $-$           & 446.4         & $0.8\pm0.2$                   & $0.5^{+0.3}_{-0.3}$       & $-$                       & $-$                       & $30.4^{+6.3}_{-6.2}$                  & $9.1^{+3.7}_{-2.7}$           & 127.72(176)   \\
   &                        &                       &                       &                       &               &               & $0.35^{+0.16}_{-0.14}$        & $-$                       & $1.5^{+0.3}_{-0.2}$       & $0.32^{+0.09}_{-0.07}$    & $21.0^{+3.4}_{-5.9}$                  & $6.3^{+2.2}_{-2.3}$           & 133.90(176)   \\  \hline
5  & 4U 1145$-$61           & $2063^{+81}_{-87}$    & $187.5\pm1.0$ [11]    & 292 [11,12]           & $-$           & 573.4         & $0.24^{+0.07}_{-0.06}$        & $1.2^{+0.2}_{-0.2}$       & $-$                       & $-$                       & $17.3^{+3.3}_{-2.9}$                  & $0.9^{+0.3}_{-0.2}$           & 280.20(257)   \\
   &                        &                       &                       &                       &               &               & $<0.03$                       & $-$                       & $0.80\pm0.06$             & $0.30^{+0.05}_{-0.04}$    & $9.3^{+0.9}_{-1.2}$                   & $0.47^{+0.09}_{-0.10}$        & 281.61(257)   \\  \hline
6  & GRO J1008$-$57         & $3540\pm150$          & $248.9\pm0.5$ [13]     & 93.5 [14]              & 90.32 [15]    & 436.1         & $1.9^{+0.9}_{-0.7}$           & $1.4^{+0.8}_{-0.7}$       & $-$                       & $-$                       & $6.6^{+2.1}_{-3.4}$                   & $1.0^{+0.4}_{-0.6}$           & 55.81(62)     \\
   &                        &                       &                       &                       &               &               & $0.9^{+0.6}_{-0.4}$           & $-$                       & $1.0^{+0.4}_{-0.2}$       & $0.24^{+0.15}_{-0.09}$    & $4.5^{+1.0}_{-2.9}$                   & $0.7^{+0.2}_{-0.5}$           & 59.49(62)     \\  \hline
7  & GX 304$-$1             & $1856^{+37}_{-46}$    & $132.189\pm0.02$ [16] & 272 [17]              & 52 [18]       & 473.7         & $1.8^{+0.6}_{-0.5}$           & $1.3\pm0.4$               & $-$                       & $-$                       & $16.2^{+3.7}_{-4.7}$                  & $0.7\pm0.2$                   & 146.50(140)   \\
   &                        &                       &                       &                       &               &               & $0.9^{+0.3}_{-0.4}$           & $-$                       & $1.1\pm0.2$               & $0.16^{+0.05}_{-0.04}$    & $11.3^{+1.9}_{-3.9}$                  & $0.5^{+0.1}_{-0.2}$           & 142.64(140)   \\  \hline
8  & IGR J06074$+$2205      & $5989^{+560}_{-600}$  & $-$                   & 373.226 [19]          & 51 [20]       & 139.8         & $<2.4$                        & $0.7^{+1.6}_{-1.3}$       & $-$                       & $-$                       & $<12.6$                               & $<6.4$                        & 15.51(15)     \\
   &                        &                       &                       &                       &               &               & $<1.4$                        & $-$                       & $1.3^{+3.1}_{-0.6}$       & $<0.7$                    & $<5.6$                                & $<2.9$                        & 16.03(15)     \\  \hline
9  & MAXI J0655$-$013       & $3620^{+280}_{-220}$  & $27.9\pm1.7$ [21]     & 1129.1 [22]           & $-$           & 161.8         & 0.555$^\star$                 & $0.9\pm1.1$               & $-$                       & $-$                       & $4.3^{+6.8}_{-2.7}$                   & $0.7^{+1.3}_{-0.5}$           & 8.61(13)      \\
   &                        &                       &                       &                       &               &               & 0.555$^\star$                 & $-$                       & $0.8^{+1.6}_{-0.3}$       & $<0.5$                    & $<2.4$                                & $<0.5$                        & 11.46(13)     \\  \hline
10 & MXB 0656$-$072         & $5720\pm480$          & $101.2$ [23]          & 160.7 [24]            & 32.8 [25]     & 176.7         & $<3.5$                        & $0.8^{+2.2}_{-1.5}$       & $-$                       & $-$                       & $<9.2$                                & $<4.2$                        & 22.02(10)     \\
   &                        &                       &                       &                       &               &               & $<2.0$                        & $-$                       & $1.1^{+3.0}_{-0.5}$       & $<0.9$                    & $<3.3$                                & $<1.5$                        & 21.49(10)     \\  \hline
11 & 1A 1118$-$61           & $2900^{+86}_{-76}$    & $24.0\pm0.4$ [26]     & 405.3 [27]            & 55 [28]       & 560.6         & $1.7^{+0.8}_{-0.6}$           & $1.6^{+0.8}_{-0.7}$       & $-$                       & $-$                       & $3.5^{+1.3}_{-2.1}$                   & $0.35^{+0.16}_{-0.20}$        & 50.48(59)     \\
   &                        &                       &                       &                       &               &               & $0.8^{+0.5}_{-0.4}$           & $-$                       & $0.9^{+0.3}_{-0.2}$       & $0.2\pm0.1$               & $2.4^{+0.5}_{-1.8}$                   & $0.24^{+0.07}_{-0.18}$        & 51.31(59)     \\  \hline
12 & IGR J08262$-$3736      & $5120^{+280}_{-230}$  & $-$                   & $-$                   & $-$           & 273.3         & $0.7^{+0.6}_{-0.5}$           & $1.3^{+0.8}_{-0.8}$       & $-$                       & $-$                       & $4.8^{+2.4}_{-2.9}$                   & $1.5^{+1.0}_{-1.0}$           & 36.71(43)     \\
   &                        &                       &                       &                       &               &               & $<0.6$                        & $-$                       & $1.0^{+0.4}_{-0.3}$       & $0.3^{+0.2}_{-0.1}$       & $3.0^{+0.8}_{-2.5}$                   & $0.9^{+0.4}_{-0.8}$           & 37.92(43)     \\  \hline
13 & IGR J11435$-$6109      & $7900^{+1400}_{-1200}$ & $52.46\pm0.06$ [29]  & 161.76 [30]           & $-$           & 505.7         & $1.6^{+1.6}_{-1.0}$           & $1.6^{+1.7}_{-1.3}$       & $-$                       & $-$                       & $<2.7$                                & $<2.7$                        & 16.66(24)     \\
   &                        &                       &                       &                       &               &               & $0.73^{+0.99}_{-0.67}$        & $-$                       & $0.9^{+0.6}_{-0.3}$       & $<1.0$                    & $<1.4$                                & $<1.4$                        & 17.46(24)     \\  \hline
14 & IGR J14331$-$6112      & $2550^{+610}_{-570}$  & $-$                   & $-$                   & $-$           & 256.7         & $0.9^{+0.9}_{-0.7}$           & $0.6^{+1.0}_{-0.7}$       & $-$                       & $-$                       & $<10.3$                               & $<1.2$                        & 13.95(27)     \\
   &                        &                       &                       &                       &               &               & $<1.0$                        & $-$                       & $1.4^{+1.2}_{-0.5}$       & $<0.2$                    & $<5.8$                                & $<0.7$                        & 15.16(27)     \\  \hline
\end{tabular}
\end{table*}

\begin{table*}
\ContinuedFloat
\caption{List of the BeXRB sources and their properties. Best-fit spectral parameters for the \text{0.5$-$8 keV} eRASS1 data fitted for the two spectral models, i.e., i) \texttt{tbabs} $\times$ \texttt{powerlaw}, and ii)  \texttt{tbabs} $\times$ \texttt{bbodyrad}. The upper-limits on flux has a one-sided confidence level of 95$\%$. (continued) }
\label{tab:spectral_model_fit_2}
\renewcommand{\arraystretch}{2.5}
\centering
\tiny
\setlength\tabcolsep{5pt}
\begin{tabular}{l|l|cccc|cccccccc}
\hline
   & Source                 & Distance              & Orbital               & Spin Period           & Cyclotron     & Exposure      & N$_H$                         & $\Gamma$                  & kT$_{bb}$                 & R$_{BB}$                  & Observed Flux                         & Luminosity                    & c-stat        \\
   &                        & (pc)                  & Period                & (s)                   & Line          & (seconds)     & ($10^{22}$ cm$^{-2}$)        &                           & (KeV)                     & (km)                      & $(0.5-10\text{ keV})$                  & $(0.5-10\text{ keV})$         & (dof)         \\
   &                        &                       & (d)                   &                       & (keV)         &               &                               &                           &                           &                           & ($10^{-12}$ erg s$^{-1}$ cm$^{-2}$)   & ($10^{34}$ erg s$^{-1}$)    & \\  \hline
15 & IGR J17200$-$3116      & 5000 [31]             & $-$                   & $\sim328$ [32]            & $-$       & 212.3         & $2.7^{+1.5}_{-1.1}$           & $1.3^{+0.9}_{-0.8}$       & $-$                       & $-$                       & $18^{+5}_{-16}$                 & $5.2^{+1.5}_{-4.7}$           & 56.38(55)     \\
   &                        &                       &                       &                           &           &               & $1.5^{+0.9}_{-0.7}$           & $-$                       & $1.2^{+0.5}_{-0.3}$       & $0.4^{+0.3}_{-0.1}$       & $12.0^{+2.5}_{-9.2}$                  & $3.6^{+0.7}_{-2.7}$           & 57.03(55)     \\  \hline
16 & SRGA J124404.1         & $5810^{+760}_{-540}$  & $\sim138$ [34]        & $538.70\pm0.05$           & $-$       & 494.1         & $4.4^{+4.2}_{-3.1}$           & $2.0^{+2.1}_{-1.7}$       & $-$                       & $-$                       & $<4.2$                                & $<2.2$                        & 29.52(28)     \\
   & $-$632232              & [33]        &                       & [33]                      &           &               & $2.5^{+2.7}_{-1.8}$           & $-$                       & $0.9^{+0.9}_{-0.3}$       & $\lesssim1$               & $<2.7$                                & $<1.4$                        & 28.12(28)     \\  \hline
17 & IGR J10101$-$5654      & $4500^{+1100}_{-1200}$ & $-$                  & $-$                       & $-$       & 431.9         & $1.9^{+2.3}_{-1.5}$           & $0.0^{+1.4}_{-1.1}$       & $-$                       & $-$                       & $<12.3$                               & $<4.6$                        & 18.22(21)     \\
   &                        &                       &                       &                           &           &               & $1.3^{+1.5}_{-1.0}$           & $-$                       & $<2.5$                    & $<0.3$                    & $<8.4$                                & $<3.1$                        & 18.81(21)     \\  \hline
\end{tabular}
\justifying
\tablecomments{ \scriptsize \\
{$\star$: The parameter was frozen.} \\
{The distance values for sources with known Gaia source identifiers are taken from Gaia early data release 3 (EDR3) distance catalogue $\citep{gaia_edr3_distance}$. For some sources, we use the distance values from the literature.} \\
{
References: [1] \cite{Finger_1994AIPC}; [2] \cite{Mukherjee2005aanda_3a0535p262}; [3] \cite{Terada2006ApJ_3a0535p262}; [4] \cite{Finger_1996AandAS}; [5] \cite{Kelley1981ApJ_2S_1417m624}; [6] \cite{Liu2024aanda_2S_1417m624}; [7] \cite{Corbet_2016ATel}; [8] \cite{Roy2020RAA_3A_0726m260}; [9] \cite{Cusumano2013MNRAS}; [10] \cite{Li2012ApJ_4U_1036m56}; [11] \cite{Watson1981MNRAS}; [12] \cite{Nagase1989PASJ}; [13] \cite{Levine_n_Corbet2006}; [14] \cite{Wilson1994AIPC_GRO_J1008m57}; [15] \cite{Ge2020ApJ_GRO_J1008m57}; [16] \cite{Sugizaki2015PASJ}; [17] \cite{McClintock1977ApJ_gx_304m1}; [18] \cite{Yamamoto2011PASJ_gx_304m1}; [19] \cite{Reig2018aanda_IGR_J06074p2205}; [20] \cite{Roy2024aanda_IGR_J06074p2205}; [21] \cite{Rai2023MNRAS}; [22] \cite{Shidatsu2022ATel15495_MAXI_J0655m013}; [23]  \cite{Yan2012ApJ}; [24] \cite{Morgan2003ATel_MXB_0656m072}; [25] \cite{McBride2006aanda_MXB_0656m072}; [26] \cite{Staubert2011aanda}; [27] \cite{Ives1975Natur_1A_1118m615}; [28] \cite{Doroshenko2010aanda_1A_1118m61}; [29] \cite{Corbet2005ATel}; [30] \cite{2004ATel_IGR_J11435M6109}; [31] \cite{Esposito2014MNRAS_IGR_J17200m3116}; [32] \cite{Nichelli_2011ATel_IGR_J17200m3116}; [33] \cite{Doroshenko2022AnA_SRGA_J124404m632232}; [34] \cite{Mroz_2021ATel14361_SRGA_J1244041m632232};
}   \\
}
\end{table*}

\begin{table*}
\caption{List of undetected X-ray sources in the eRASS1 survey and their properties. Flux and luminosity upper-limit of X-ray sources from their coordinates for an assumed spectral shape according to the method given in \cite{Tubin_2024AnA_erosita_upperlimit}. For these sources, the upper-limit on flux has a one-sided confidence level of 99.87$\%$.}
\label{tab:luminosity_upper_limit}
\renewcommand{\arraystretch}{2.0}
\centering
\tiny
\setlength\tabcolsep{5pt}
\begin{tabular}{l|l|cccc|ccccccc}
\hline
   &                            &                               & {\scriptsize Source Properties}   &       &               &           &                                       &                           \\  \hline
   & Source                     & Distance                      & Orbital               & Spin Period       & Cyclotron     & Exposure  & Observed Flux                         & Luminosity                \\
   &                            & (pc)                          & Period                & (s)               & Line          & (seconds) & $(0.5-10\text{ keV})$                 & $(0.5-10\text{ keV})$     \\
   &                            &                               & (d)                   &                   & (keV)         &           & ($10^{-12}$ erg s$^{-1}$ cm$^{-2}$)   & ($10^{34}$ erg s$^{-1}$)  \\  \hline
1   & MAXI J1409$-$619          & $14.5\times10^{3}$ [1]        & 14.7 [2]              & $503\pm10$ [3]    & 44 [1]        & 137.1     & $<0.55$                               & $<1.37$       \\  \hline
2   & RX J0812.4$-$3114         & $6660^{+530 }_{-420}$         & 81.3 [4]              & 31.8851 [5]       & $-$           & 120.6     & $<0.64$                               & $<0.40$       \\
    & $/$ LS 992                &                               &                       &                   &               &           &                                       &               \\  \hline
3   & Swift J1626.6$-$5156      & $9100^{+2900}_{-1800}$        & 132.9 [6]             & 15.36 [7]         & {4.9} [8]     & 127.9     & $<0.83$                               & $<1.44$       \\  \hline
4   & IGR J11305$-$6256         & $1730^{+140}_{-110}$          & 120.83 [9]            & $-$               & $-$           & 258.9     & $<0.45$                               & $<0.019$      \\  \hline
5   & 2RXP J130159.6            & $6000^{+2400}_{-1300}$        & $-$                   & 643 [10]          & $-$           & 254.7     & $<0.74$                               & $<0.61$       \\  
    & $-$635806                 &                               &                       &                   &               &           &                                       &               \\  \hline
6   & AX J1700.2$-$4220         & $1496^{+50}_{-54}$            & 44 [11]               & 54 [11]           & $-$           & 111.2     & $<0.70$                               & $<0.02$       \\  \hline
7   & GS 0834$-$43              & $6700^{+2900}_{-1800}$        & $105.8\pm0.4$ [12]    & 12.293 [13]       & $-$           & 153.2     & $<0.76$                               & $<0.84$       \\  \hline
8   & 2S 1553$-$542             & $20000\pm4000$ [14]           & $30.6\pm2.2$ [14]     & 9.28 [15]     & $23.5\pm0.4$ [14] & 139.9     & $<0.33$                               & $<2.31$       \\  \hline
9   & SAX J0635.2$+$0533        & $6290^{+660}_{-490}$          & $11.2\pm0.5$ [16]     & $33.8\times10^{-3}$ [17]  & $-$   & 72.0      & $<0.53$                               & $<0.31$       \\  \hline
10  & HD 249179                 & $1600^{+96}_{-73}$            & $-$                   & $-$               & $-$           & 93.7      & $<0.75$                               & $<0.026$      \\  \hline
11  & HD 63666                  & $649\pm10$                    & $-$                   & $-$               & $-$           & 248.8     & $<0.22$                               & $<0.0011$     \\  \hline
12  & HR 4804                   & $209^{+1}_{-1}$               & $-$                   & $-$               & $-$           & 196.2     & $<0.24$                               & $<0.00015$    \\  
    & $/$ HD 109857             &                               &                       &                   &               &           &                                       &               \\  \hline
13  & IGR J16318$-$4848 $\dagger$   & $6700^{+4100}_{-2000}$    & $80.09\pm0.012$ [18, 19]  & $-$           & $-$           & 122.5     & $<0.46$                               & $<0.63$       \\  \hline
14  & IGR J16283$-$4838 $\dagger$   & $13.6-21.6\times10^3$ [20]    & $287.6\pm1.7$ [21]    & $-$           & $-$           & 123.4     & $<0.38$                               & $<2.13$       \\  \hline
15  & AX J163904$-$4642         & $10.6\times10^{3}$ [22]       & 4.24 [23]             & 908.79 [24]       & $29.3^{+1.1}_{-1.3}$ [25] & 120.5     & $<1.60$                   & $<2.15$       \\  \hline
16  & XTE J1543$-$568       & $10-26\times10^{3}$ [26]     & $75.56\pm0.25$ [26]   & $27.1225\pm0.0043$ [26]   & $-$   & 136.6         & $<0.28$               & $<2.28$  \\  \hline
17  & IGR J13186$-$6257     & $0.7-5.4\times10^{3}$ [27]   & $19.99\pm0.01$ [28]   & $-$                       & $-$   & 205.4         & $<0.60$               & $<0.21$  \\  \hline
\end{tabular}
\justifying
\tablecomments{ \scriptsize \\
{The distance values for sources with known Gaia source identifiers are taken from Gaia early data release 3 (EDR3) distance catalogue $\citep{gaia_edr3_distance}$. For some sources, we use the distance values from the literature.} \\
{ $\dagger$: These sources have an optical companion of spectral type - IGR J16283-4838: sgBe and IGR J16318-4848: B0$-$5 sgBe \citep{Kretschmar_2019NewAR_sgBe_2_sources}. } \\
{
References: [1] \cite{Orlandini2012ApJ}; [2] \cite{Donmez2020MNRAS_MAXI_J1409m619}; [3] \cite{Kennea_2010ATel_MAXI_J1409m619}; [4] \cite{Corbet_2000ApJ_RX_J08124m3114}; [5] \cite{Reig_1999MNRAS_RX_J08124m3114}; [6] \cite{Baykal_2010ApJ_Swift_J1626m5156}; [7] \cite{Staubert_2019AnA}; [8] \cite{Molkov_2021ApJ_Swift_J16266m5156_crsf}; [9] \cite{LaParola_2013arXiv_IGR_J11305m6256}; [10] \cite{Krivonos_2015ApJ_2RXPJ1301596m635806}; [11] \cite{Markwardt2010ATel_AXJ17002m4220}; [12] \cite{Wilson_1997ApJ_GS_0834m430}; [13] \cite{Miyasaka_2013ApJ_GS_0834m430}; [14] \cite{Tsygankov2016MNRAS}; [15] \cite{Malacaria_2022ApJ_2S_1553m542}; [16] \cite{Kaaret_2000ApJ_SAX_J0635p0533}; [17] \cite{Cusumano_2000ApJ_SAX_J0635p0533}; [18] \cite{Jain_2009RAA_IGR_J16318m4848}; [19] \cite{Iyer_2017MNRAS_IGR_J16318m4848}; [20] \cite{Pellizza_2011AnA_IGR_J16283m4838}; [21] \cite{Cusumano_2013ApJ_IGR_J16283m4838}; [22] \cite{Chaty2008AnA}; [23] \cite{Corbet_2010ATel_IGR_J16393m4643}; [24] \cite{Islam_2015MNRAS_IGR_J16393m4643}; [25] \cite{Bodaghee_2016ApJ_IGR_J16393m4643}; [26] \cite{intZand_2001ApJ_XTE_J1543m568_dist}; [27] \cite{Fortin_2018AnA_IGR_J13186m6257}; [28] \cite{DAI_2011AnA_J13186_6257};  \\}   \\
}
\end{table*}

\section{Observations} \label{sec:observation}

The SRG space observatory \citep{Sunyaev2021AnA} launched in 2019 orbits about the L2 Lagrange point of the Sun-Earth system. The SRG space observatory consists of two X-ray telescopes, i.e., the eROSITA X-ray telescope \citep{Predehl2021AnA} which consists of seven identical and co-aligned X-ray mirror assemblies of the Wolter-I geometry sensitive in the energy range of 0.2$-$8 keV, and the Mikhail Pavlinsky Astronomical Roentgen Telescope $-$ X-ray Concentrator (ART-XC) \citep{Pavlinsky_2021AnA_ARTmXC_paper}.

In this study, we use data from the first eROSITA survey eRASS1 which began in December 2019 and completed on June 2020 after six months. The first data release (DR1) consists of X-ray sources detected in the soft (0.2$-$2.3 keV) and hard (2.3$-$5 keV) energy bands placed into the eRASS1 main and eRASS1 hard catalogues \citep{Merloni2024AnA}, respectively.

The eROSITA instrument has an angular resolution with half-energy width of $30''$ \citep{Sunyaev2021AnA}. In this work, we use the data products from the eRASS1 main catalogue \citep{Merloni2024AnA} through the catalogue search page\footnote{\url{https://erosita.mpe.mpg.de/dr1/erodat/catalogue/search/}}. The data products were obtained for BeXRBs which were detected in the eRASS1 main catalogue within a angular distance of $\leq5''$. The BeXRB sources for this study are taken from the Galactic HMXB catalogues by \cite{Fortin_2023AnA__catalogue}\footnote{\url{https://binary-revolution.github.io/HMXBwebcat/}} and \cite{Neumann_2023AnA_catalogue}\footnote{\url{http://astro.uni-tuebingen.de/~xrbcat/HMXBcat.html}}. From these we have the eRASS1 data products available for 17 X-ray sources, for which the properties are given in ${\text{Table \ref{tab:spectral_model_fit}}}$. In this work, the source coordinates used are taken from the SIMBAD database \citep{Simbad_Wenger_2000AnA}.

SRG has a scanning strategy where it rotates to scan the sky with a period of 4 hours, and observes each point for upto 40 seconds in each rotation \citep{Merloni2024AnA}. The direction of the rotation axis shifts along the ecliptic by approximately one degree every day. Depending on the ecliptic coordinate of a source, six or more source exposures are obtained for each source over a duration of one or more days, respectively. Therefore, the observations are carried out within a span of 1$-$3 days. The sources for which the long-term lightcurves are available are plotted in \text{Figure \ref{fig:Outbursts}} of the appendix, and during the eROSITA observations, the BeXRB sources seem not to be in outburst.

\section{Data Analysis and Results} \label{sec:results}

The spectral analysis is done using XSPEC v12.14.0h \citep{Arnaud_1996ASPC_XSPEC}. The elemental abundance of the absorbing material in the line of sight used are from \cite{Wilms_2000ApJ_abund} and the photoionization cross-sections are taken from \cite{Verner_1996ApJ_crosssection}. The data products of the sources given in $\text{Table \ref{tab:spectral_model_fit}}$ are available through the catalogue search page and are downloaded for the 17 X-ray sources. These data products include the spectra and ligthcurves files for these sources. The spectrum obtained from these sources are fitted with two spectral models, ${\text{i) } \texttt{tbabs $\times$ powerlaw}}$, and ${\text{ii) } \texttt{tbabs $\times$ bbodyrad}}$ in the ${\text{0.2$-$8 keV}}$ energy band. We use \texttt{grppha}, an \texttt{FTOOLS} task \citep{Blackburn1999_FTOOLS} to rebin all the spectra such that each spectral bin has at least two counts. Since, the counts per bin is low for many of the spectra, we use cash-statistics (cstat) for fitting \citep{Cash_1979ApJ_fitting}. We have kept all the spectral parameters free while fitting for all the sources, except for MAXI J0655$-$013, for which, due to poor statistics the absorption column density, $N_{\text{H}}$, is fixed to the Galactic absorption column density along the line of sight \citep{HI4PI_2016AmA_NH}.

\begin{figure*}
    \centering
    \begin{subfigure}[b]{0.33\textwidth}
        \centering
        \includegraphics[width=\textwidth]{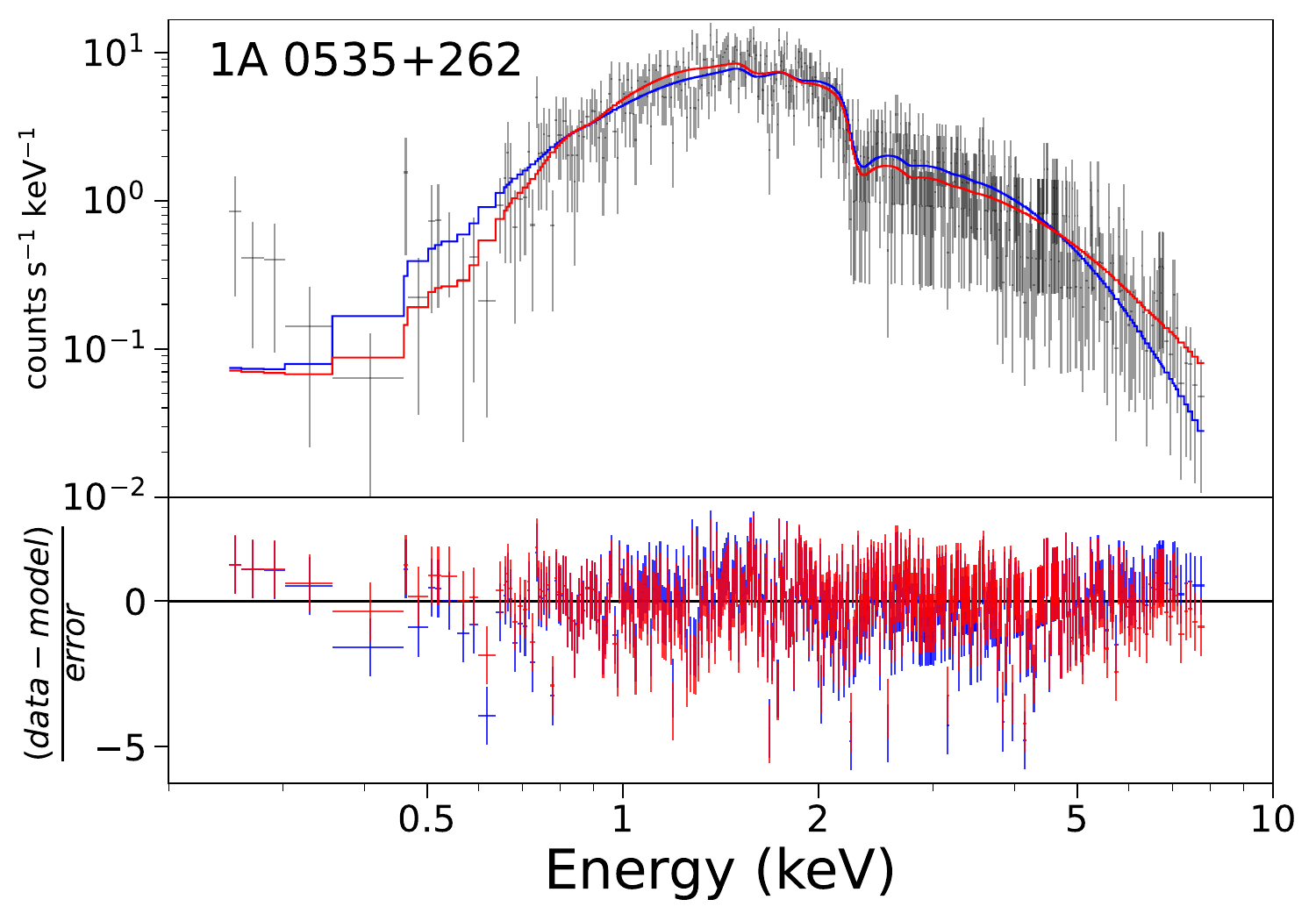}
        \caption{1A 0535$+$262}
    \end{subfigure}
    \begin{subfigure}[b]{0.33\textwidth}
        \centering
        \includegraphics[width=\textwidth]{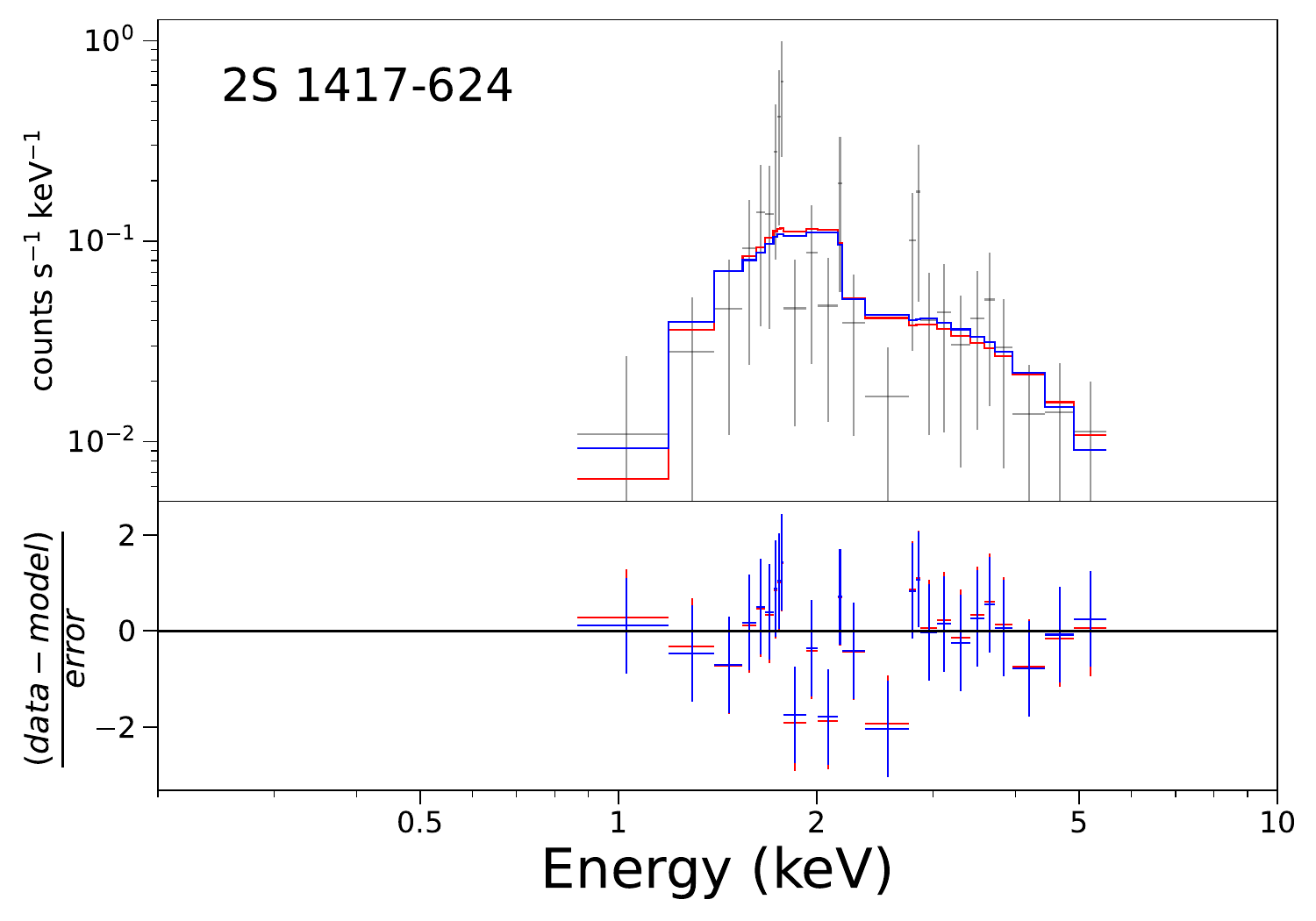}
        \caption{2S 1417$-$624}
    \end{subfigure}
    \begin{subfigure}[b]{0.33\textwidth}
        \centering
        \includegraphics[width=\textwidth]{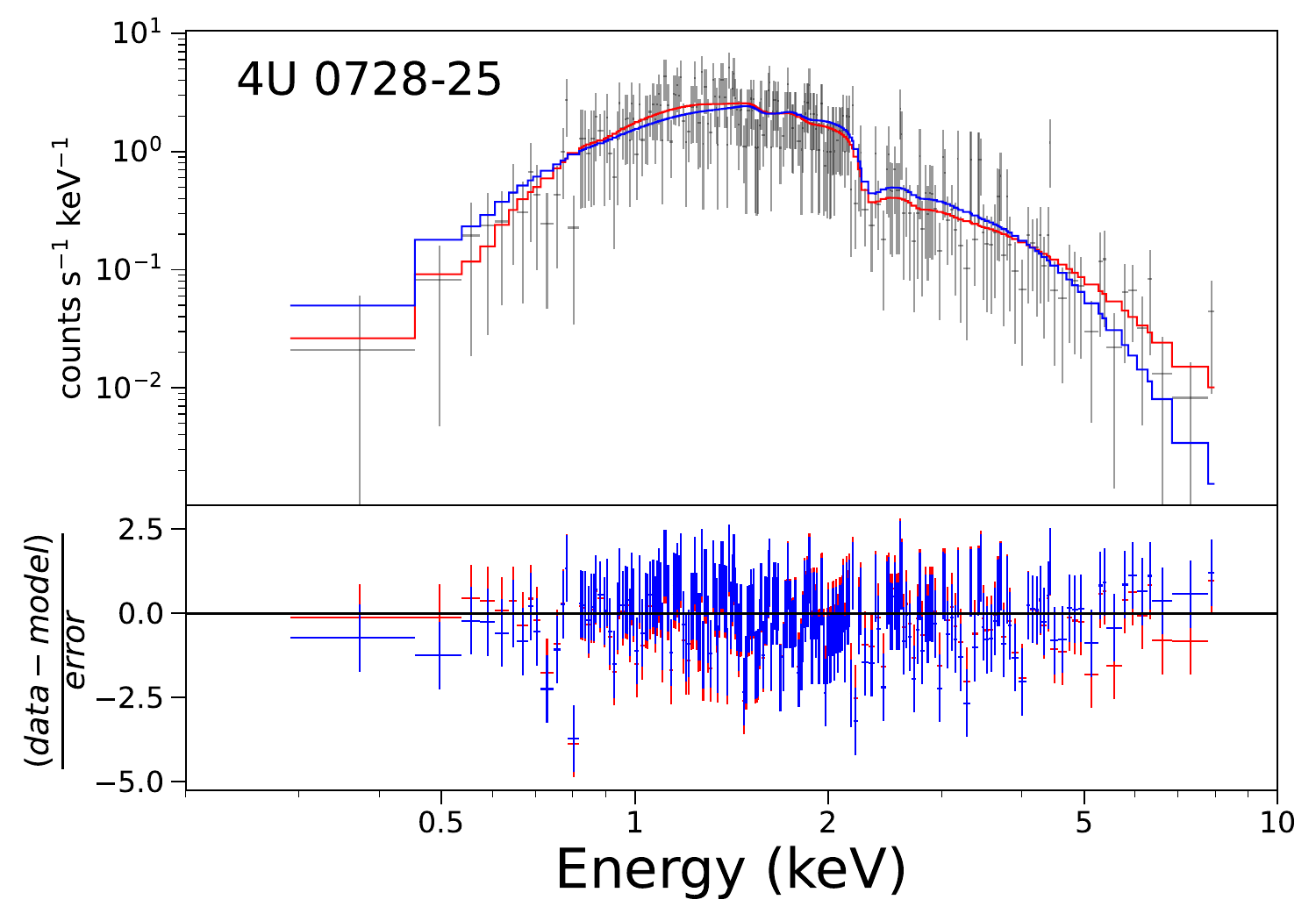}
        \caption{4U 0728$-$25}
    \end{subfigure}
    \begin{subfigure}[b]{0.33\textwidth}
        \centering
        \includegraphics[width=\textwidth]{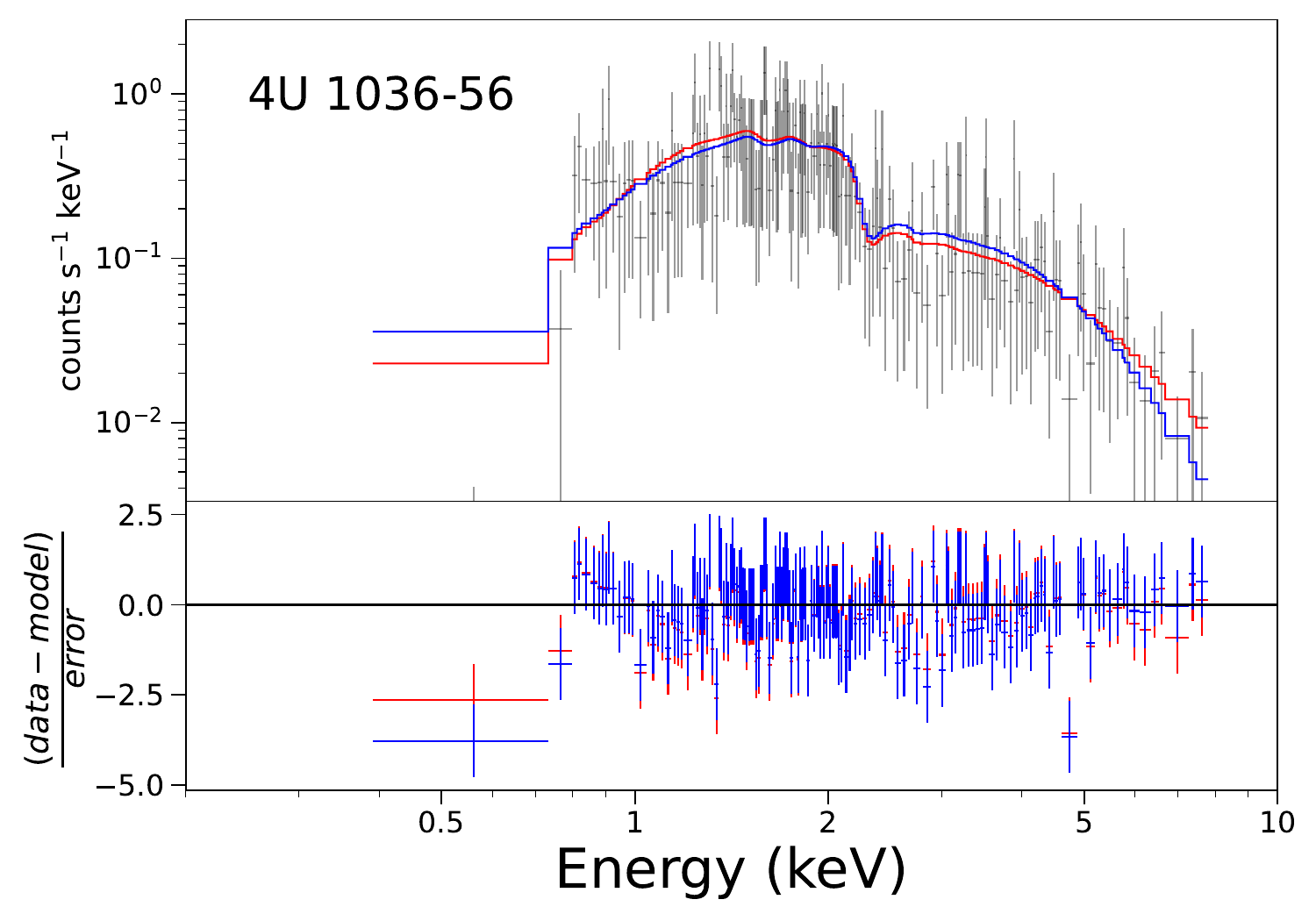}
        \caption{4U 1036$-$56}
    \end{subfigure}
    \begin{subfigure}[b]{0.33\textwidth}
        \centering
        \includegraphics[width=\textwidth]{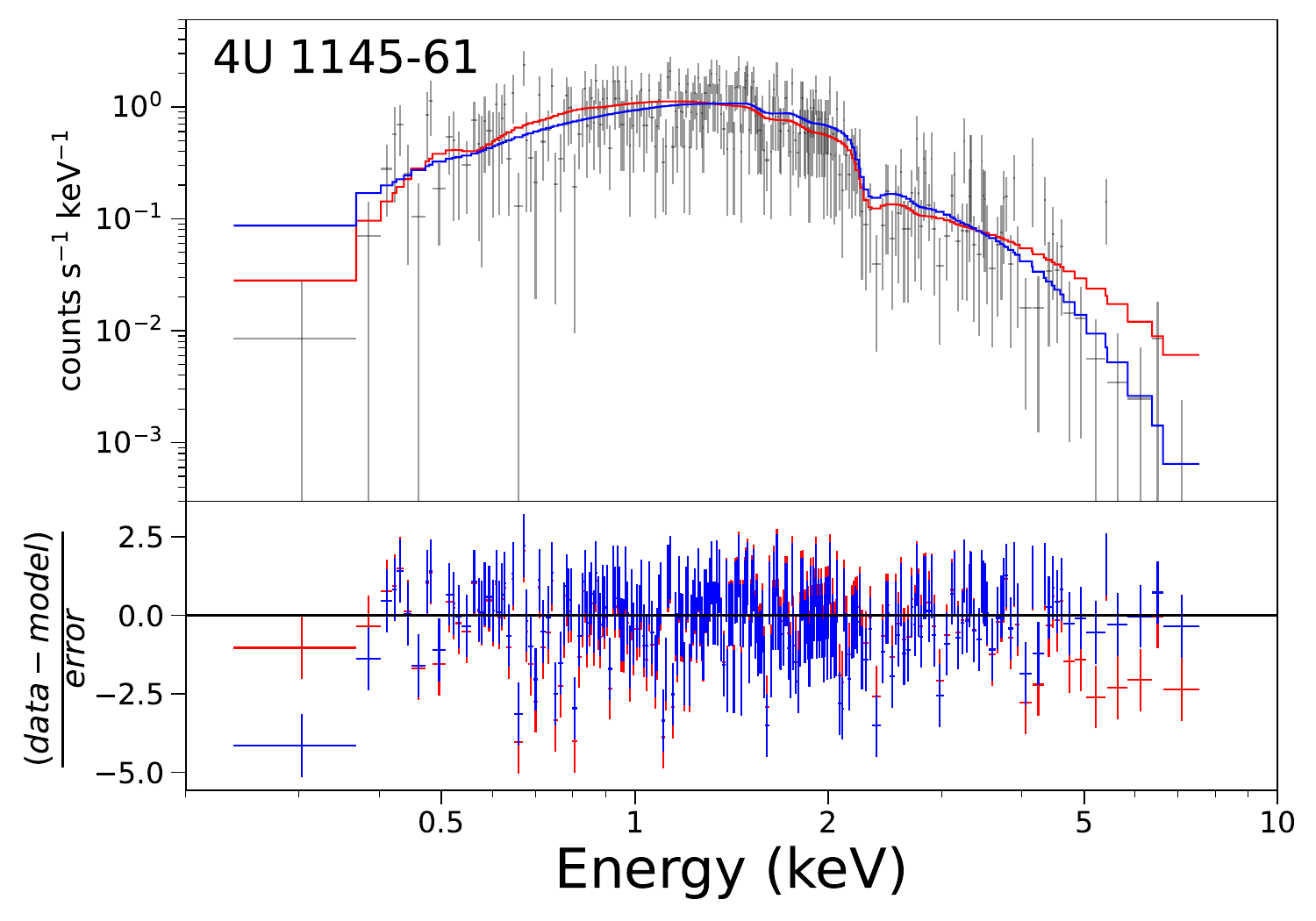}
        \caption{4U 1145$-$61}
    \end{subfigure}
    \begin{subfigure}[b]{0.33\textwidth}
        \centering
        \includegraphics[width=\textwidth]{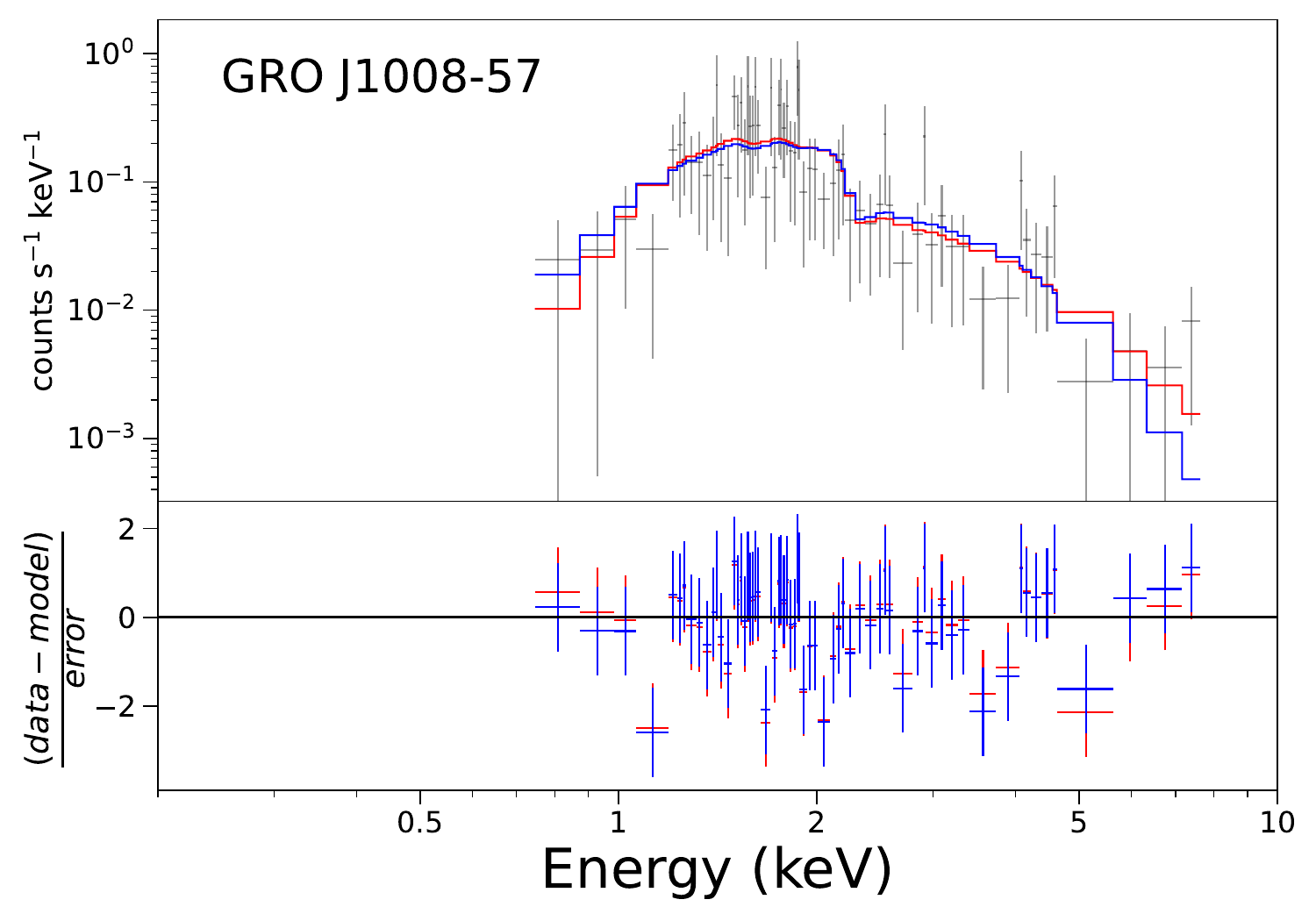}
        \caption{GRO J1008$-$57}
    \end{subfigure}
    \begin{subfigure}[b]{0.33\textwidth}
        \centering
        \includegraphics[width=\textwidth]{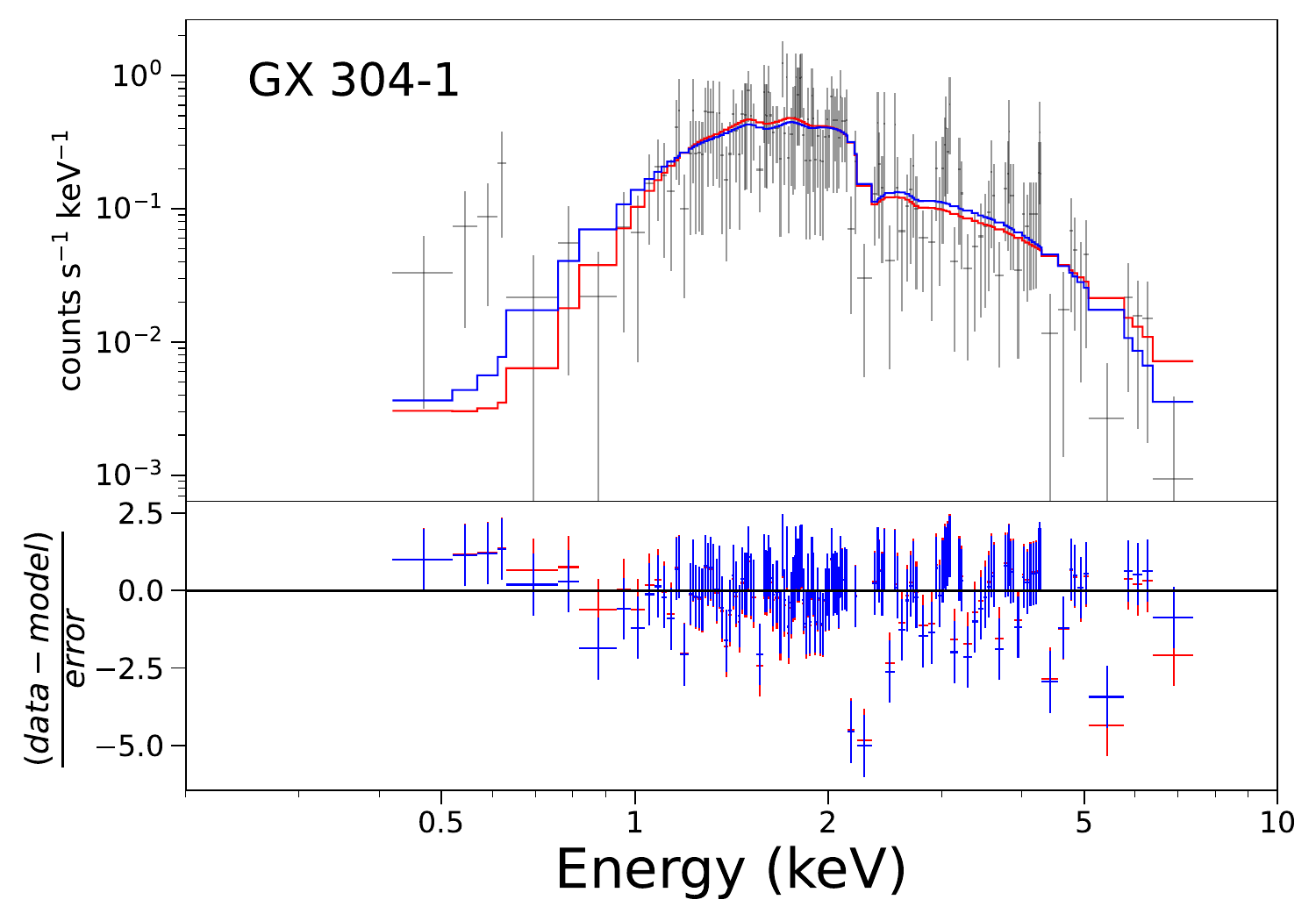}
        \caption{GX 304$-$1}
    \end{subfigure}
    \begin{subfigure}[b]{0.33\textwidth}
        \centering
        \includegraphics[width=\textwidth]{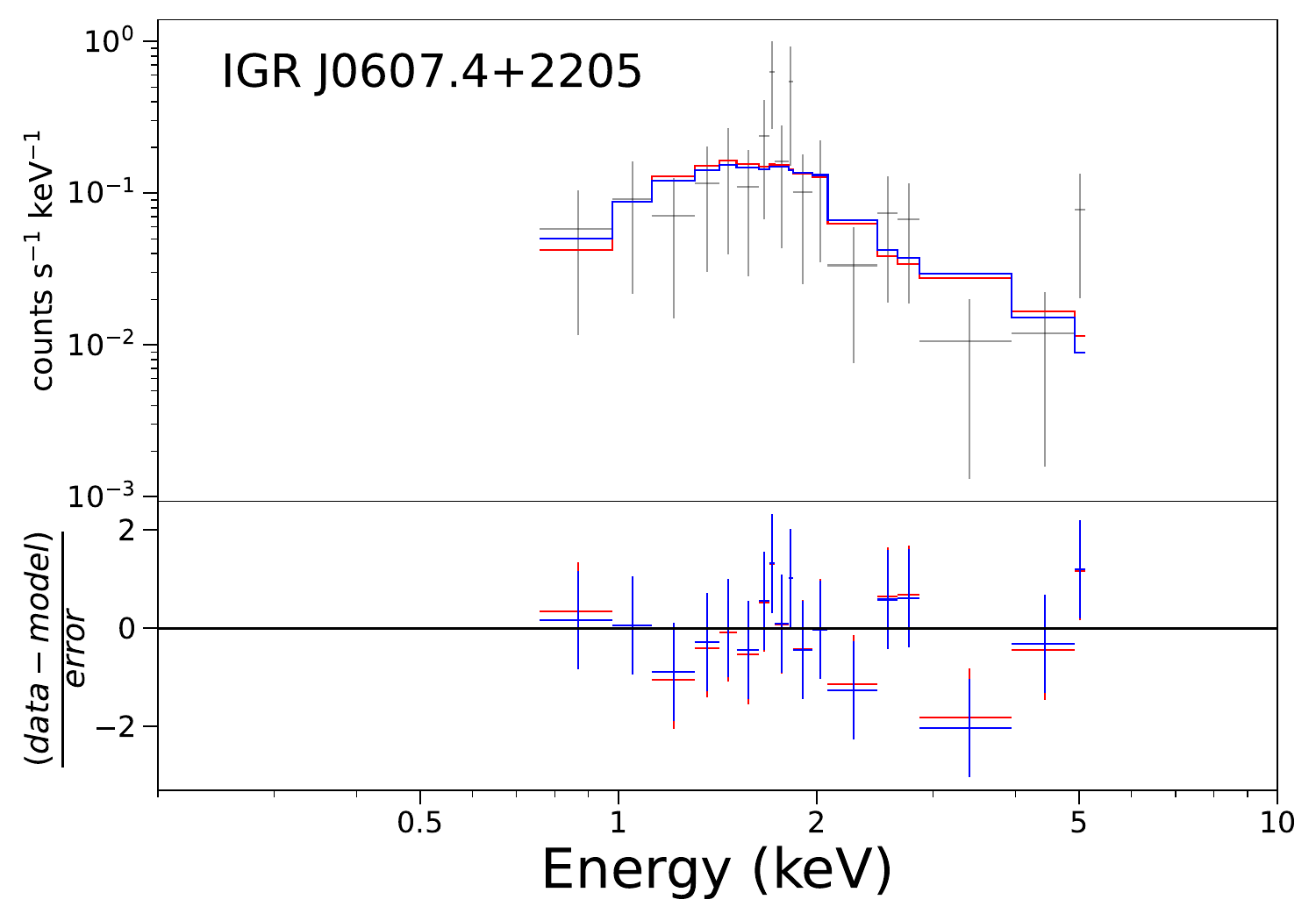}
        \caption{IGR J06074$+$2205}
    \end{subfigure}
    \begin{subfigure}[b]{0.33\textwidth}
        \centering
        \includegraphics[width=\textwidth]{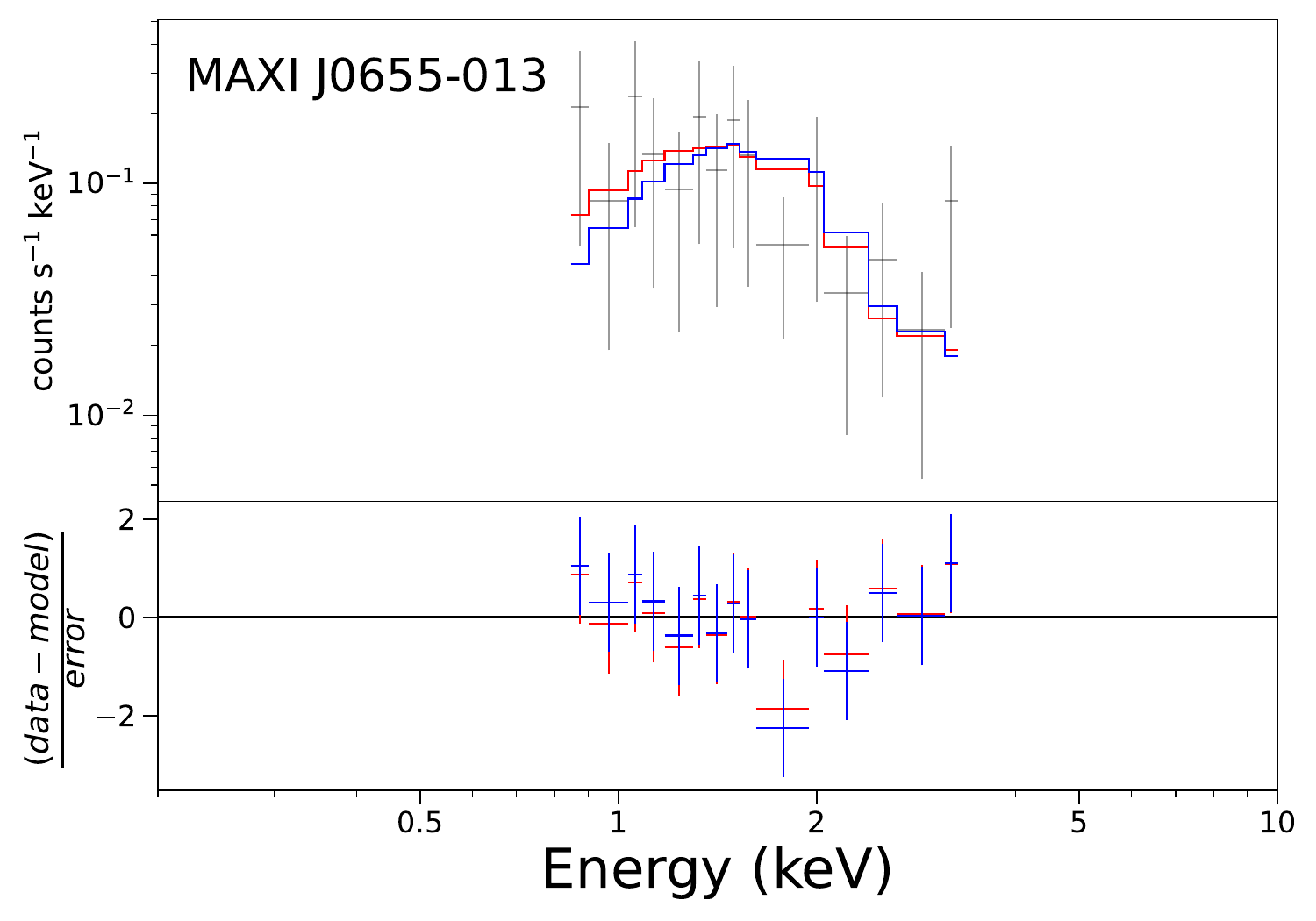}
        \caption{MAXI J0655$-$013}
    \end{subfigure}
    \begin{subfigure}[b]{0.33\textwidth}
        \centering
        \includegraphics[width=\textwidth]{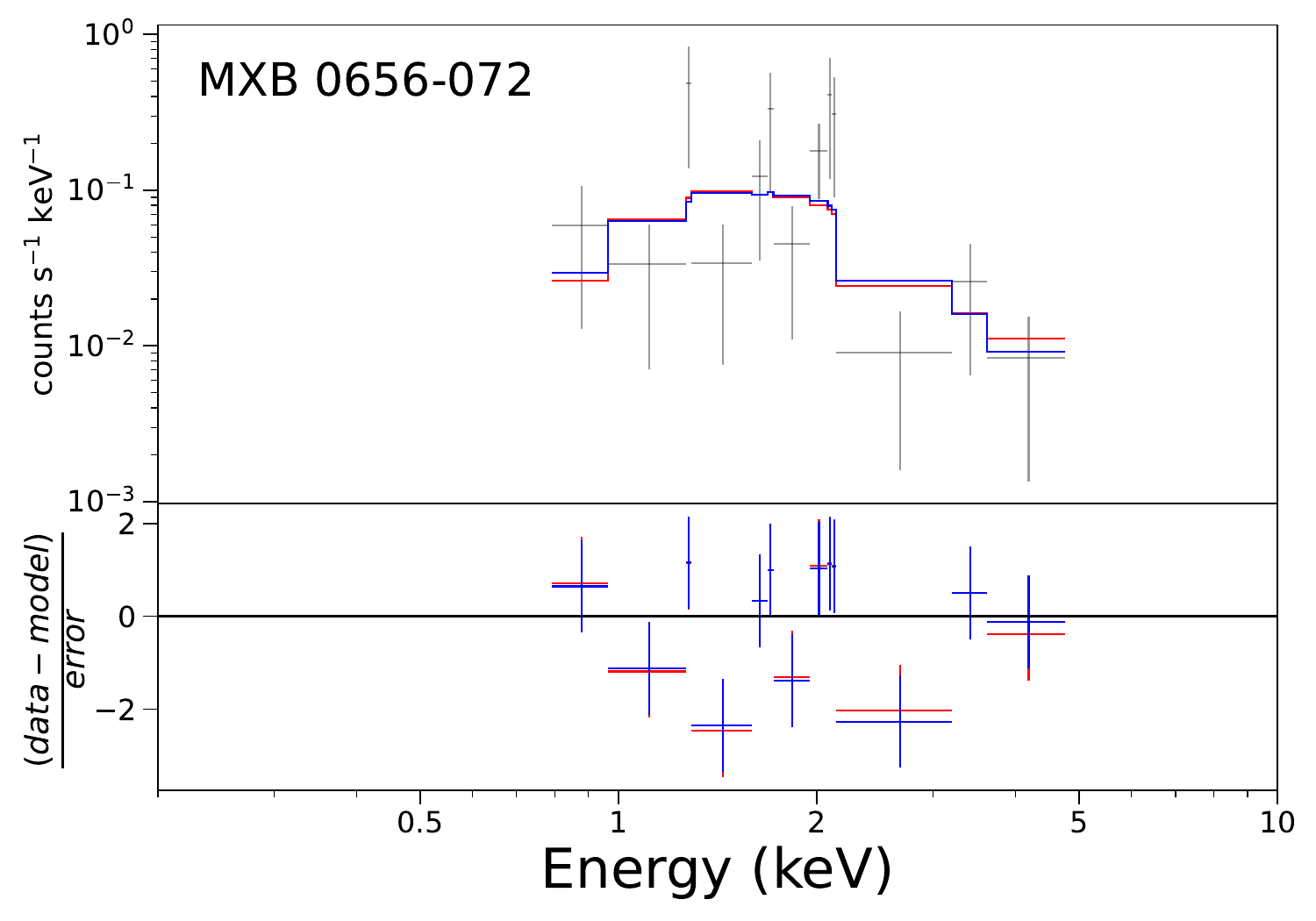}
        \caption{MXB 0656$-$072}
    \end{subfigure}
    \begin{subfigure}[b]{0.33\textwidth}
        \centering
        \includegraphics[width=\textwidth]{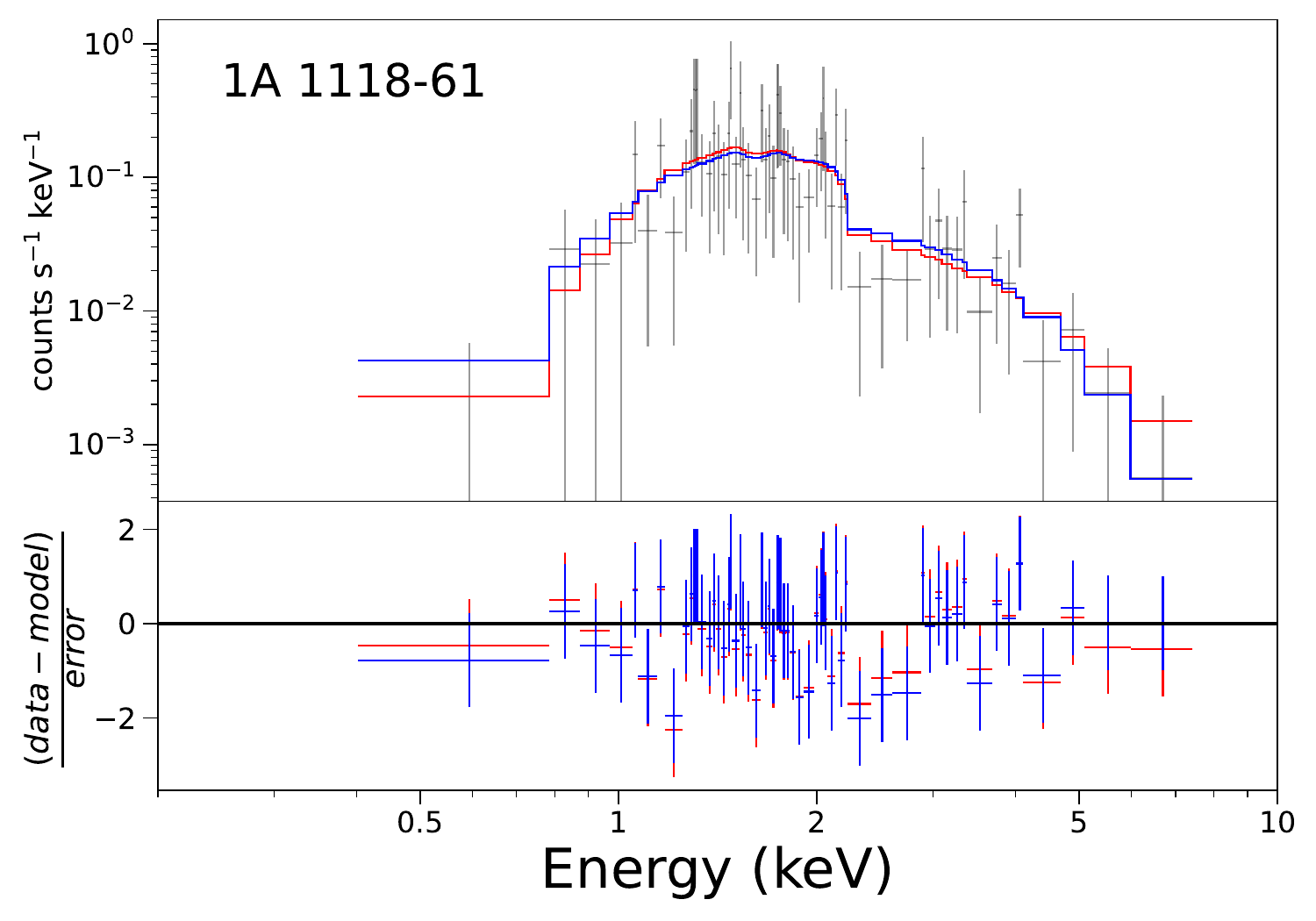}
        \caption{1A 1118$-$61}
    \end{subfigure}
    \begin{subfigure}[b]{0.33\textwidth}
        \centering
        \includegraphics[width=\textwidth]{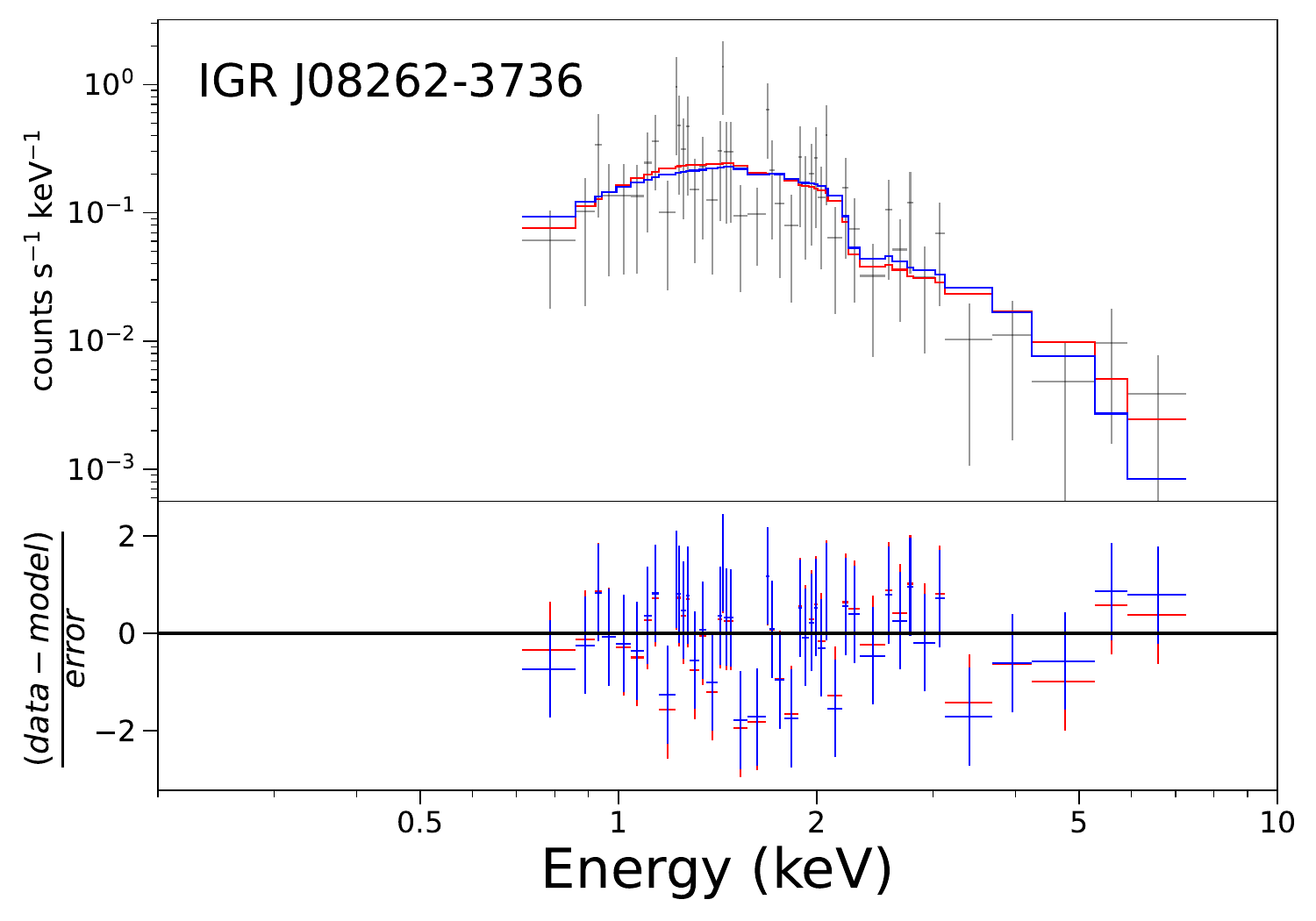}
        \caption{IGR J08262$-$3736}
    \end{subfigure}
    \caption{Best-fit spectra for BeXRB sources for two spectral models, i) \texttt{tbabs$\times$powerlaw}, and ii) \texttt{tbabs$\times$bbodyrad} where the model and its residual is plotted in red and blue, respectively. }
    \label{fig:spectral_best_fit}
\end{figure*}

\begin{figure*}
\ContinuedFloat
    \centering
    \begin{subfigure}[b]{0.33\textwidth}
        \centering
        \includegraphics[width=\textwidth]{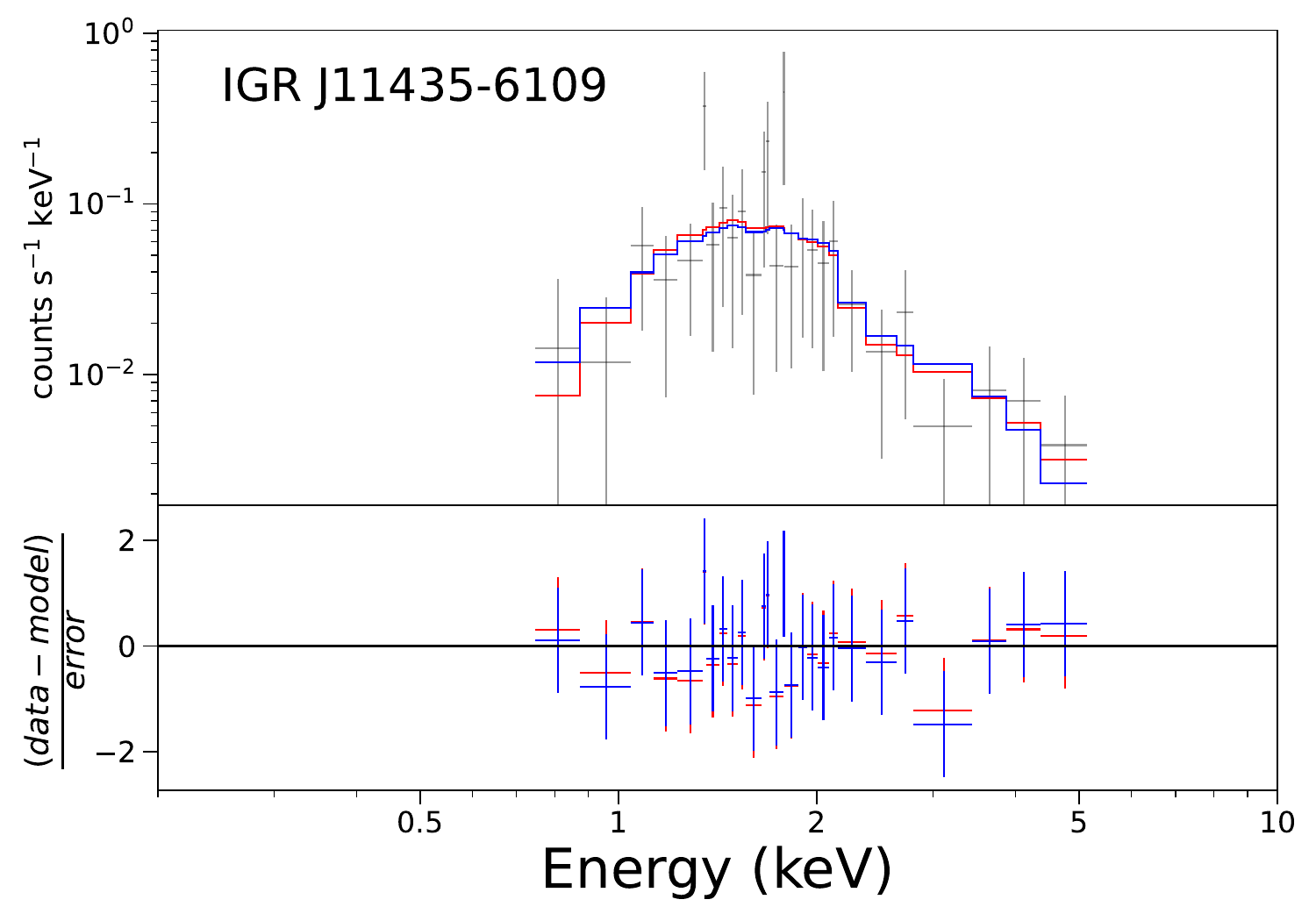}
        \caption{IGR J11435$-$6109}
    \end{subfigure}
    \begin{subfigure}[b]{0.33\textwidth}
        \centering
        \includegraphics[width=\textwidth]{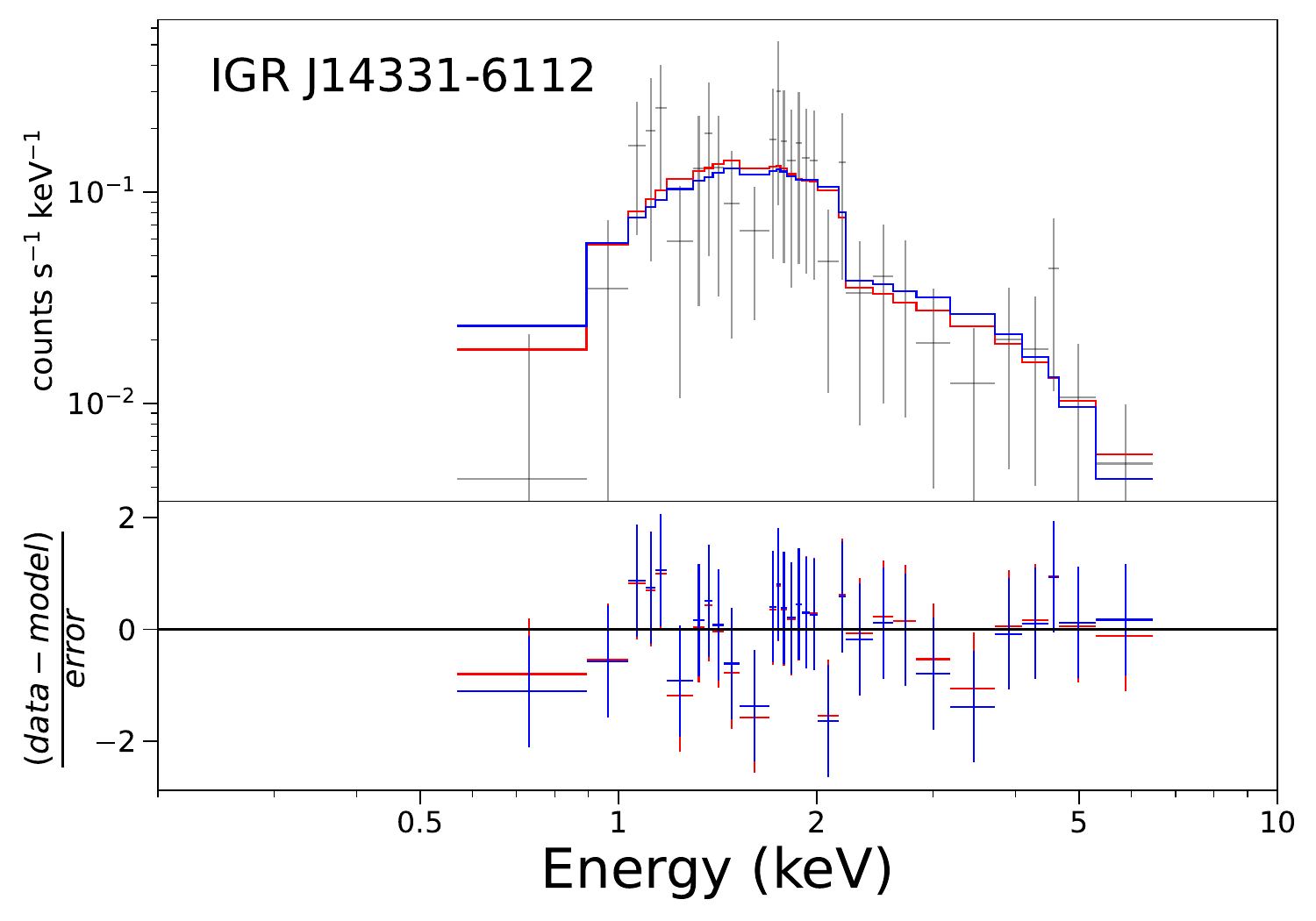}
        \caption{IGR J14331$-$6112}
    \end{subfigure}
    \begin{subfigure}[b]{0.33\textwidth}
        \centering
        \includegraphics[width=\textwidth]{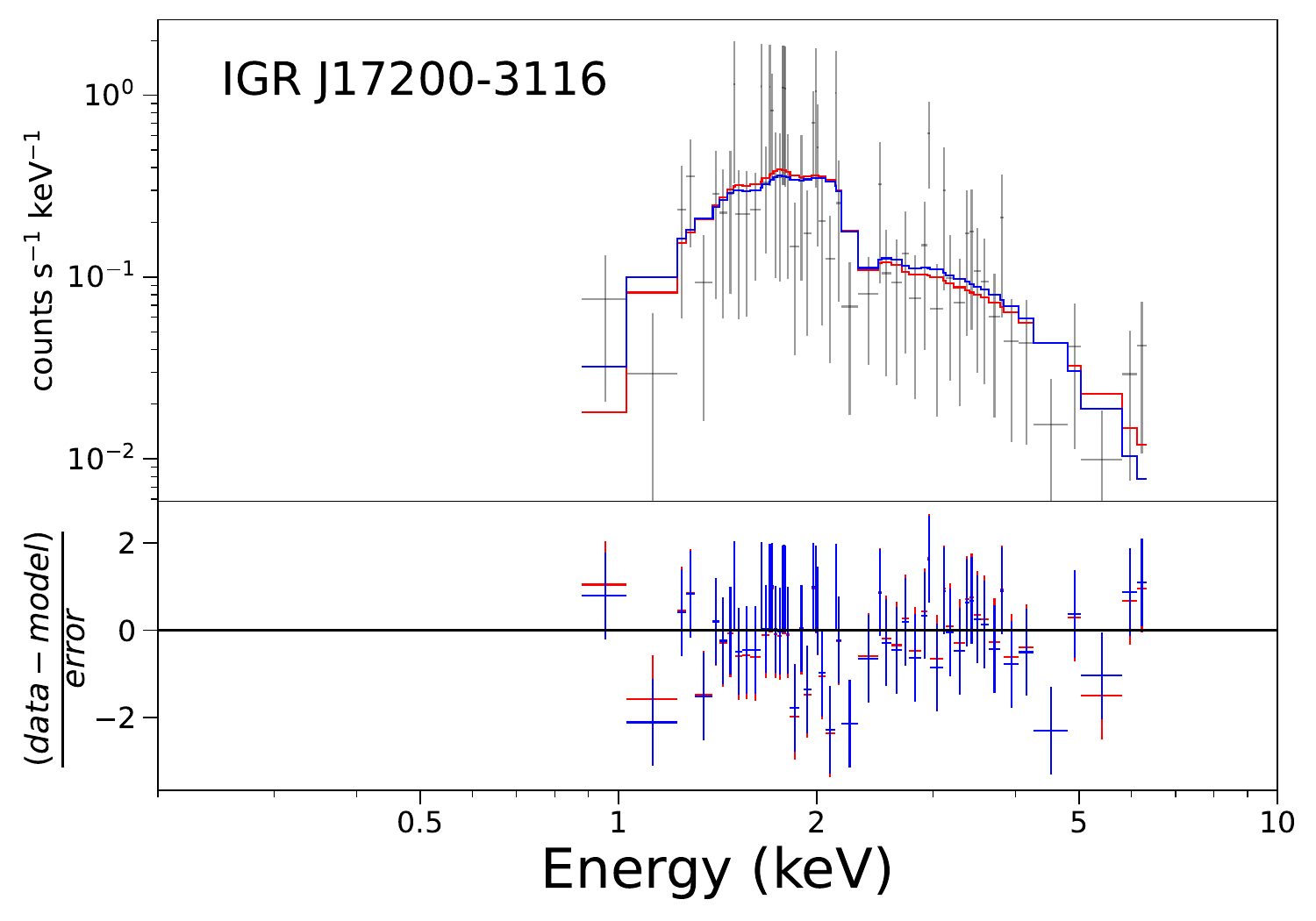}
        \caption{IGR J17200$-$3116}
    \end{subfigure}
    \begin{subfigure}[b]{0.33\textwidth}
        \centering
        \includegraphics[width=\textwidth]{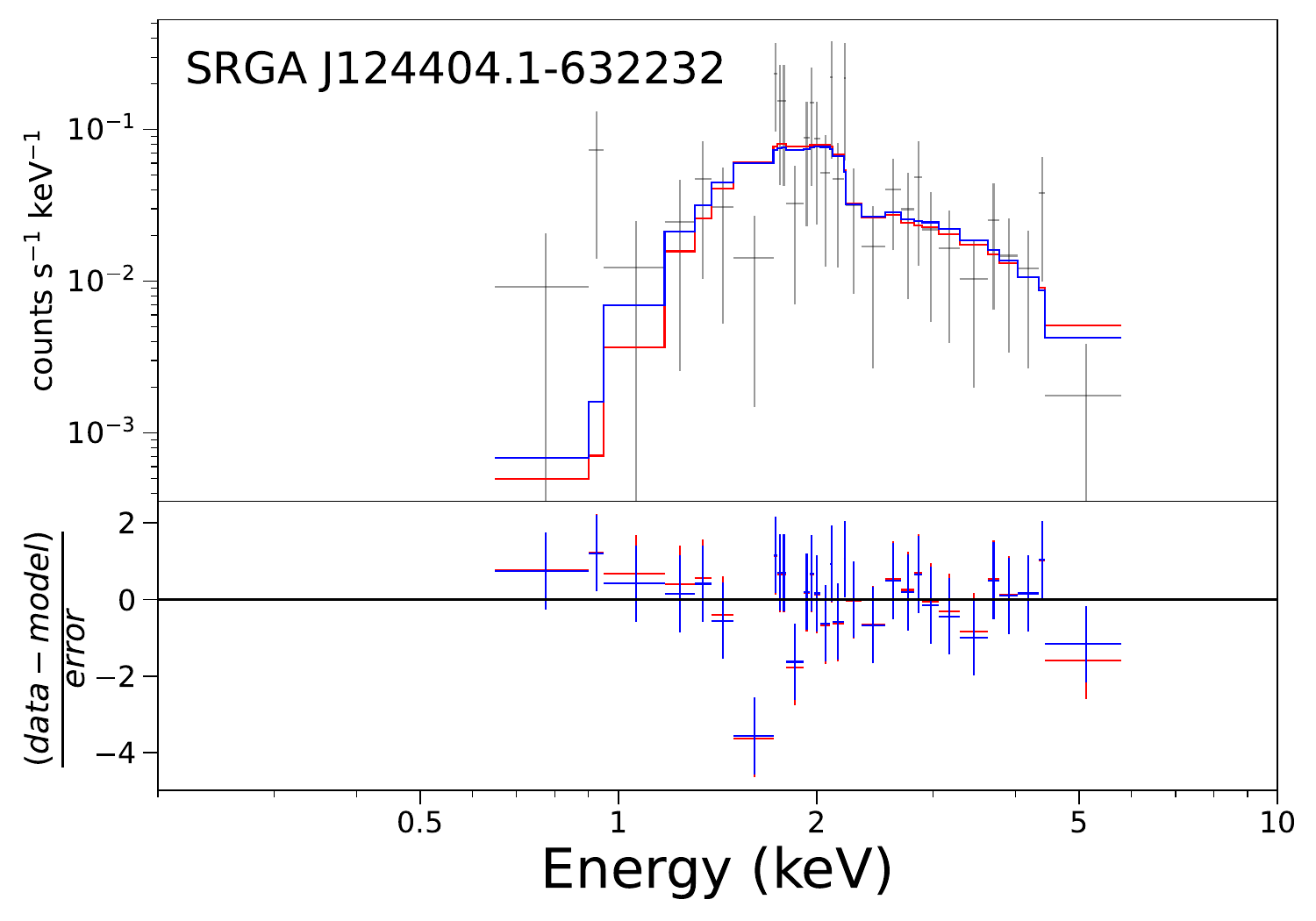}
        \caption{SRGA J124404.1$-$632232}
    \end{subfigure}
    \begin{subfigure}[b]{0.33\textwidth}
        \centering
        \includegraphics[width=\textwidth]{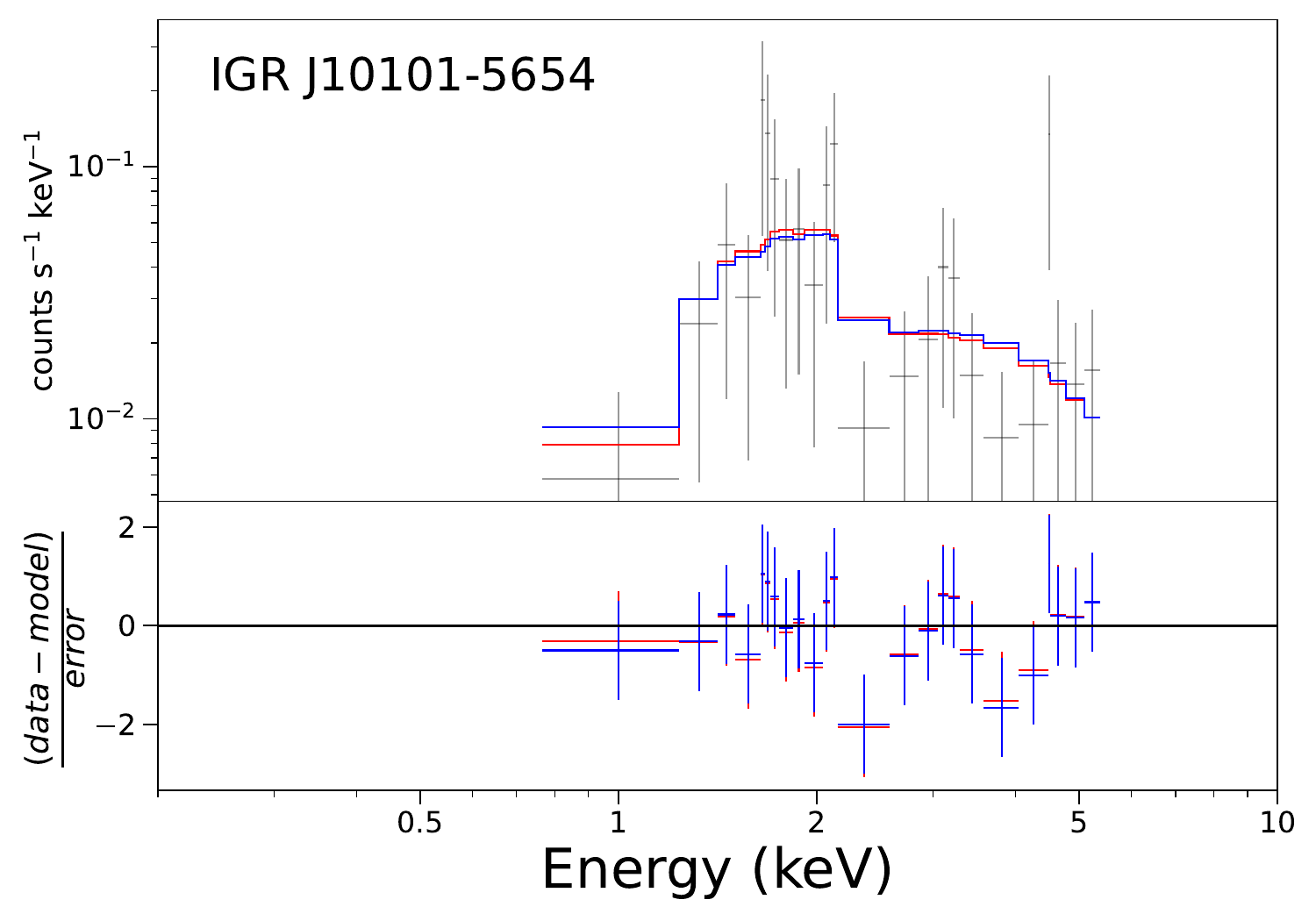}
        \caption{IGR J10101$-$5654}
    \end{subfigure}
    \caption{Best-fit spectra for BeXRB sources for two spectral models, i) \texttt{tbabs$\times$powerlaw}, and ii) \texttt{tbabs$\times$bbodyrad} where the model and its residual is plotted in red and blue, respectively. (continued)}
    \label{fig:spectral_best_fit_2}
\end{figure*}

\begin{figure}
    \centering
    \includegraphics[width=\linewidth]{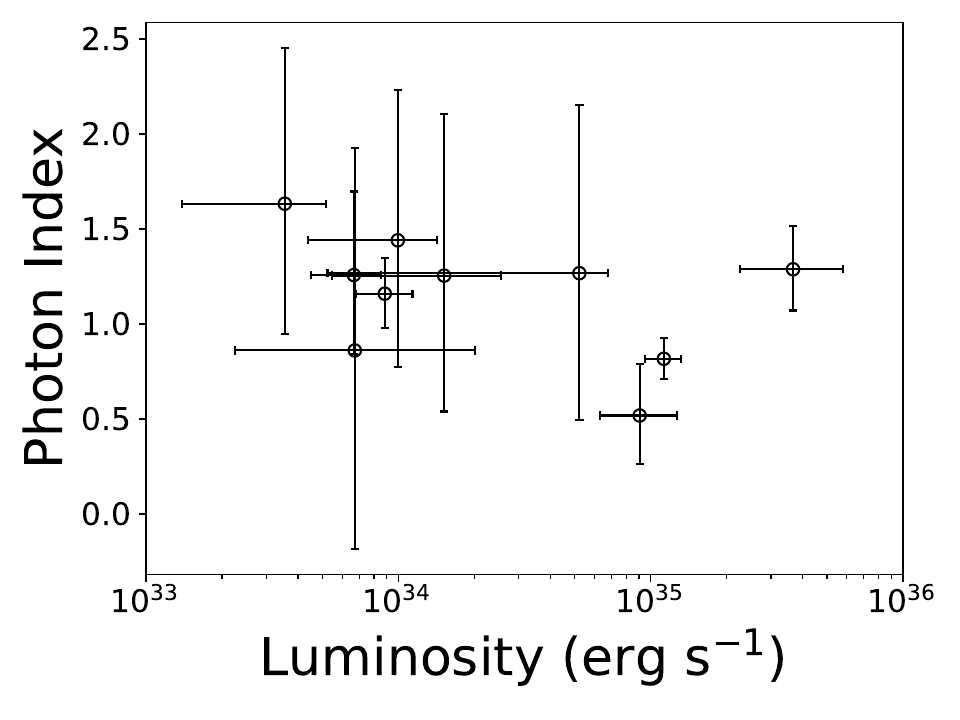}
    \caption{Variation of photon-index, $\Gamma$, with luminosity for the absorbed powerlaw spectral model using the eRASS1 data of BeXRB sources.}
    \label{fig:photon_index_vs_luminosity}
\end{figure}

The best-fit spectral parameters for both the spectral models are shown in ${\text{Table \ref{tab:spectral_model_fit}}}$, and their best-fit spectra are shown in ${\text{Figure \ref{fig:spectral_best_fit}}}$. The errors in spectral parameters are quoted at $90\%$ confidence interval in $\text{Table \ref{tab:spectral_model_fit}}$. Similarly, the best-fit observed flux values are estimated in the 0.5$-$10 keV energy range for both models. The luminosity and radius of emission region is then calculated for these X-ray sources using the geometric distance values for sources with known Gaia source identifiers in the Gaia early data release 3 (EDR3) distance catalog \citep{gaia_edr3_distance}, or the distance values from previous studies of these sources are used.

\begin{figure*}
    \centering
    \begin{subfigure}[b]{0.3\textwidth}
        \centering
        \includegraphics[width=\textwidth]{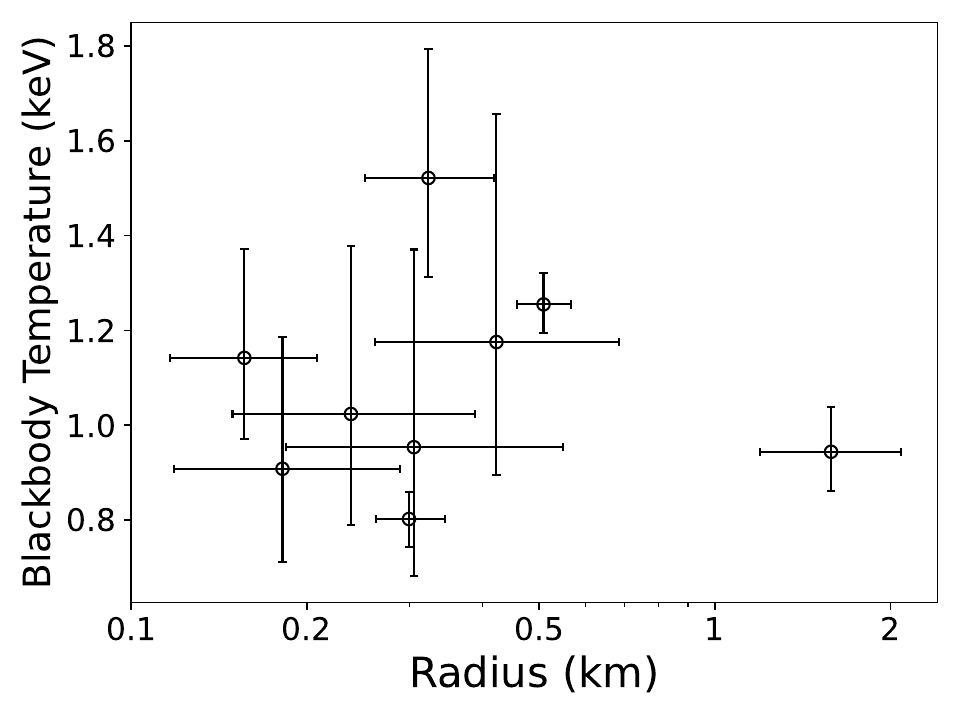}
        \caption{Variation in blackbody temperature against the radius of the emission region of the source.}
        \label{fig:radius_vs_temperature}
    \end{subfigure}
    \begin{subfigure}[b]{0.3\textwidth}
        \centering
        \includegraphics[width=\textwidth]{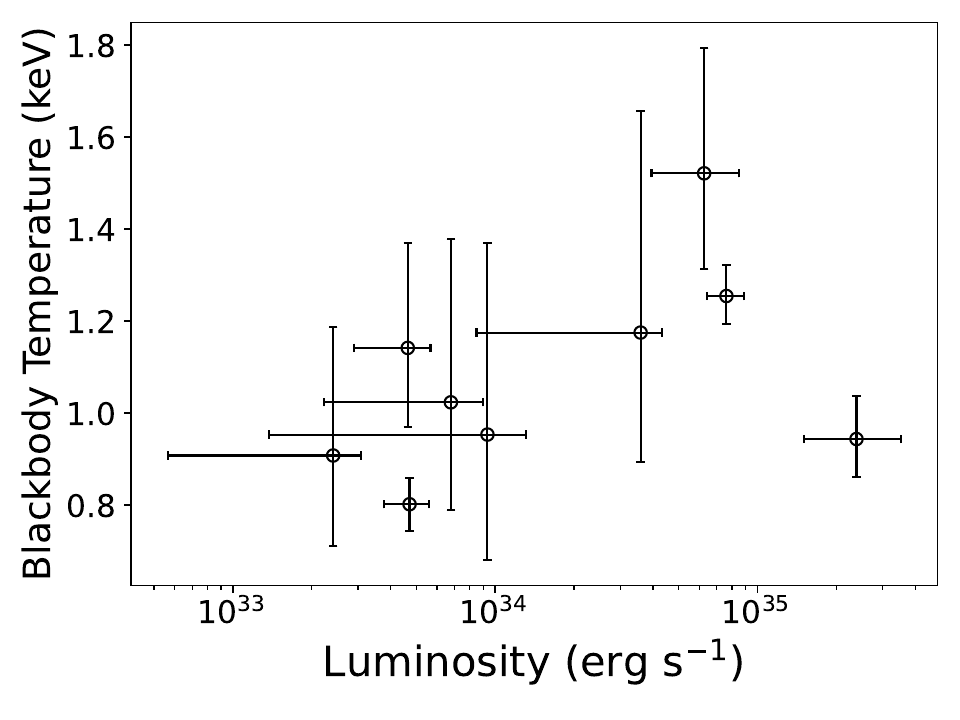}
        \caption{Variation in the blackbody temperature with the 0.5$-$10 keV luminosity of the source.}
        \label{fig:temperature_vs_luminosity}
    \end{subfigure}
    \begin{subfigure}[b]{0.3\textwidth}
        \centering
        \includegraphics[width=\textwidth]{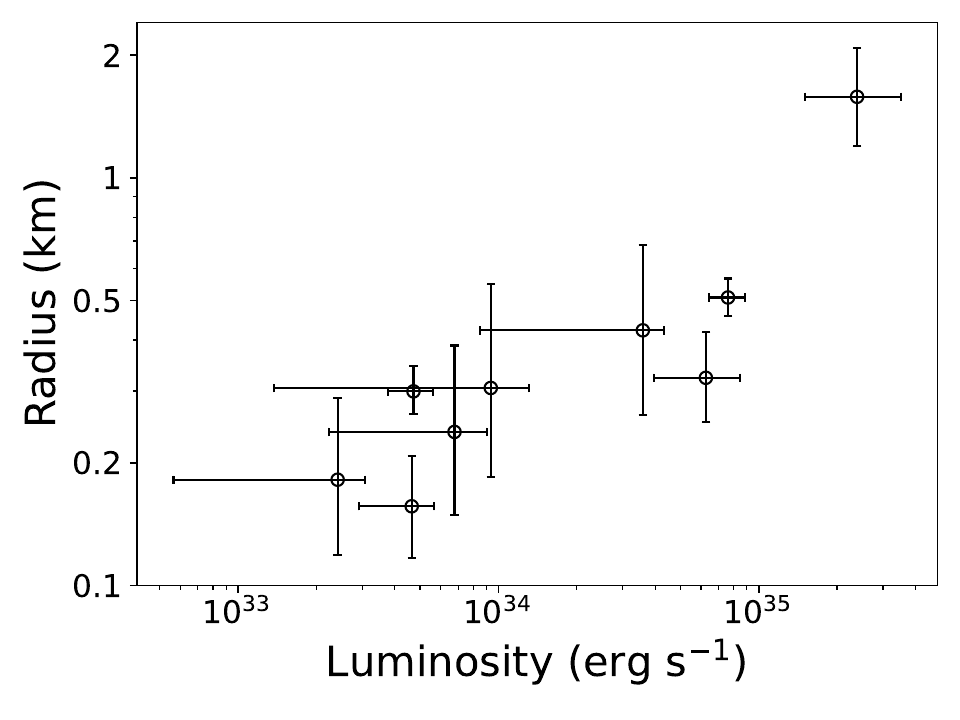}
        \caption{Variation in the radius of the blackbody emission region with the 0.5$-$10 keV source luminosity.}
        \label{fig:radius_vs_luminosity}
    \end{subfigure}
    \caption{The best-fit spectral parameters for the eRASS1 data of BeXRB sources for the absorbed blackbody spectral model are plotted against each other.}
    \label{fig:blackbody_parameter_correlation}
\end{figure*}

\begin{figure}
    \centering
    \includegraphics[width=\linewidth]{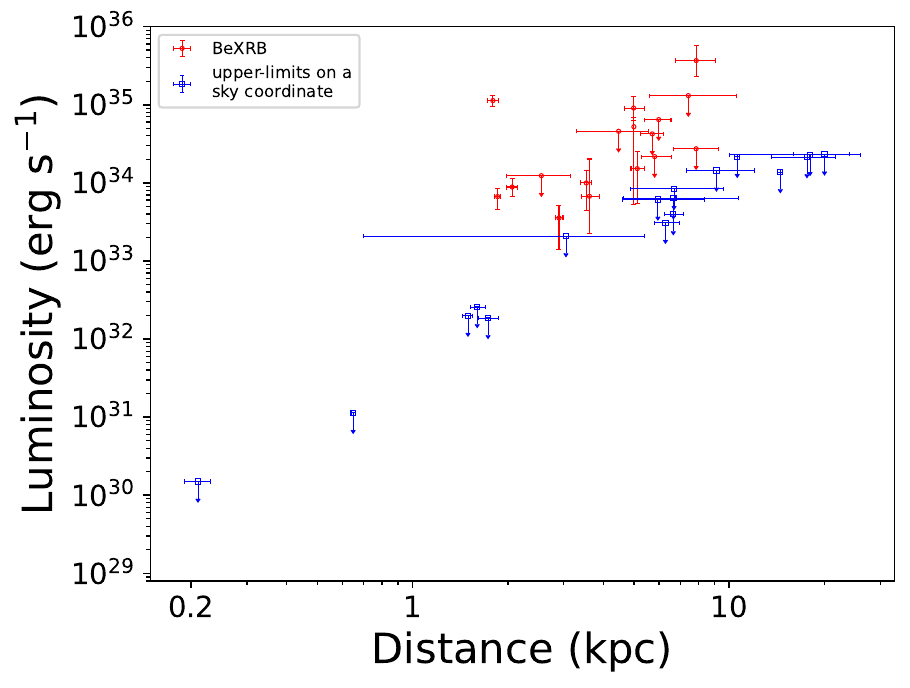}
    \caption{Variation  of luminosity with distance for the eRASS1 data of BeXRB sources for the absorbed powerlaw model. The luminosity values calculated from the best-fit spectra for the BeXRB sources are plotted in red. The luminosity upper limits found for the coordinate of BeXRBs are plotted in blue \citep{Tubin_2024AnA_erosita_upperlimit}.}
    \label{fig:eROSITA_dist_lumin_plot}
\end{figure}

We note that the difference in the c-stat value between the two models are generally very small and is not significant enough to distinguish between them. Similar to \cite{Tsygankov2017MNRAS_pulsars_in_quiescence}, we use the Akaike information criterion \citep{1974ITAC_Akaike_aic}, $\text{AIC}=2k+C$, where $C$ is the Cash-statistics of the spectral fit and $k$ is the total model parameters which are same for both the spectral models. From the spectral fits, we have two sources (1A 0535$+$262 and 4U 0728$-$25) that have a lower c-stat value for the absorbed powerlaw model compared to the absorbed blackbody model and show a c-stat difference of more than ten ($\Delta C\gtrsim10$) between the two spectral models. Thus the absorbed powerlaw model is preferred for these two sources. We see that the c-stat value for the absorbed blackbody model is lower than the absorbed powerlaw model for only three sources (GX 304$-$1, MXB 0656$-$072, and SRGA J124404.1$-$632232), but the difference in c-stat value is not significant to prefer one of the models.

The photon index of the sources with different luminosities are shown in \text{Figure \ref{fig:photon_index_vs_luminosity}} when both the photon index and source luminosity are constrained. It is seen that the photon index is typically centered about $\Gamma\sim1$ though the errors are large for some sources. The photon-index for the BeXRB sources is in the range of $\Gamma\sim0.5-2.0$ for two and a half order of magnitude in X-ray luminosity as shown in \text{Figure \ref{fig:photon_index_vs_luminosity}}.

For the absorbed blackbody model, the temperature is found to be in the range of $kT_{bb}\sim0.8-1.6\text{ keV}$ for the BeXRB sources. The radius of the emission region was calculated from the \texttt{bbodyrad:} norm using the known source distances. The radius of the thermal emission region for the BeXRB sources is in the range of \text{$\sim0.1-2$ km}.

We checked for any correlations between model parameters, i.e., a) blackbody temperature against the radius of the emission region (\text{Figure \ref{fig:radius_vs_temperature}}), b) blackbody temperature against source luminosity (\text{Figure \ref{fig:temperature_vs_luminosity}}), and c) the emission region size against the source luminosity (\text{Figure \ref{fig:radius_vs_luminosity}}). In these plots, we ignore sources for which the spectral parameters could not be constrained. We then check for any correlation between the blackbody parameters for BeXRBs. The blackbody temperature and radius seem crowded at a region for most BeXRBs in \text{Figure \ref{fig:radius_vs_temperature}}. The blackbody temperature does not vary much with luminosity in \text{Figure \ref{fig:temperature_vs_luminosity}}, but a clear rise in emission radius calculated from blackbody normalization with increasing luminosity can be seen in \text{Figure \ref{fig:radius_vs_luminosity}}. These are discussed below in \text{Section \ref{sec:discussion}}.

The \text{0.5$-$10 keV} observed flux could only be constrained for 10 sources for the absorbed powerlaw model and 9 sources for the absorbed blackbody model (see, \text{Table \ref{tab:spectral_model_fit}}). Also, the luminosity calculated for these sources using the known distances given in \text{Table \ref{tab:spectral_model_fit}} for the energy band \text{0.5$-$10 keV} falls within the range of $L\sim10^{33}-10^{36}$ erg s$^{-1}$. The distribution of the \text{0.5$-$10 keV} luminosity with distance of all the X-ray sources from \text{Table \ref{tab:spectral_model_fit}} are plotted in \text{Figure \ref{fig:eROSITA_dist_lumin_plot}}; the luminosity values used in this plot are for the powerlaw model as more luminosity values were constrained with the powerlaw model compared to the blackbody model. For BeXRB sources present in the Western Galactic hemisphere in the eRASS1 survey, but with no available data products, an upper limit on the flux with one-sided confidence level of $99.87\%$ of the X-ray source is found in the energy range of \text{0.2$-$5.0 keV} for its coordinates\footnote{\url{https://erosita.mpe.mpg.de/dr1/erodat/upperlimit/single/}} from an assumed spectral model of absorbed powerlaw with $\Gamma=2.0$ and galactic absorption of $N_{\text{H}}=3\times10^{20}$ cm$^{-2}$ as given in \cite{Tubin_2024AnA_erosita_upperlimit}. From this, we calculate the \text{0.2$-$5.0 keV} upper limit for the case of absorbed powerlaw with $\Gamma=1.0$ and $N_{\text{H}}=3\times10^{20}$ cm$^{-2}$, using a multiplication factor of 1.67 after calculating the energy-to-count conversion factor for both cases of the absorbed powerlaw model as given in \cite{Tubin_2024AnA_erosita_upperlimit}. We then calculate the upper limit on luminosity for the \text{0.5$-$10 keV} energy range for this spectral model and plot them against their distances (see, $\text{Table \ref{tab:luminosity_upper_limit}}$ and \text{Figure \ref{fig:eROSITA_dist_lumin_plot}}).

Many BeXRBs in quiescence have been shown to contain multiple spectral components like a thermal blackbody and a non-thermal powerlaw spectral component \citep{Zhao2019MNRAS}, so we also considered a multi-component model, like \texttt{tbabs} $\times$ ( \texttt{powerlaw} + \texttt{bbodyrad} ). We fitted this model and found that more than one continuum model components are not required to fit the eROSITA spectra due to the poorly constrained spectral parameters.

We also check for any possible outburst for the BeXRB sources during the eROSITA observation (see, ${\text{Figure \ref{fig:Outbursts}}}$) using the MAXI/GSC \citep{ Matsuoka2009PASJ_MAXI, Mihara_2011PASJ_MAXI_GSC} and Swift/BAT \citep{Markwardt_2005ApJ_Swift_BAT} long-term light-curve with a time bin of 1 to 5 days. In the outburst plots (see, ${\text{Figure \ref{fig:Outbursts}}}$), the eROSITA observation is marked by a green vertical line with its width calculated as the difference between the start and stop time of the interval when the region around a source was observed,  taken from the eROSITA lightcurve files. This exercise suggests that the BeXRBs, for which long term monitoring is available, are not in outburst during the eROSITA observation. The observations are carried out in a span of about $1-3$ days, which is shorter than typical durations of BeXRB outbursts \citep{PabloReig2011Ap}.

In \text{Table \ref{tab:spectral_model_fit}}, we have few persistent BeXRB sources, i.e., 4U 0728$-$25, 4U 1036$-$56, and AX J163904$-$4642 $/$ IGR J16393$-$4643. Persistent BeXRBs show less X-ray variability compared to significant outbursts observed in transient BeXRBs \citep{PabloReig2011Ap}. The long-term lightcurves do not show much variability in X-ray emission for the detected persistent BeXRB sources \text{4U 0728$-$25} and \text{4U 1036$-$56} during the eRASS1 observation (${\text{Figure \ref{fig:Outbursts}}}$). The best-fit spectral parameters for few sources that are not typical BeXRBs ($\gamma$ Cas-type or gamma-ray binaries) are given in the Appendix (see \text{Table \ref{tab:appendix_spectral_fit}} and \text{Figure \ref{fig:appendix_spectral_best_fit_plot}}).

The BeXRBs have been studied previously using dedicated X-ray observatories in their quiescence phase, or for a period after an outburst. We compare previous reports of 42 BeXRB sources detected during their low luminosity state. This survey of BeXRB sources include multiple X-ray observatories, namely, Uhuru \citep{Forman_1978ApJS_UHURU}, HEAO$-$1 \citep{Schwartz_1980AJ_HEAO, Wood_1984ApJS_HEAO1}, EXOSAT \citep{Taylor_1981SSRv_EXOSAT}, RXTE \citep{Bradt_1993AnAS_RXTE}, \textit{BeppoSAX} \citep{Boella_1997AnAS_BeppoSAX}, \textit{Chandra} \citep{Weisskopf_2000SPIE_chandrapaper}, \textit{XMM$-$Newton} \citep{Jansen_2001AnA_XMM_newton}, the Swift X-ray telescope (\textit{Swift$-$XRT}) \citep{Burrows_2005SSRv_SWIFT_XRT}, Suzaku \citep{Mitsuda_2007PASJ_suzaku}, and the Nuclear Spectroscopic Telescope Array (\textit{NuSTAR}) \citep{NuSTAR_paper_2013ApJ}.

The details of the BeXRBs observations from different X-ray observatories as reported in literature are compiled in \text{Table \ref{tab:literature_flux_luminosity_1}}, where we mention their best fitting spectral models along with the observed flux and luminosity for the \text{0.5$-$10 keV} energy band. We directly use the observed flux in literature if it is already given in the \text{0.5$-$10 keV} energy band, otherwise is it calculated. In case only the observed flux or unabsorbed flux is given for a different energy bands for a source, we calculate the expected \text{0.5$-$10 keV} observed flux from the best-fit spectral parameters for the preferred spectral model in literature. We thus convert the absorbed$/$unabsorbed flux given in literature to the absorbed flux expected in the 0.5-10 keV energy range. For any errors on the flux value, it is scaled as the ratio between the old and new flux. The multiplicative component model \texttt{Tbabs} from \texttt{XSPEC v12.14.0h} uses photoionization cross-sections from \cite{Verner_1996ApJ_crosssection}. We use this to get the dependence of the photoionization cross-section with energy. This is used to calculate the observed flux of a spectral model for a given hydrogen column density. For some references in our survey only the luminosity of a source is given and it is used to calculate its respective flux from the source distance used in the same reference, which is then converted to the \text{0.5$-$10 keV} observed flux and the luminosity is calculated using the source distances given in \text{Table \ref{tab:literature_flux_luminosity_1}}.


\begin{table*}
\caption{Luminosity of BeXRB sources in their quiescent state calculated from literature for their best-fit spectral models and multiple X-ray observatories. The spectral models used in these studies include, i) powerlaw: \texttt{po}, ii) blackbody: \texttt{bb}, iii) powerlaw $+$ blackbody: \texttt{po$+$bb}, iv) powerlaw, high energy exponential rolloff: \texttt{cutoffpl}, v) powerlaw with high-energy cutoff $+$ blackbody: \texttt{po$\times$highecut$+$bb}, vi) partial covering fraction absorption powerlaw: \texttt{pcf$\times$po},  vii) powerlaw with high-energy cutoff: \texttt{po$\times$highecut}. }
\label{tab:literature_flux_luminosity_1} 
\renewcommand{\arraystretch}{1.55}
\centering
\tiny
\begin{threeparttable}
\setlength\tabcolsep{3pt}
\begin{tabular}{|l|l|c|c|c|c|c|c|c|l|}
\hline
   & Source             & Instr.            & Model     & Absorbed Flux                         & Distance                      & Luminosity                                & Ref.      \\
   &                    &                   &           & (0.5$-$10 keV)                        & (pc)                          & (0.5$-$10 keV)                            &           \\
   &                    &                   &           & (erg s$^{-1}$ cm$^{-2}$)              &                               & (erg s$^{-1}$)                            &           \\  \hline
1  & MXB 0656$-$072     & \textit{NuStar}   & \texttt{po}         & $2.67\pm0.27\times10^{-12}$           & $5720\pm480$          & $1.0\pm0.3\times10^{34}$                  & [1]       \\
   &                    & \textit{Chandra}  & \texttt{po}         & $2.61^{+0.18}_{-0.15}\times10^{-12}$  &                               & $1.0^{+0.3}_{-0.2}\times10^{34}$          & [2]       \\  \hline
2  & RX J0812.4$-$3114  & \textit{XMM}      & \texttt{po$+$bb} & $1.35\pm0.08\times10^{-13}$           & $6660^{+530}_{-420}$         & $7.2^{+1.7}_{-1.3}\times10^{32}$          & [3]       \\
   &                    & \textit{Chandra}  & \texttt{po}         & $2.37^{+0.88}_{-0.37}\times10^{-14}$  &                               & $1.3^{+0.8}_{-0.3}\times10^{32}$          & [2]       \\  \hline
3  & GS 0834$-$430      & \textit{Chandra}  & \texttt{po}         & $1.15^{+0.74}_{-0.52}\times10^{-13}$  & $6700^{+2900}_{-1800}$        & $6.2^{+14.7}_{-4.4}\times10^{32}$         & [2]       \\  \hline
4  & GRO J1008$-$57     & \textit{Chandra}  & \texttt{po}         & $1.84^{+0.05}_{-0.05}\times10^{-11}$  & $3540\pm150$          & $2.8\pm0.3\times10^{34}$                  & [2]       \\  \hline
5  & 2S 1417$-$624      & \textit{Chandra}  & \texttt{po}         & $5.43^{+0.99}_{-0.98}\times10^{-13}$  & $7400^{+3100}_{-1800}$        & $3.6^{+4.9}_{-1.9}\times10^{33}$          & [2]       \\  \hline
6  & GS 1843$+$00       & \textit{Chandra}  & \texttt{po}         & $9.91^{+6.55}_{-5.79}\times10^{-14}$  & $5100^{+4200}_{-1900}$        & $3.1^{+13.9}_{-2.6}\times10^{32}$         & [2]       \\  \hline
7  & KS 1947$+$300      & \textit{Chandra}  & \texttt{po}         & $3.42^{+0.19}_{-0.24}\times10^{-12}$  & $15100^{+3100}_{-2600}$       & $9.4^{+5.1}_{-3.4}\times10^{34}$          & [2]       \\  \hline
8  & Cep X$-$4          & \textit{Chandra}  & \texttt{po}         & $8.67^{+0.95}_{-1.33}\times10^{-13}$  & $7450^{+580}_{-490}$          & $5.8^{+1.7}_{-1.5}\times10^{33}$          & [2]       \\  \hline
9  & SAX J2239.3$+$6116 & \textit{Chandra}  & \texttt{po}         & $5.77^{+1.47}_{-1.28}\times10^{-13}$  & $7390^{+860}_{-680}$          & $3.8^{+2.1}_{-1.3}\times10^{33}$          & [2]       \\  \hline
10 & IGR J01363$+$6610  & \textit{XMM}      & \texttt{bb}  & $1.0^{+0.2}_{-0.1}\times10^{-13}$     & $5820\pm410$          & $4.0^{+1.5}_{-0.9}\times10^{32}$          & [4]       \\  \hline
11 & GX 304-1           & \textit{XMM}      & \texttt{po$+$bb} & $4.3\pm1.1\times10^{-12}$             & $1856^{+37}_{-46}$            & $1.8\pm0.5\times10^{33}$                  & [5]       \\  \hline
12 & 1A 0535$+$262      & \textit{XMM}      & \texttt{cutoffpl}  & $2.44\pm0.01\times10^{-11}$           & $1793^{+76}_{-72}$            & $9.4^{+0.9}_{-0.8}\times10^{33}$          & [6]       \\  \hline
13 & IGR J06074$+$2205  & \textit{Chandra}  & \texttt{po}         & $2.4^{+0.5}_{-0.3}\times10^{-12}$     & $5990^{+560}_{-600}$          & $1.0^{+0.5}_{-0.3}\times10^{34}$          & [7]       \\  \hline
14 & IGR J10101$-$5654  & \textit{Chandra}  & \texttt{po}         & $1.52^{+0.26}_{-0.19}\times10^{-12}$  & $4500^{+1100}_{-1200}$        & $3.6^{+3.0}_{-1.9}\times10^{33}$          & [7]      \\  \hline
15 & IGR J11305$-$6256  & \textit{Chandra}  & \texttt{po}         & $4.3^{+1.9}_{-3.3}\times10^{-11}$     & $1730^{+140}_{-110}$          & $1.5^{+1.1}_{-1.2}\times10^{34}$          & [7]      \\  \hline
16 & IGR J11435$-$6109  & \textit{Chandra}  & \texttt{po}         & $4.7^{+2.5}_{-0.9}\times10^{-12}$     & $7900^{+1400}_{-1200}$        & $3.5^{+3.8}_{-1.5}\times10^{34}$          & [7]      \\  \hline
17 & SAX J1324.4$-$6200 & \textit{XMM}      & \texttt{po}         & $\sim4.5\pm1.3\times10^{-12}$         & $1.5-8.0$ kpc [8]             & $1.2^{+3.2}_{-1.1}\times10^{34}$          & [8]      \\  \hline
18 & GRO J1750$-$27     & \textit{Chandra}  & \texttt{po}         & $2.7^{+1.2}_{-0.8}\times10^{-14}$     & $14-22$ kpc [9]           & $1.0^{+1.2}_{-0.6}\times10^{33}$          & [10]      \\
   &                    & $''$              & \texttt{po}         & $0.66^{+0.50}_{-0.24}\times10^{-14}$  &                               & $2.6^{+4.2}_{-1.6}\times10^{32}$          &           \\
   &                    & $''$              & \texttt{po}         & $1.44^{+0.55}_{-0.38}\times10^{-14}$  &                               & $5.6^{+5.9}_{-3.1}\times10^{32}$          &           \\  \hline   
19 & IGR J18214$-$1318  & \textit{Chandra}  & \texttt{po}         & $39^{+206}_{-23}\times10^{-12}$       & $4200^{+1700}_{-1300}$        & $8.3^{+94.4}_{-6.6}\times10^{34}$         & [7]      \\  \hline
20 & XTE J1946$+$274    & \textit{Chandra}  & \texttt{po}         & $9.92^{+1.05}_{-0.85}\times10^{-13}$  & $13100^{+3400}_{-2300}$       & $2.0^{+1.5}_{-0.8}\times10^{34}$          & [2]       \\
   &                    & $''$              & \texttt{po}         & $1.0\pm0.3\times10^{-12}$             & $''$                          & $2.1^{+2.1}_{-1.1}\times10^{34}$          & [11]      \\  \hline
21 & SAX J2103.5$+$4545 & \textit{Chandra}              & \texttt{po}         & $3.89^{+0.96}_{-0.49}\times10^{-13}$  & $6220^{+580}_{-450}$          & $1.8^{+0.9}_{-0.5}\times10^{33}$          & [2]       \\ \hline
22 & MWC 656 $^{\dagger}$ (Be/BH) & \textit{Chandra} & \texttt{po} & $3.5^{+2.0}_{-1.3}\times10^{-15}$    & $1985^{+81}_{-77}$            & $1.7^{+1.2}_{-0.7}\times10^{30}$          & [12]      \\  \hline
23 & RX J0440.9$+$4431  & \textit{XMM}      & \texttt{po$+$bb}  & $5.2^{+0.2}_{-0.3}\times10^{-11}$     & $2444^{+61}_{-77}$            & $3.7^{+0.3}_{-0.4}\times10^{34}$          & [13]      \\  \hline
24 & IGR J20006$+$3210 $/$ & \textit{Suzaku} & \texttt{po$\times$highecut$+$bb}  & $\sim3.20\pm0.01\times10^{-11}$       & $8400^{+1800}_{-1200}$        & $2.7^{+1.3}_{-0.7}\times10^{35}$          & [14]      \\
   & SW J2000.6$+$3210  & $''$          & $''$  & $\sim5.70\pm0.01\times10^{-11}$       & $''$                          & $4.8^{+2.2}_{-1.3}\times10^{35}$          &           \\ 
   &                    & \textit{Swift}    & \texttt{po}         & $\sim3.40\pm0.01\times10^{-11}$       & $''$                          & $2.8^{+1.3}_{-0.8}\times10^{35}$          &           \\
   &                    & $''$              & \texttt{po}         & $\sim4.20\pm0.01\times10^{-11}$       & $''$                          & $3.5^{+1.7}_{-0.9}\times10^{35}$          &           \\  \hline
25 & CXOU J225355.1$+$624336 & \textit{XMM} & \texttt{pcf$\times$po} & $3.8^{+0.5}_{-0.4}\times10^{-12}$ & $9700^{+1700}_{-1200}$        & $4.3^{+2.4}_{-1.3}\times10^{34}$        & [15]      \\  \hline
26 & XTE J1906$+$090    & \textit{Chandra}  & \texttt{po}         & $2.8^{+0.5}_{-1.8}\times10^{-12}$     & $6400^{+3500}_{-2000}$        & $1.4^{+2.5}_{-1.1}\times10^{34}$          & [16]      \\  \hline
27 & 4XMM J182531.5$-$  & \textit{XMM +}    & \texttt{po}         & $1.58\pm0.02\times10^{-12}$           & $1-7$ kpc                  & $3.0^{+6.4}_{-2.8}\times10^{33}$      & [17]      \\ 
   & 144036 $/$         & \textit{Chandra}  &           &                                       &  [17]                         &                                           &           \\  \hline
28 & 4U 0728$-$25       & \textit{XMM}      & \texttt{cutoffpl}  & $1.48\pm0.02\times10^{-11}$           & $7900^{+1200}_{-1100}$        & $1.1^{+0.4}_{-0.3}\times10^{35}$          & [18]      \\
   & (3 Obs.)           & \textit{Swift}    & \texttt{po}         & $2.09^{+0.13}_{-0.16}\times10^{-12}$      & $''$        & $1.5^{+0.6}_{-0.5}\times10^{34}$   & [2]       \\
   & (1 Obs.)           & \textit{Swift}    & \texttt{po}         & $1.07^{+0.54}_{-0.31}\times10^{-12}$      & $''$        & $7.9^{+7.9}_{-3.7}\times10^{33}$  & [2]       \\  \hline
29 & 4U 0115$+$63       & \textit{XMM}      & \texttt{po}         & $3.85^{+0.53}_{-0.48}\times10^{-14}$      & $5790^{+820}_{-450}$ & $1.5^{+0.7}_{-0.4}\times10^{32}$  & [2]       \\  \hline
30 & V 0332$+$53        & \textit{BeppoSAX} & \texttt{po}         & $<2\times10^{-13}$                        & $5580^{+730}_{-510}$          & $<9.5\times10^{32}$               & [19]      \\
   &                    & \textit{XMM}      & \texttt{po}         & $3.91^{+0.62}_{-0.52}\times10^{-14}$      & $''$                          & $1.5^{+0.7}_{-0.4}\times10^{32}$  & [2]       \\
   &                    & \textit{Chandra}  & \texttt{po}         & $2.55^{+0.69}_{-0.72}\times10^{-14}$      & $''$                          & $9.5^{+5.9}_{-3.9}\times10^{31}$  & $''$      \\  \hline
\end{tabular}
\end{threeparttable}
\end{table*}
\begin{table*}
\ContinuedFloat
\caption{Luminosity of BeXRB sources in their quiescent state calculated from literature for their best-fit spectral models and multiple X-ray observatories. (Continued)}
\label{tab:literature_flux_luminosity_2} 
\renewcommand{\arraystretch}{1.55}
\centering
\tiny
\begin{threeparttable}
\setlength\tabcolsep{3pt}
\begin{tabular}{|l|l|c|c|c|c|c|l|l|l|}
\hline
   & Source             & Instr.            & Model     & Absorbed Flux                             & Distance                      & Luminosity                        & Ref.      \\
   &                    &                   &           & (0.5$-$10 keV)                            & (pc)                          & (0.5$-$10 keV)                    &           \\
   &                    &                   &           & (erg s$^{-1}$ cm$^{-2}$)                  &                               & (erg s$^{-1}$)                    &           \\  \hline
31 & 4U 0352$+$309 / X$-$Per & \textit{RXTE}     & \texttt{po$+$bb} & $1.71^{+0.08}_{-0.06}\times10^{-10}$      & $596^{+17}_{-14}$   & $7.3^{+0.8}_{-0.6}\times10^{33}$  & [20]      \\  \hline
32 & SAX J0635.2$+$0533 & \textit{Swift}    & \texttt{po}         & $1.08^{+0.04}_{-0.08}\times10^{-12}$      &  $6290^{+660}_{-490}$  & $5.1^{+1.4}_{-1.1}\times10^{33}$  & [21]      \\  \hline
33 & MAXI J1409$-$619   & \textit{NuSTAR}   & \texttt{cutoffpl}  & $\sim1.34\pm0.01\times10^{-12}$           & $14.5\times10^{3}$ [22]  & $3.37\pm0.03\times10^{34}$        & [23]      \\  \hline
34 & HD 34921           & \textit{Uhuru}    & $-$         & $\sim7.7\pm1.7\times10^{-11}$             & $1318^{+55}_{-52}$            & $1.6\pm0.5\times10^{34}$          & [24]      \\
   & 4U 0515$+$38       & \textit{HEAO1-A1} &           & $\sim1.2\pm0.3\times10^{-11}$             &                               & $2.5^{+0.9}_{-0.8}\times10^{33}$  &           \\
   & 1H 0521$+$373      & \textit{EXOSAT ME Ar} &       & $\lesssim0.29\times10^{-11}$              &                               & $<6.5\times10^{32}$               &           \\
   & EXO 051910$+$3737.7& \textit{EXOSAT ME Xe} &       & $\lesssim0.73\times10^{-11}$              &                               & $<1.7\times10^{33}$               &           \\  \hline
35 & 2S 1553$-$542      & \textit{Chandra}  & \texttt{po}         & $1.71^{+1.16}_{-0.99}\times10^{-14}$      & $20\pm4\times10^3$ [25]    & $8.2^{+11.6}_{-6.0}\times10^{32}$ & [2]       \\  \hline
36 & Swift J1626.6$-$5156  (Obs. 1) & \textit{Chandra}  & \texttt{po} & $1.19^{+0.06}_{-0.07}\times10^{-12}$ & $9100^{+2900}_{-1800}$        & $1.2^{+1.0}_{-0.5}\times10^{34}$  & [2]       \\
   & (Obs. 2)           & \textit{Chandra}  & \texttt{po}         & $2.91^{+0.85}_{-0.62}\times10^{-13}$      & $"$                           & $2.9^{+3.6}_{-1.4}\times10^{33}$  & $''$      \\ \hline
37 & HD 249179          & \textit{BeppoSAX} & $-$        & $\lesssim0.19\times10^{-12}$              & $1600^{+95}_{-73}$            & $\lesssim6.5\times10^{31}$        & [26]      \\
   &                    &                   &           &                                           &                               &                                   &           \\  \hline
38 & IGR J17200$-$3116  & \textit{XMM}      & \texttt{po$+$bb} & $3.12\pm0.02\times10^{-11}$               & $5\times10^{3}$ [27]     & $9.33\pm0.06\times10^{34}$        & [27]      \\
   &                    & \textit{Chandra}  & \texttt{po}         & $22^{+33}_{-13}\times10^{-12}$            & $''$                          & $6.6^{+9.9}_{-3.9}\times10^{34}$  & [7]      \\  \hline
39 & IGR J18179$-$1621  & \textit{Chandra}  & \texttt{po}         & $2.40\pm0.12\times10^{-11}$               & $\sim8\times10^3$ [28]        & $1.84\pm0.09\times10^{35}$        & [28]      \\
   &                    & $+$ \textit{Swift}    &       &                                           &                               &                                   &           \\  \hline
40 & LS I +61 303       & \textit{Chandra}  & \texttt{po}         & $7.1^{+1.8}_{-1.4}\times10^{-12}$         & $2504^{+72}_{-68}$            & $5.3^{+1.7}_{-1.3}\times10^{33}$  & [29]      \\  \hline
41 & 
2RXP J130159.6$-$635806
 /  & \textit{NuSTAR} & \texttt{cutoffpl}     & $2.0\pm0.1\times10^{-12}$        & $6000^{+2400}_{-1300}$        & $8.5^{+9.0}_{-3.6}\times10^{33}$      & [30]  \\
   & IGR J13020$-$6359  & \textit{NuSTAR} & \texttt{po$\times$highecut} & $2.25^{+0.12}_{-0.40}\times10^{-11}$ & $''$                        & $9.5^{+10.1}_{-4.8}\times10^{34}$     & [31]  \\
   &                    & $''$              & $''$      & $3.41^{+0.27}_{-1.53}\times10^{-11}$ & $''$                           & $1.4^{+1.6}_{-1.0}\times10^{35}$      & $''$  \\
   &                    & $''$              & $''$      & $2.19^{+0.08}_{-0.14}\times10^{-11}$ & $''$                           & $9.3^{+9.6}_{-4.1}\times10^{34}$      & $''$  \\
   &                    & $''$              & $''$      & $3.93^{+0.05}_{-0.10}\times10^{-11}$ & $''$                           & $1.7^{+1.6}_{-0.7}\times10^{35}$      & $''$  \\
   &                    & \textit{Swift}    & \texttt{po} & $2.15^{+0.67}_{-0.39}\times10^{-11}$ & $''$                         & $9.1^{+14.3}_{-4.6}\times10^{34}$     & [32]  \\
   &                    & $''$              & \texttt{po} & $2.42^{+0.90}_{-0.64}\times10^{-11}$ & $''$                         & $1.0^{+1.7}_{-0.6}\times10^{35}$      & $''$  \\
   &                    & $''$              & \texttt{po} & $1.98^{+1.10}_{-0.67}\times10^{-11}$ & $''$                         & $8.4^{+17.2}_{-5.1}\times10^{34}$     & $''$  \\
   &                    & $''$              & \texttt{po} & $1.90^{+1.59}_{-0.80}\times10^{-11}$ & $''$                         & $8.1^{+20.9}_{-5.2}\times10^{34}$     & $''$  \\
   &                    & \textit{XMM}      & \texttt{po} & $2.29^{+0.11}_{-0.20}\times10^{-11}$ & $''$                         & $9.7^{+10.2}_{-4.4}\times10^{34}$     & $''$  \\
   &                    & $''$              & \texttt{po} & $3.50^{+0.06}_{-0.09}\times10^{-11}$ & $''$                         & $1.5^{+1.5}_{-0.6}\times10^{35}$      & $''$  \\
   &                    & $''$              & \texttt{po} & $2.97^{+0.10}_{-0.23}\times10^{-11}$ & $''$                         & $1.3^{+1.3}_{-0.6}\times10^{35}$      & $''$  \\  \hline
42 & MAXI J0655$-$013   & \textit{NuSTAR}   & \texttt{cutoffpl} & $2.0\pm0.2\times10^{-11}$     &  $3620^{+280}_{-220}$         & $3.1^{+0.9}_{-0.6}\times10^{34}$      & [33]  \\  \hline
\end{tabular}
    \begin{tablenotes}
        \small
        \item[$\text{a}$] {The distance values for most of the sources are taken from Gaia early data release 3 (EDR3) distance catalogue $\citep{gaia_edr3_distance}$. For some sources, we use the distance values from the literature.}
        \item[$\text{c}$] {$\dagger$ The X-ray source MWC 656 may be a Be/BH, i.e., a Be black-hole binary \citep{Casares_2014Nature_MWC_656}. Though, \cite{Janssens_2023AnA_MWC656_BeBH_or_not} claim that it may in fact not be a Be/BH binary.}
        \item[$\text{d}$] {References: [1] \cite{Raman2023MNRAS_quiescent}; [2] \cite{Tsygankov2017MNRAS_pulsars_in_quiescence}; [3] \cite{Zhao2019MNRAS}; [4] \cite{Tomsick2011ApJ_IGR_J01363p6610}; [5] \cite{Chandra_2023arXiv_GX_304m1}; [6] \cite{Doroshenko2014AnA_1A0535p262}; [7] \cite{Tomsick2008ApJ_Chandra}; [8] \cite{Kaur2009MNRAS_SAX_J13244m6200}; [9] \cite{Lutovinov2019MNRAS_GRO_J1750m27}; [10] \cite{Rouco_2019AnA_GROJ1750m27_lowlevel_acc}; [11] \cite{Arabac_2015AnA_XTE_J1946p274}; [12] \cite{Ribo2017ApJ_MWC_656}; [13] \cite{Palombara_RX_J04409p4431}; [14] \cite{Pradhan2013MNRAS_SW_J20006p3210}; [15] \cite{Palombara2021AnA_CXOU_J2253551p624336}; [16] \cite{Ersin2005ApJ_XTE_J1906p090}; [17] \cite{Mason2024PASA_4XMM_J1825315m144036}; [18] \cite{Palombara_2025AnA_4U_0728m25}; [19] \cite{Campana2002ApJ_3_sources}; [20]  \cite{Coburn_2001ApJ_X_Persei}; [21] \cite{La_Palombara_2017AnA_SAX_J0635p0533}; [22] \cite{Orlandini2012ApJ}; [23] \cite{Ghimiray2024MNRAS}; [24] \cite{Polcaro2001AdSpR}; [25] \cite{Tsygankov2016MNRAS}; [26] \cite{Torrej2001AnA}; [27] \cite{Esposito2014MNRAS_IGR_J17200m3116}; [28] \cite{Nowak2012ApJ}; [29] \cite{Paredes2007ApJ}; [30] \cite{Salganik2025AnA}; [31] \cite{Krivonos_2015ApJ_2RXPJ1301596m635806}; [32] \cite{Chernyakova2009MNRAS}; [33] \cite{Rai2023MNRAS}.}
    \end{tablenotes}
\end{threeparttable}
\end{table*}

\begin{figure*}
    \centering
    \centering
    \includegraphics[width=\textwidth]{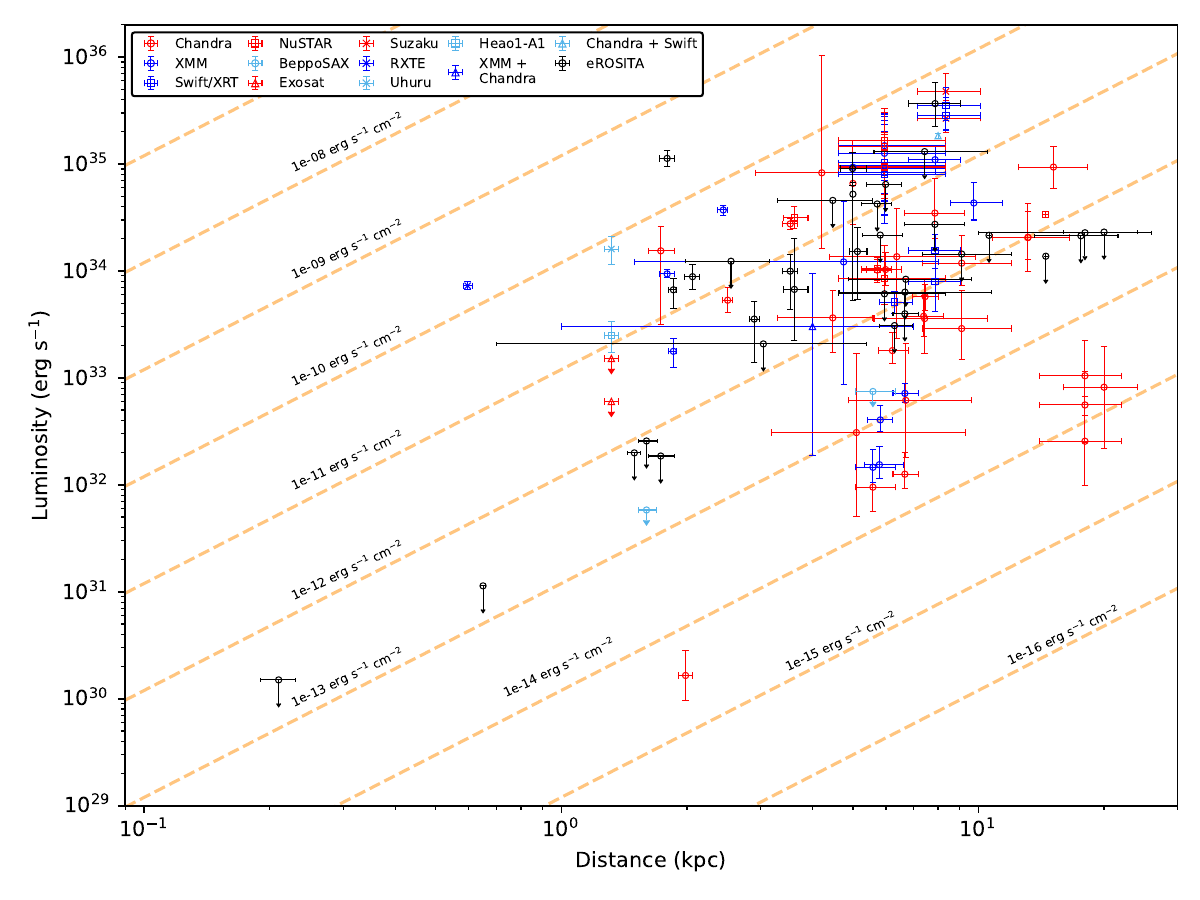}
    \caption{Distribution of the \text{0.5$-$10 keV} luminosity of BeXRB sources in quiescence with source distance, which includes both the BeXRB sources from the previous reports and the X-ray sources from the eRASS1 survey in \text{Table \ref{tab:spectral_model_fit} \& \ref{tab:luminosity_upper_limit} }. The dashed yellow line is for flux at different orders of magnitude.}
    \label{fig:dist_lumin_plot}
\end{figure*}

For some sources the spectral parameters are not provided in literature (e.g. HD 249179 and HD 34921), so we directly use their flux or luminosity values as provided in the corresponding paper in \text{Table \ref{tab:literature_flux_luminosity_1}}. These luminosity values of BeXRBs from previous studies are plotted along with the luminosity values calculated from the eRASS1 data in \text{Figure \ref{fig:dist_lumin_plot}}.

\section{Discussion} \label{sec:discussion}


\begin{table*}
\caption{Luminosities of BeXRBs calculated for two accretion regimes from known parameters of BeXRBs.}
\label{tab:basic_info}
\renewcommand{\arraystretch}{1.3}
\centering
\tiny
\setlength{\tabcolsep}{8pt}
\begin{tabular}{l|l|c|c|c|c|c|c|c}
\hline
   & Source                             & Spin Period               & Cyclotron     & L$_{\text{PL}}$$^{\text{\tiny a}}$ & L$_{\text{cold}}$$^{\text{\tiny b}}$ & L$_{\text{prop}}$$^{\text{\tiny c}}$ & \text{L}$_{\text{prop, min (r$_{\text{corr}}$)}}$$^{\text{\tiny c}}$ & $P^{*}$   \\
   &                                    & (s)                       & Line          & (0.5$-$10 keV)&(10$^{34}$ erg s$^{-1}$) & (10$^{34}$ erg s$^{-1}$) & (10$^{34}$ erg s$^{-1}$) & (s)  \\ 
   &                                    &                           & (keV)         & (10$^{34}$ erg s$^{-1}$)&    &               &                   & \\  \hline
1  & 1A 0535$+$262                      & 103.41 [1]                & 45.5 [2]      & $12.4^{+2.1}_{-1.8}$  & 1.0   & 0.11          & $3\times10^{-5}$           & 39.5              \\  \hline
2  & 2S 1417$-$624                      & 17.64 [3]                 & 100 [4]       & $6.7^{+16.2}_{-4.1}$  & 2.0   & 32.4          & $3\times10^{-2}$             & 58.1              \\  \hline
3  & GRO J1008$-$57                     & 93.5 [5]                  & 90.32 [6]     & $1.4^{+0.6}_{-0.4}$   & 1.9   & 0.54          & $2\times10^{-4}$           & 55.2              \\  \hline
4  & GX 304$-$1                         & 272 [7]                   & 52 [8]        & $0.9\pm0.2$           & 1.2   & 0.01          & $2\times10^{-6}$          & 42.1              \\  \hline
5  & IGR J06074$+$2205                  & 373.226 [9]               & 51 [10]       & $3.3^{+7.7}_{-2.0}$   & 1.1   & 0.007         & $8\times10^{-7}$         & 41.7              \\  \hline
6  & MXB 0656$-$072                     & 160.7 [11]                & 32.8 [12]     & $2.1^{+6.7}_{-1.4}$   & 0.8   & 0.02          & $4\times10^{-6}$          & 33.6              \\  \hline
7  & 1A 1118$-$61                       & 405.3 [13]                & 55 [14]       & $0.5^{+0.2}_{-0.1}$   & 1.2   & 0.007         & $7\times10^{-7}$         & 43.3              \\  \hline
8  & MAXI J1409$-$619                   & $503\pm10$ [15]           & 44 [16]       & $<1.37$       & 1.0           & 0.003         & $2\times10^{-7}$         & 38.8              \\  \hline
9  & Swift J1626.6$-$5156               & 15.36 [17]                & 4.9 [18]      & $<1.44$       & 0.15          & 0.11          & $1\times10^{-4}$           & 13.2              \\  \hline
10 & 2S 1553$-$542                      & 9.28 [19]                 & $23.5\pm0.4$ [20] & $<2.31$   & 0.6           & 8.0           & $1\times10^{-2}$              & 28.6              \\  \hline
11 & AX J163904$-$4642 $/$              & 908.79 [21]   & $29.3^{+1.1}_{-1.3}$ [22] & $<2.15$       & 0.7           & 0.0003        & $2\times10^{-8}$        & 31.8              \\
   & IGR J16393$-$4643                  &                           &               &               &               &               &                   &                   \\  \hline
12 & 4U 0728$-$25                       & 103.145 [23]              & $-$           & $43.5^{+22.3}_{-14.8}$& 0.82  & 0.06          & $2\times10^{-5}$          & 34.6              \\  \hline
13 & 4U 1036$-$56                       & 854.75 [24]               & $-$           & $9.7^{+4.0}_{-2.7}$   & $''$  & 0.0005        & $3\times10^{-8}$        & $''$              \\  \hline
14 & 4U 1145$-$61                       & 292 [25,26]               & $-$           & $0.9^{+0.3}_{-0.2}$   & $''$  & 0.006         & $8\times10^{-7}$         & $''$              \\  \hline
15 & MAXI J0655$-$013                   & 1129.1 [27]               & $-$           & $0.7^{+2.3}_{-0.5}$   & $''$  & 0.0002        & $1\times10^{-8}$       & $''$              \\  \hline
16 & IGR J11435$-$6109                  & 161.76 [28]               & $-$           & $1.8^{+2.8}_{-1.0}$   & $''$  & 0.02          & $5\times10^{-6}$         & $''$              \\  \hline
17 & IGR J17200$-$3116                  & $\sim328$ [29]            & $-$           & $7.2^{+2.3}_{-1.6}$   & $''$  & 0.004         & $5\times10^{-7}$        & $''$              \\  \hline
18 & SRGA J124404.1                     & $538.70\pm0.05$           & $-$           & $2.8^{+20.2}_{-1.4}$  & $''$  & 0.001         & $1\times10^{-7}$         & $''$              \\
   & $-$632232                          & [30]                      & $-$           &               & $''$          &               &                   &                   \\  \hline
19 & RX J0812.4$-$3114                  & 31.8851 [31]              & $-$           & $<0.40$       & $''$          & 1.0           & $6\times10^{-4}$            & $''$              \\  \hline
20 & 2RXP J130159.6                     & 643 [32]                  & $-$           & $<0.61$       & $''$          & 0.001         & $7\times10^{-8}$        & $''$              \\
   & $-$635806                          &                           &               &               & $''$          &               &                   &                   \\  \hline
21 & AX J1700.2$-$4220                  & 54 [33]                   & $-$           & $<0.02$       & $''$          & 0.3           & $1\times10^{-4}$           & $''$              \\  \hline
22 & GS 0834$-$43                       & 12.293 [34]               & $-$           & $<0.84$       & $''$          & 9.1           & $1\times10^{-2}$              & $''$              \\  \hline
23 & XTE J1543$-$568   & $27.1225\pm0.0043$ [35]   & $-$   & $<2.28$  & $''$      & 1.44          & $1\times10^{-3}$             & $''$              \\  \hline
\end{tabular}
\justifying
\tablecomments{\scriptsize \\
{$^{\text{a}}$ Unabsorbed luminosity for the powerlaw model. For X-ray sources with no data products, the upper limits on flux were found as per \cite{Tubin_2024AnA_erosita_upperlimit} for their position in sky (see, Table \ref{tab:luminosity_upper_limit}).} \\
{$^{\text{b}}$ Limiting luminosity in cold disc regime. For sources with no known CRSFs, we assume a magnetic field strength of $3\times10^{12}$ G.} \\
{$^{\text{c}}$ Limiting luminosities in propeller regime. For sources with no known CRSFs, we assume a magnetic field strength of $3\times10^{12}$ G.} \\
{References: [1] \cite{Mukherjee2005aanda_3a0535p262}; [2] \cite{Terada2006ApJ_3a0535p262}; [3] \cite{Kelley1981ApJ_2S_1417m624}; [4] \cite{Liu2024aanda_2S_1417m624}; [5] \cite{Wilson1994AIPC_GRO_J1008m57}; [6] \cite{Ge2020ApJ_GRO_J1008m57}; [7] \cite{McClintock1977ApJ_gx_304m1}; [8] \cite{Yamamoto2011PASJ_gx_304m1}; [9] \cite{Reig2018aanda_IGR_J06074p2205}; [10] \cite{Roy2024aanda_IGR_J06074p2205}; [11] \cite{Morgan2003ATel_MXB_0656m072}; [12] \cite{McBride2006aanda_MXB_0656m072}; [13] \cite{Ives1975Natur_1A_1118m615}; [14] \cite{Doroshenko2010aanda_1A_1118m61}; [15] \cite{Kennea_2010ATel_MAXI_J1409m619}; [16] \cite{Orlandini2012ApJ}; [17] \cite{Staubert_2019AnA}; [18] \cite{Molkov_2021ApJ_Swift_J16266m5156_crsf}; [19] \cite{Malacaria_2022ApJ_2S_1553m542}; [20] \cite{Tsygankov2016MNRAS}; [21] \cite{Islam_2015MNRAS_IGR_J16393m4643}; [22] \cite{Bodaghee_2016ApJ_IGR_J16393m4643}; [23] \cite{Roy2020RAA_3A_0726m260}; [24] \cite{Li2012ApJ_4U_1036m56}; [25] \cite{Watson1981MNRAS}; [26] \cite{Nagase1989PASJ}; [27] \cite{Shidatsu2022ATel15495_MAXI_J0655m013}; [28] \cite{2004ATel_IGR_J11435M6109}; [29] \cite{Nichelli_2011ATel_IGR_J17200m3116}; [30] \cite{Doroshenko2022AnA_SRGA_J124404m632232}; [31] \cite{Reig_1999MNRAS_RX_J08124m3114}; [32] \cite{Krivonos_2015ApJ_2RXPJ1301596m635806}; [33] \cite{Markwardt2010ATel_AXJ17002m4220}; [34] \cite{Miyasaka_2013ApJ_GS_0834m430}; [35] \cite{intZand_2001ApJ_XTE_J1543m568_dist}; [36] \cite{Corbet_2017ApJ_pulse_period}  \\
}
}
\end{table*}

We conducted a study of BeXRB sources in quiescence using the eRASS1 data products available from the eRASS1 sky survey. We also carried out a thorough literature survey for observations of BeXRBs during quiescence from multiple X-ray observatories to have an overall picture of their quiescence luminosity, and combine it with the results of the current study to better understand the detection space across multiple instruments for BeXRB sources in quiescence. We have some sources with better lower luminosity constraints compared to the literature. We checked for possible outbursts of the BeXRB sources for which the eRASS1 data products are available. Sources for which long term monitoring data are available show that they were not in outburst during the eROSITA observations, and since  their luminosity was $\lesssim10^{36}$ erg s$^{-1}$, which is expected from BeXRBs in quiescence \citep{PabloReig2011Ap, Tsygankov2017MNRAS_pulsars_in_quiescence}.

BeXRBs in their quiescent state exhibit luminosity in the range $10^{33}-10^{36}$ erg s$^{-1}$ \citep{PabloReig2011Ap, Tsygankov2017MNRAS_pulsars_in_quiescence}. The NS is away from the highly dense circumbinary region of the optical companion when the BeXRB is in its quiescent state, which is useful to probe the various emission mechanisms of the NS at low mass accretion rates when not undergoing an outburst. The popular emission mechanisms usually used to explain the quiescent luminosity in BeXRBs are i) emission due to magnetospheric accretion in the propeller regime \citep{Illarionov_1975AnA_propeller_regime, Campana2002ApJ_3_sources}, ii) stable accretion from a cold disc at temperature below $\sim$6500\text{ K} in highly magnetized NSs \citep{Tsygankov2017AnA_cold_disc}  and iii) deep crustal heating \citep{Brown1998ApJ_Quiescent}. The X-ray spectra of some BeXRBs in quiescence was observed to be softer when the emission mechanism shifts to the propeller regime in BeXRBs \citep{Tsygankov2017MNRAS_pulsars_in_quiescence}.

\subsection{Spectral analysis of eRASS1 data}

The spectral analysis of quiescent state spectra of BeXRBs from eRASS1 could be modeled with two single component models, i.e., the absorbed powerlaw and the absorbed blackbody (\text{Table \ref{tab:spectral_model_fit}}). Both models give satisfactory fit to the observed data with comparable fit statistics for most sources, and are thus not very distinguishable. The powerlaw model fits sufficiently well for most sources and is able to provide a better constrain on the source flux. We also checked the spectral parameters of the models for any correlations between themselves or with the luminosities of the BeXRBs.

For the absorbed powerlaw model, the powerlaw photon index is not constrained well for some sources. The photon index is consistent with a value of $\sim1.0$. For the blackbody spectral model, the emission region size and the temperature are clustered together for BeXRBs in \text{Figure \ref{fig:radius_vs_temperature}}, this cluster corresponds to a region of size $\sim100-500\text{ m}$ and blackbody temperature of $\sim0.8-1.6\text{ keV}$. From this, it may be inferred that the emission is coming from hot matter of small size comparable to the NS polar cap for the BeXRBs. This is interestingly consistent with many persistent BeXRBs in \cite{La_Palombara_2025AN_persistent_BeXRB}, where they use a two component model to study the origin of the thermal component in persistent BeXRBs. The blackbody temperature falls between $\sim0.8-1.6\text{ keV}$ for most BeXRBs in \text{Figure \ref{fig:temperature_vs_luminosity}} for a range of X-ray luminosity, it may be inferred that they emit at similar temperature over a large range of luminosity. We can observe a clear positive correlation between the radius of blackbody emitting region and the X-ray luminosity for BeXRBs (\text{Figure \ref{fig:radius_vs_luminosity}}). We fit a powerlaw between the luminosity and the radius of emitting region for the BeXRB sources, and get $\alpha=2.3\pm0.2$ for $L\propto R^{\alpha}$.

We note that the spectral properties alone is not enough to certainly determine if the emission is of thermal origin from the surface of a cooling NS or due to a low-level accretion onto the NS \citep{Tsygankov_2016AnA_prop_eff_4U_0115p63_N_V_0332p53, Tsygankov2017MNRAS_pulsars_in_quiescence}, as in the case of disc accretion at a low-level the thermal photons across the accretion channel are unable to Comptonize at low accretion luminosity \citep{Mushtukov_2015MNRAS_optical_thickness}.

From the more sensitive observations of BeXRBs in quiescence made with pointed observations using \textit{Chandra}, \textit{XMM-Newton}, \textit{BeppoSAX}, \textit{NuSTAR}, and more (see, \text{Table \ref{tab:literature_flux_luminosity_1}}), show that single component spectral models like powerlaw, blackbody, cutoff-powerlaw, etc., are sufficient in modeling the observed quiescent spectrum and very rarely multiple component models are used. In cases where multiple spectral components, are used for BeXRBs, the origin of the `soft' thermal component has been explained either as blackbody emission at a temperature of $\sim0.1\text{ keV}$ from a region size that is consistent with NS surface \citep{Zhao2019MNRAS}, or blackbody emission at a temperature of $\sim1\text{ keV}$ with small region size comparable to the NS polar caps \citep{Coburn_2001ApJ_X_Persei, Palombara_RX_J04409p4431, Chandra_2023arXiv_GX_304m1}. We have also checked for the simultaneous presence of both thermal and non-thermal components in the eROSITA spectrum of our sources. We use the spectral model \texttt{tbabs} $\times$ ( \texttt{powerlaw} + \texttt{bbodyrad} ) to fit the eROSITA spectra of quiescent BeXRB. The spectral model gives a decent fit, but with the c-stat values comparable to the powerlaw or blackbody models from above. The spectral fit for this model gives very poor constrain on the spectral parameters and seems to suggest a large region of emission at low temperature $\sim0.1\text{ keV}$, with only upper limits on temperature and emitting region size for many sources in \text{Table \ref{tab:spectral_model_fit}}. Therefore, we conclude that, from the survey data obtained with eROSITA, which has a few hundred seconds of typical exposure, presence of two spectral components could not be established for the quiescent spectra of the BeXRBs.

\subsection{Accretion models for BeXRBs in quiescence}

Primarily two models are considered to explain the accretion process for BeXRBs for low mass accretion rate, i.e., i) magnetospheric accretion when matter does not reach the NS surface due to centrifugal inhibition of the accretion process called the \text{`propeller'} regime and ii) the stable accretion from a cold disc. The limiting luminosity for the onset of propeller regime is given as \citep{Campana2002ApJ_3_sources}
\begin{equation}
    L<L_{\text{prop}}\simeq4\times10^{37}k^{7/2}B_{12}^{2}P^{-7/3}M_{1.4}^{-2/3}R_6^{5} \text{ erg s}^{-1},
\end{equation}
where the factor $k=R_m/R_A$ accounts for the interaction between the magnetosphere and the accretion flow with $R_A$ and $R_m$ being the Alfv\'{e}n and magnetospheric radius, respectively. We use $k\approx0.5$ assuming disc accretion as the accretion flow geometry \citep{Ghosh1978ApJ_accretion_flow_factor}. $B_{12}$ is the magnetic field strength in units of $10^{12}$ G, $P$ is the pulse period in seconds, $M_{1.4}$ is the mass of NS in units of 1.4 M$_\odot$, and $R_6$ is the radius of NS in units of \text{$10^{6}$ cm}. However, once the accretion rate drops below the value corresponding to L$_{\text{prop}}$, the luminosity will jump down to \citep{Campana2002ApJ_3_sources}
\begin{equation}
    L<L_{\text{min}}(r_{\text{cor}})\simeq2.4\times10^{35}k^{7/2}B_{12}^{2}P^{-3}M_{1.4}^{-1}R_6^{6}\text{ erg s}^{-1}.
\end{equation}
where $r_{\text{cor}}$ is the corotation radius. However, this is the bolometric luminosity in the propeller regime and is therefore treated as an upper limit for the X-ray luminosity in the propeller regime.

Another possibility is that, a source may undergo stable accretion from a cold accretion disc at a temperature lower than \text{6500 K} for $R>R_{\text{m}}$ below the limiting luminosity of \citep{Lasota_1997ASPC_cold_disc_temp, Tsygankov2017AnA_cold_disc}
\begin{equation}
    L<L_{\text{cold}}\simeq9\times10^{33}k^{1.5}M_{1.4}^{0.28}R_6^{1.57}B_{12}^{0.86} \text{ erg s}^{-1}.
\end{equation}

Additionally, a critical value for the spin period \citep{Tsygankov2017AnA_cold_disc} calculated from the NS magnetic field is given as
\begin{equation}
    P^{*}=36.6\,k^{6/7}B_{12}^{0.49}M_{1.4}^{-0.17}R_{6}^{1.22} \text{ s},
\end{equation}
where after an outburst, if spin period $P<P^{*}$ ($L_{\text{cold}}<L_{\text{prop}}$), the source will go into the propeller regime, and for $P>P^{*}$ ($L_{\text{cold}}>L_{\text{prop}}$) the source will go through stable accretion from a cold disc.

In the cold disc regime, the matter near the inner-disc radius is around the temperature of $\sim6500\text{ K}$, and is able to interact and follow the magnetic field lines towards the NS poles \citep{Tsygankov2017MNRAS_pulsars_in_quiescence}. Presence of pulsation for a quiescent BeXRB is expected in the cold disc regime. Though, such an analysis is not possible with the current eRASS1 data.

The threshold luminosities for the two accretion regimes are calculated and shown in \text{Table \ref{tab:basic_info}} for all the sources for which the spin period and magnetic field values from cyclotron line measurements are known. For sources where only the spin period is known, we assume a magnetic field strength value of \text{$3\times10^{12}$ G} for completeness in \text{Table \ref{tab:basic_info}}. A source such as \text{SAX J0635.2+0533} with very small pulse period is not included in \text{Table \ref{tab:basic_info}} as it gives extremely large limiting luminosity above $L>10^{38}$ erg s$^{-1}$ for the propeller regime for the assumed magnetic field value of \text{$3\times10^{12}$ G}. The limiting luminosities values calculated for an assumed magnetic field are not to be used very strictly. These limiting luminosities for the two accretion regimes are calculated and shown in \text{Table \ref{tab:basic_info}}, along with the 0.5$-$10 keV unabsorbed luminosity for the powerlaw model for BeXRBs detected in eRASS1, or the luminosity upper limit found for the position of BeXRBs in sky (see, \text{Table \ref{tab:luminosity_upper_limit}}).

From the results reported for the eRASS1 data, we compare the unabsorbed luminosity (or upper limits) of X-ray sources against their limiting luminosity for the propeller and cold disc accretion regimes to examine whether these sources are present in either of these accretion regimes (\text{Table \ref{tab:basic_info}}). We  have 11 X-ray sources for which both the spin period and the magnetic field values from direct measurement of cyclotron line are known. Similarly, we find possible accretion regimes for sources with no known CRSFs for an assumed magnetic field of $3\times10^{12}$ G in \text{Table \ref{tab:basic_info}}. The accretion regime for sources with bounds on their unabsorbed luminosity and for sources with upper-limits on their luminosity known from their position in sky are discussed below.

\subsubsection{Sources with bounds on luminosities}
\label{subsec:source_with_bound_lum}

The BeXRB sources with measured unabsorbed X-ray luminosities (\text{1A 0535$+$262}, \text{4U 0728$-$25}, \text{4U 1036$-$56}, and \text{IGR J17200$-$3116}) have $L_{X}>L_{\text{cold}}$ and $L_{\text{prop}}$, therefore they are undergoing normal accretion. These four sources have $P>P^{*}$, so if the luminosity decreases further, they are likely to transition into stable accretion in the cold disc regime.

The condition for the accretion disc to transition into cold state for stable accretion at low mass accretion rate (i.e. $ L_{\text{cold}}>L_{X}> L_{\text{prop}}$) is fulfilled for 9 sources (GRO J1008$-$57, GX 304$-$1, IGR J06074$+$2205, MXB 0656$-$072, 1A 1118$-$61, 4U 1145$-$61, MAXI J0655$-$013, IGR J11435$-$6109, and SRGA J124404.1$-$632232). Therefore, these sources are likely to be in cold disc accretion regime during the eRASS1 observations.

The source 2S 1417$-$624, has $L_{\text{prop}}>L_{X}>L_{\text{cold}}$ so it fulfills the condition for transition onto propeller regime. But, its luminosity has not fallen below the limiting luminosity  ($L_{\text{min}}(r_{\text{cor}})$), which is expected once the source reaches the propeller regime. The source is also unlikely to transition to the cold disc regime at lower luminosities, as it has $P<P^{*}$. Therefore, the accretion mode is difficult to identify in 2S 1417$-$624.

\subsubsection{Sources with upper-limits on luminosities}

For the sources, 2S 1553$-$542, RX J0812.4$-$3114, AX J1700.2$-$4220 and GS 0834$-$43, their upper-limits on luminosity is less than $L_{\text{prop}}$. Thus, none of these sources are undergoing a normal accretion process. Among these, AX J1700.2$-$4220 has $L_{\text{cold}}>L_{\text{prop}}$, which makes it likely to be in the cold disc accretion regime and not go into the propeller regime. The remaining three sources, 2S 1553$-$542, RX J0812.4$-$3114 and GS 0834$-$43, are therefore likely to be in the propeller regime.

For the source 2RXP J130159.6$-$635806, its upper-limit on luminosity has $L_{X}<L_{\text{cold}}$, along with $L_{\text{cold}}>L_{\text{prop}}$ ($P>P^{*}$), which makes it likely that it is in cold disc accretion regime. For the sources MAXI J1409$-$619, Swift J1626.6$-$5156, AX J163904$-$4642 and XTE J1543$-$568, their upper-limits on luminosity is higher than both the limiting luminosities ($L_{\text{cold}}$ and $L_{\text{prop}}$). The sources MAXI J1409$-$619, Swift J1626.6$-$5156 and AX J163904$-$4642 have $L_{\text{cold}}>L_{\text{prop}}$ ($P>P^{*}$). So, they may either be in normal accretion mode or the cold disc accretion regime. On the other hand, XTE J1543$-$568 has $L_{\text{cold}}<L_{\text{prop}}$ ($P<P^{*}$), so it may either be in normal accretion mode or transitioning into propeller regime.

For very low upper limits on luminosity for undetected BeXRB sources (\text{HD 249179}, \text{IGR J11305$-$6256}, \text{AX J1700.2$-$4220}, HD 63666, or \text{HR 4804}$/$\text{HD 109857}) in \text{Table \ref{tab:luminosity_upper_limit}} and \text{Figure \ref{fig:eROSITA_dist_lumin_plot}}, other emission mechanisms may be possible, such as such as deep-crustal heating \citep{Brown1998ApJ_Quiescent}, where after outbursts, part of the heat conducted to the NS core is then emitted as thermal photon from the NS surface. The luminosity for this model is expected to be around \text{$\sim10^{33}$ erg s$^{-1}$}. Thus, deep-crustal heating is disfavored for the detected BeXRB sources with \text{$L_{X}>10^{33}$ erg s$^{-1}$}.

\subsection{Time since previous X-ray outburst and the eRASS1 X-ray luminosity measurements}

There are different timescales related to the outburst and subsequent decay in BeXRBs. The outbursts often last for 1$-$4 weeks and have a slow decay with a timescale of several tens of days \citep{Paul_Naik_2011BASI_transient_hmxb, Tsygankov_2016AnA_prop_eff_4U_0115p63_N_V_0332p53, Tsygankov2017AnA_cold_disc, Tsygankov_2026_real_time_observations_of_the_transition}. On the other hand, a transition to a propeller regime with a sharp decline in the light-curve usually happens at a time-scale of several tens of hours or less \citep{Tsygankov_2016AnA_prop_eff_4U_0115p63_N_V_0332p53}, whereas the timescale of decline in source flux for the cold-disc accretion regime may last several months to over a year \citep{Tsygankov2017AnA_cold_disc, Rouco_2018AnA_colddisc_quasi_stable_low_acc}. Following the giant outbursts, decay with an e-folding time of around 55 and 48 days (several weeks) has been observed in a few sources like 4U 0115+63 and V0332+53, which may be the crust cooling timescale of the NS or due to the meta-stable low-level accretion state \citep{Wijnands2016MNRAS_V0332p53_n_4U0115p63, Rouco2017MNRAS_4U0115p63, Rouco_2020AnA_4U_0115p63_crust_cooling_low_level_accretion}.

In the case of crust cooling, the cooling seems faster for systems with high magnetic field strength ($B\sim10^{12}$ G) than compared with systems of low magnetic field strength ($B\sim10^{8-9}$ G) (\cite{Rouco_2020AnA_4U_0115p63_crust_cooling_low_level_accretion} referring to the NS LMXBs in \text{Table 7} of \cite{Homan_2014ApJ_decay_timescales_low_mag_field_LMXBs}). Compared to BeXRBs, some NS LMXB systems have very long outbursts heating up the crust out of thermal equilibrium, which can be followed up by very late quiescent cooling of the crust spanning from few years to a decade \citep{Cackett_2008ApJ_MXB1659m29, Cackett_2010ApJ_LMXB_KS1731m260_crustal_cooling, Ootes_2016MNRAS_KS1731m260_crustcool, Deibel_2017ApJ_MXB1659m29_n_KS1731m260, Parikh_2019AnA_LMXB_MXB1659m29_crustcool}.

There have been some rare cases amongst the BeXRBs with unusual cooling patterns. A prolonged quasi-stable low-luminosity state in the cold disc accretion regime for the source GX 304-1 has been explained by replenishment of the cold disc during the periastron passages, which somehow did not cause a full type-I outburst or a noticeable increase in luminosity \citep{Rouco_2018AnA_colddisc_quasi_stable_low_acc}. The quiescence X-ray emission of the BeXRB GRO J1750$-$27 was seen to be variable at very low luminosities due to very low-level accretion onto the fast rotating magnetised NS \citep{Rouco_2019AnA_GROJ1750m27_lowlevel_acc}.

\begin{table}[h]
\centering
\caption{Time interval after the last major outburst for sources from the long-term lightcurve plots.}
\label{tab:outburst_time}
\renewcommand{\arraystretch}{1.0}
\setlength{\tabcolsep}{2pt}
\begin{tabular}{l l r r}
\toprule
No. & Source        & Time Interval after   & Luminosity                \\
    &               & major outburst peak   & ($10^{34}$ erg s$^{-1})$  \\
    &               & ($\sim$days)          &                           \\  \hline
1   & 1A 0535+262   & 102                   & $11.3$                    \\
2   & 2S 1417-624   & 662                   & $<13.1$                   \\
3   & GRO J1008-57  & 187                   & $1.0$                     \\
4   & GX 304-1      & 1350                  & $0.7$                     \\
5   & MXB 0656-072  & 4216                  & $<4.2$                    \\
6   & 1A 1118-61    & 4014                  & $0.35$                    \\
\hline
\end{tabular}
\end{table}

We have examined the eROSITA observations in the context of the previous outbursts for the sources in \text{Table \ref{tab:outburst_time}}. In \text{Table \ref{tab:outburst_time}}, we provide the time since the peak of last outburst for six sources. For sources without an observable major outburst in their long-term MAXI/GSC or Swift/BAT lightcurves, we can say that they have an lower limit of about $\sim15$ years since their last outburst before the eRASS1 observations (considering time from the start of Swift/BAT observation). The BeXRB sources 4U 1036-56 and MAXI J0655-013 have a lower-limit of $\sim8$ and $\sim11$ years, respectively, for time since outburst before their eRASS1 observation from their long-term lightcurves.

These source observations in the eRASS1 survey was carried out long after their previous outbursts. After the outburst is over, the sources have to transition from normal accretion to propeller regime, or normal accretion to a cold disc regime and the path depends on the combination of magnetic field strength and spin period of the NS, as mentioned in \text{Section 4.2}. In the absence of any appreciable accretion, the crust-cooling phenomena may be observable. The ongoing emission mechanisms for the individual sources in the eROSITA data presented here have been discussed in the previous sections.

The source 1A 0535$+$262 seems to still be undergoing normal accretion and has not settled into quiescence. Similarly, 2S 1417$-$624 may not have settled into quiescence with its high upper-limit on luminosity even after about $\sim600$ days. From earlier discussion, 2S 1417$-$624 is possibly moving towards propeller regime though it is unclear. It may still be undergoing normal accretion after the outburst. The BeXRB GRO J1008$-$57 is observed at a luminosity of $10^{34}$ erg s$^{-1}$ about $\sim6$ months after outburst and seems to be cold disc accretion regime from its luminosity from the discussion above. This is in agreement with the timescale of decay expected for cold-disc regime \citep{Tsygankov2017AnA_cold_disc, Rouco_2018AnA_colddisc_quasi_stable_low_acc}. \text{GX 304$-$1} is observed at a luminosity of $0.7\times10^{34}$ erg s$^{-1}$ about $\sim3.6$ years after the outburst. Previously, \text{GX 304$-$1} was observed by NuSTAR to be in the cold-disc accretion regime after nearly 2 years from its last outburst \citep{Rouco_2018AnA_colddisc_quasi_stable_low_acc}. The eRASS1 observation was made about $\sim1.6$ years after the NuSTAR observation and the source seems to still be in the cold-disc accretion regime from earlier discussion. Also, \text{GX 304$-$1} was suggested to be in a prolonged low-luminosity state for about five years during which the eRASS1 observation was made \citep{GX304m1_2025PASA_cold_disc_regime_5yrs}. The BeXRBs MXB 0656$-$072 was observed at a luminosity upper-limit of $<4.2\times10^{34}$ erg s$^{-1}$ after $\sim11.5$ years from the last outburst, and 1A 1118$-$61 was observed at a luminosity of $0.35\times10^{34}$ erg s$^{-1}$ about $\sim11$ years from its last outburst. After such long time from outbursts, crustal cooling may be ignored due to faster cooling decay in BeXRBs sources \citep{Rouco_2020AnA_4U_0115p63_crust_cooling_low_level_accretion}. The discussion earlier had put these sources in the cold disc regime, it could be that they are sustained because of residual accretion by the NS.

\subsection{Luminosity and Detection space of BeXRBs}

As listed in \text{Table \ref{tab:literature_flux_luminosity_1}}, extensive investigations have been performed on the quiescent characteristics (spectral parameters and pulsations) using deep observations of some well known BeXRBs carried out with very sensitive observatories. The eRASS1 data on the other hand, offers a different detection space for the BeXRBs in quiescence.

For understanding the detection space of quiescent BeXRBs with eRASS1 we consider the powerlaw fluxes as it gave better constraints compared to the thermal model (see, \text{Table \ref{tab:spectral_model_fit}}). The \text{0.5$-$10 keV} luminosity of the sources are plotted against their distances for the absorbed powerlaw model in  \text{Figure \ref{fig:eROSITA_dist_lumin_plot}}. The X-ray luminosity of the detected BeXRBs with a NS companion in the eRASS1 main catalogue, are in the range of $\sim10^{33}-10^{36}$ erg s$^{-1}$ and spread across a distance of $\sim2-10\text{ kpc}$, and plotted in red in \text{Figure \ref{fig:eROSITA_dist_lumin_plot}}. Also, the upper-limit on luminosity for the undetected BeXRB sources as given in \text{Table \ref{tab:luminosity_upper_limit}}, are plotted in blue in \text{Figure \ref{fig:eROSITA_dist_lumin_plot}}.

These upper limits on luminosity are found to be very low for some of the undetected BeXRB sources. We see that for the source \text{HR 4804}$/$\text{HD 109857}, the luminosity upper limit is as low as $\sim1.5\times10^{30}$ erg s$^{-1}$, which puts this in the expected luminosity range of $L_{\text{X}}\sim10^{29}-10^{33}$ erg s$^{-1}$ for accretion onto a white dwarf (WD) in a BeXRB indicating that the source \text{HR 4804} likely has a WD as its compact object \citep{Waters_1989AnA_Be_p_WD}. \text{HR 4804} is a known candidate for a BeXRB system with a WD as its compact object companion (see Sect. 4, \cite{Waters_1989AnA_Be_p_WD}). Similarly, \text{HD 63666} shows very low luminosity upper limit at $\sim1.1\times10^{31}$ erg s$^{-1}$, which is within the Be$+$WD luminosity range. The source \text{HD 63666} is cataloged as a BeXRB in \cite{Liu_2006AnA_catalogue}, but the nature of its compact object is unknown. For the remaining sources, the upper limits on luminosity for source position are within the range of $\sim10^{32}-10^{34}$ \text{erg s$^{-1}$}. The eROSITA detection space of the instrument for the BeXRB sources in quiescence has a net exposure mostly ranging in $\sim200-600\text{ s}$ for most sources.

We plot the luminosity values from \text{Table \ref{tab:spectral_model_fit} \& \ref{tab:luminosity_upper_limit} } for the eRASS1 data against the source distance in \text{Figure \ref{fig:dist_lumin_plot}}, along with the luminosity values found from previous observations of BeXRB sources in quiescence (see, \text{Table \ref{tab:literature_flux_luminosity_1}}). For flux $\lesssim10^{-12}$ erg cm$^{-2}$ s$^{-1}$ we could only get an upper-limit on luminosity from the eRASS1 data (\text{Figure \ref{fig:dist_lumin_plot}}). Compared to that, the \textit{Chandra} data is seen to be well constrained even for flux values going as low as $\sim10^{-14}$ erg cm$^{-2}$ s$^{-1}$. Legacy X-ray instruments like Chandra are capable of performing deep observation for dim sources, however they have to use long observation durations and are able to observe only a small part of the sky at a time. The advantage of a survey instrument like eRosita is the ability to cover large part of the sky with only a small fraction of observation duration of legacy instruments. This allows us to get a strong luminosity constrains on nearby sources which is a unique advantage of eRosita.

\subsection{Future prospects}

In this study, along with sources with good constrain on luminosity, the cases with low upper limits on luminosity are of similar physical significance as they may yet explain the possible nature of the compact object companion for a BeXRBs without NS or the possible accretion regime for a BeXRB with NS. In this study, we clearly see the effectiveness of the survey capabilities of eRosita to observe and understand the properties of relatively nearby BeXRBs in quiescence. Future observation of BeXRBs previously undetected by eRASS1 may help detect them and constrain their spectra and quiescence luminosity. These have been undetected due to low flux coming from BeXRB in some accretion regime during quiescence.

With further long-term surveys of BeXRBs with eROSITA we may be able to study the changes in the quiescence luminosity after possible outbursts, which may also further our understanding of BeXRBs in quiescence. This may also help us see how the mass accretion regime of BeXRBs at low accretion luminosities evolve after outbursts in these sources or in absence of them.

\begin{acknowledgments}

The authors appreciate the referee for their useful and detailed suggestions. Their comments have been helpful in improving the quality of the manuscript and make the necessary revisions accordingly. The authors are sincerely grateful for the valuable discussions and insights provided by Chandreyee Maitra to this work. Their feedback was essential in improving the quality of this study. This work is based on data from eROSITA, the soft X-ray instrument aboard SRG, a joint Russian-German science mission supported by the Russian Space Agency (Roskosmos), in the interests of the Russian Academy of Sciences represented by its Space Research Institute (IKI), and the Deutsches Zentrum für Luft- und Raumfahrt (DLR). The SRG spacecraft was built by Lavochkin Association (NPOL) and its subcontractors, and is operated by NPOL with support from the Max Planck Institute for Extraterrestrial Physics (MPE). The development and construction of the eROSITA X-ray instrument was led by MPE, with contributions from the Dr. Karl Remeis Observatory Bamberg \& ECAP (FAU Erlangen-Nuernberg), the University of Hamburg Observatory, the Leibniz Institute for Astrophysics Potsdam (AIP), and the Institute for Astronomy and Astrophysics of the University of Tübingen, with the support of DLR and the Max Planck Society. The Argelander Institute for Astronomy of the University of Bonn and the Ludwig Maximilians Universität Munich also participated in the science preparation for eROSITA.

\end{acknowledgments}

\restartappendixnumbering

\newpage 

\appendix

\section{Long-term lightcurve plots}

\begin{figure*}
    \centering
    \begin{subfigure}[b]{0.33\textwidth}
        \centering
        \includegraphics[width=\textwidth]{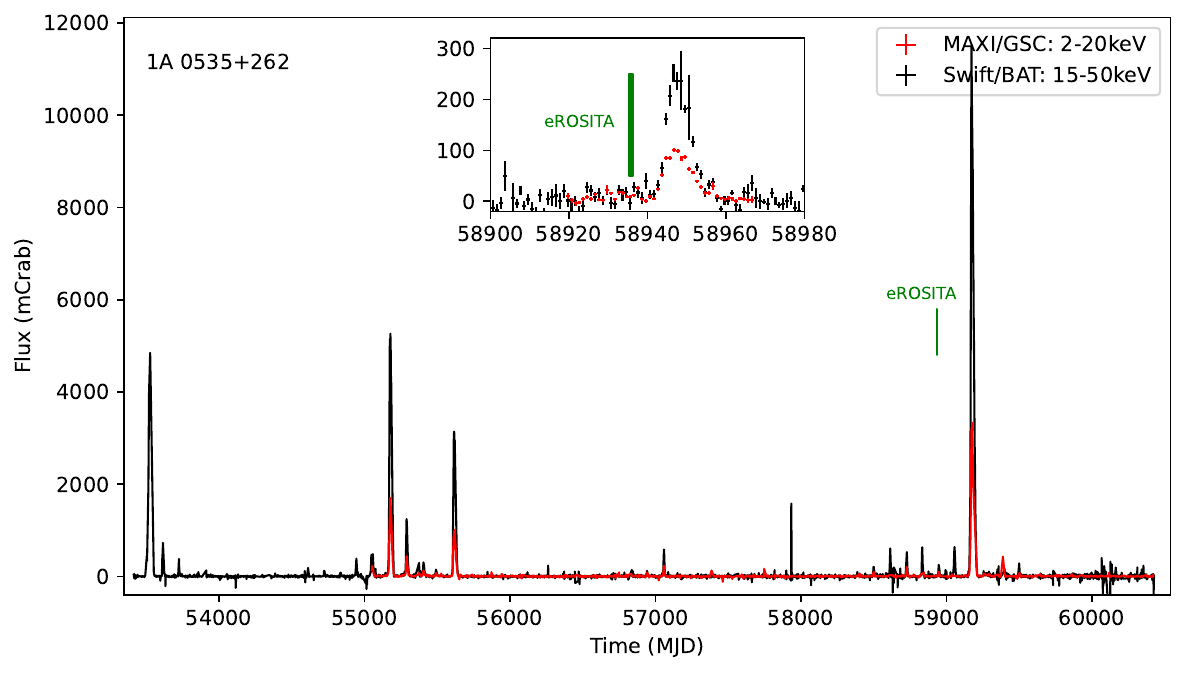}
        \caption{1A 0535$+$262}
        \label{fig:1A_0535p262}
    \end{subfigure}
    \begin{subfigure}[b]{0.33\textwidth}
        \centering
        \includegraphics[width=\textwidth]{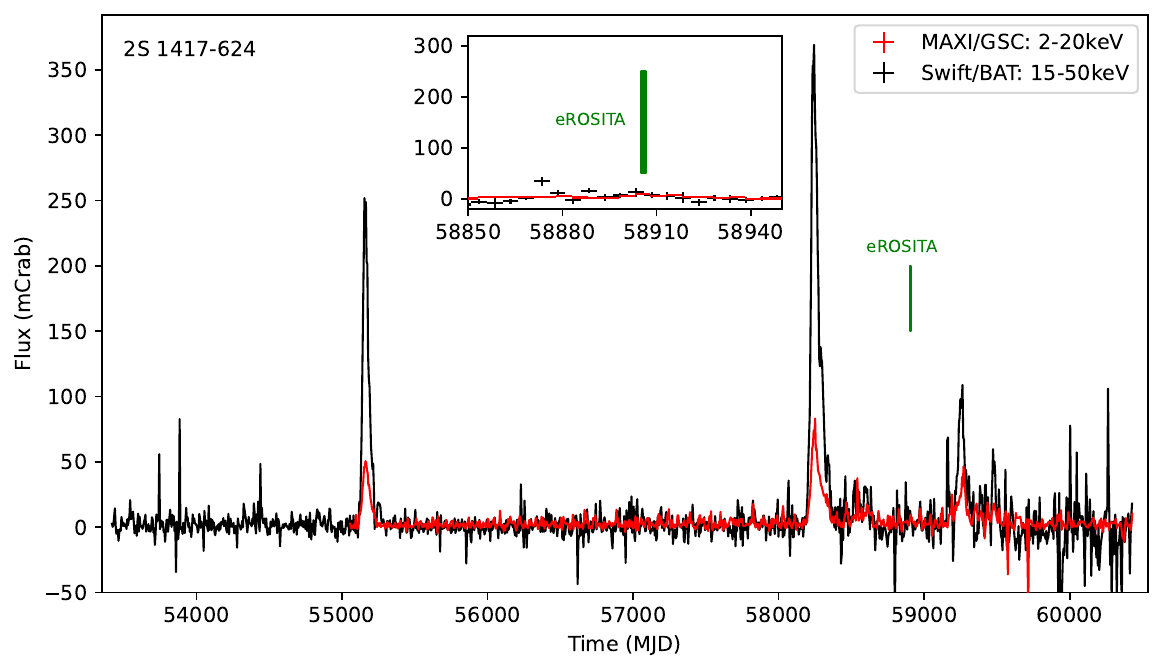}
        \caption{2S 1417$-$624}
        \label{fig:2S_1417m624}
    \end{subfigure}
    \begin{subfigure}[b]{0.33\textwidth}
        \centering
        \includegraphics[width=\textwidth]{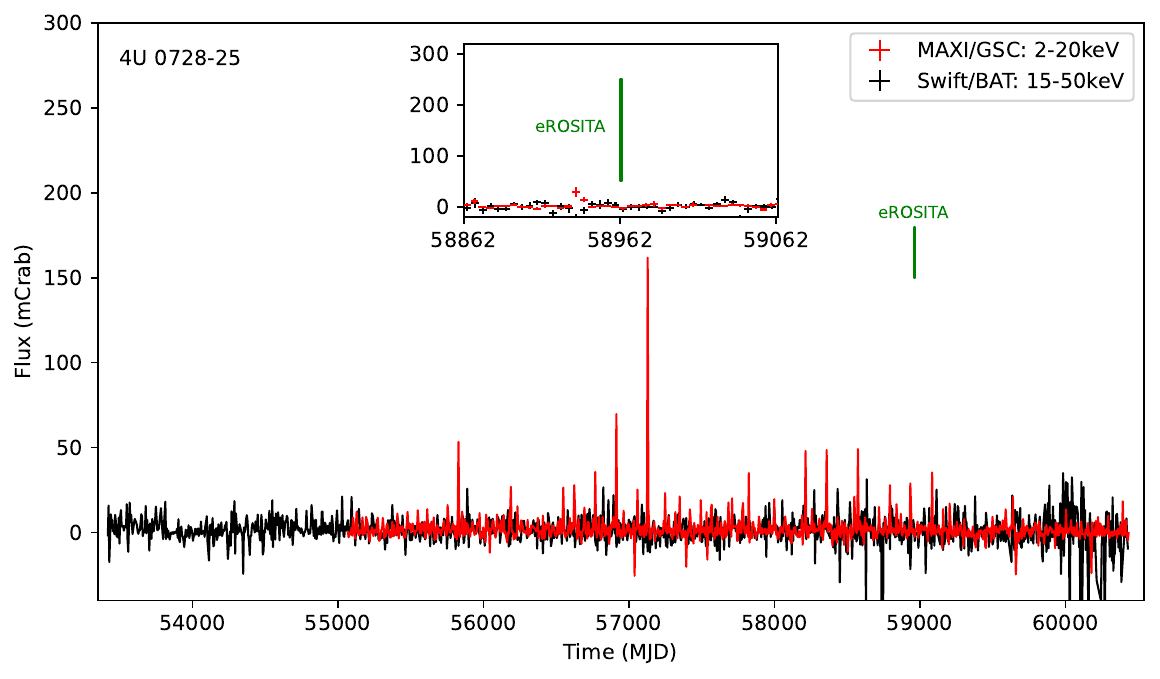}
        \caption{4U 0728$-$25}
        \label{fig:4U_0728m25}
    \end{subfigure}
    \begin{subfigure}[b]{0.33\textwidth}
        \centering
        \includegraphics[width=\textwidth]{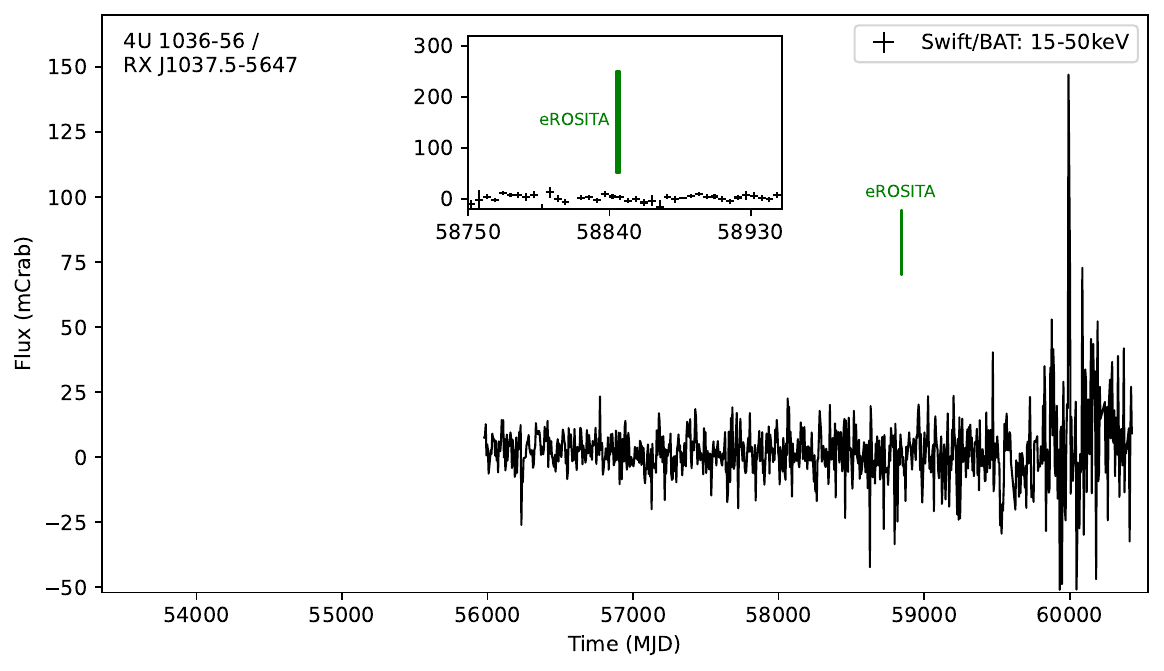}
        \caption{4U 1036$-$56}
        \label{fig:4U_1036m56}
    \end{subfigure}
    \begin{subfigure}[b]{0.33\textwidth}
        \centering
        \includegraphics[width=\textwidth]{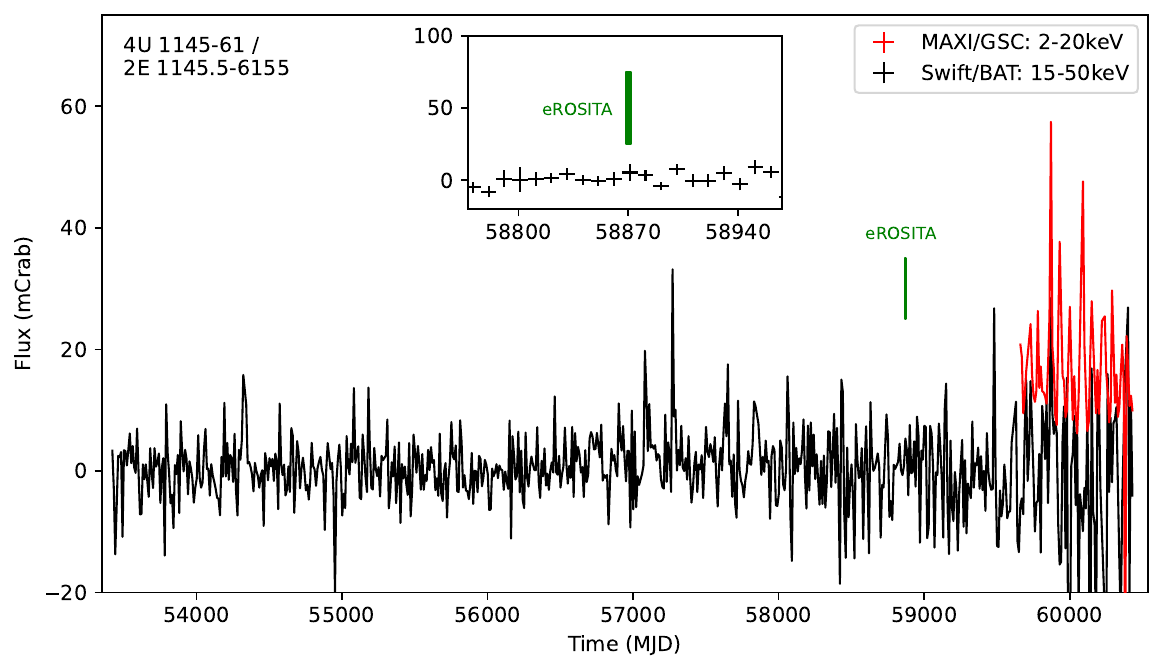}
        \caption{4U 1145$-$61}
        \label{fig:4U_1145m61}
    \end{subfigure}
    \begin{subfigure}[b]{0.33\textwidth}
        \centering
        \includegraphics[width=\textwidth]{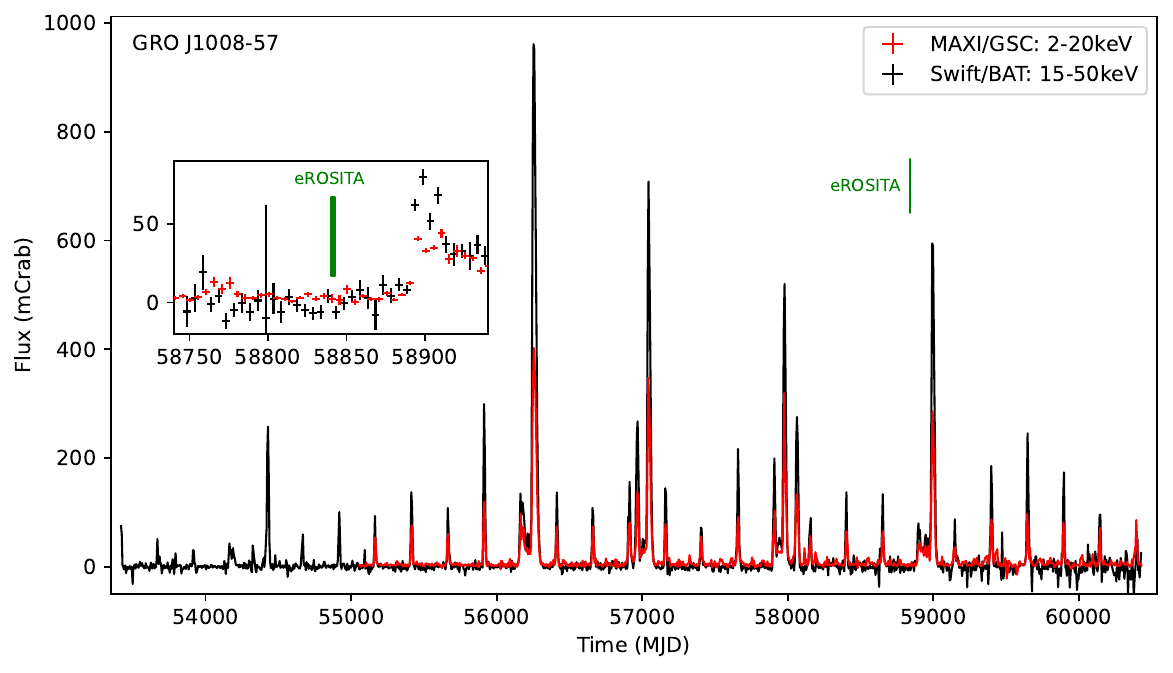}
        \caption{GRO J1008$-$57}
        \label{fig:GRO_J1008m57}
    \end{subfigure}
    \begin{subfigure}[b]{0.33\textwidth}
        \centering
        \includegraphics[width=\textwidth]{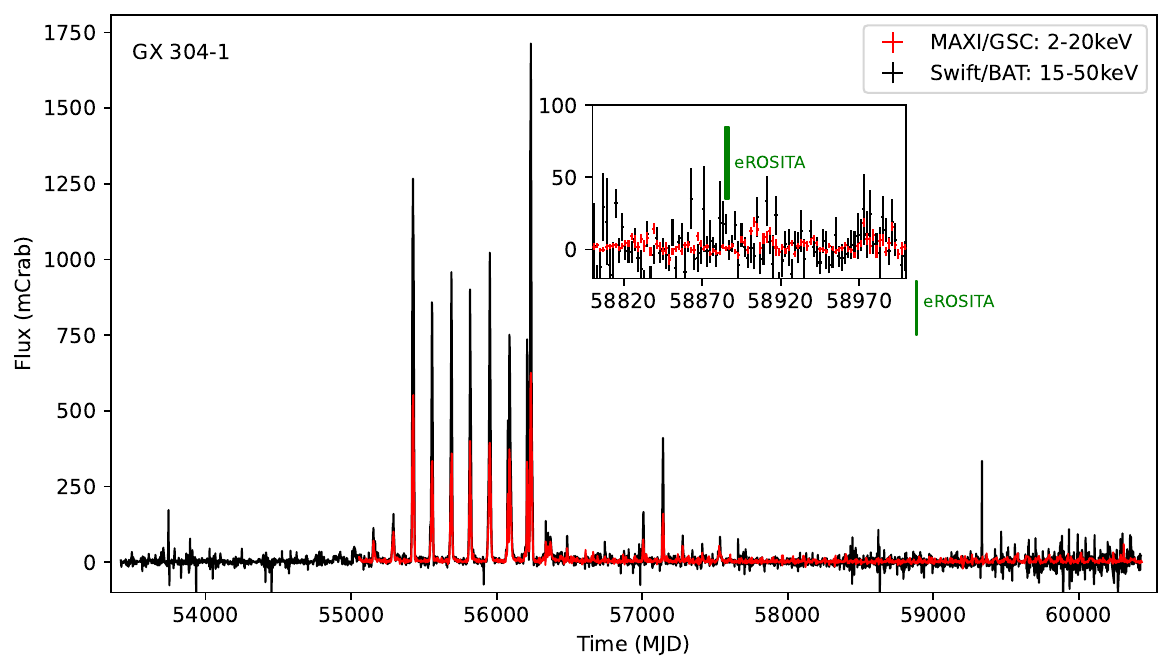}
        \caption{GX 304$-$1}
        \label{fig:GX_304m1}
    \end{subfigure}
    \begin{subfigure}[b]{0.33\textwidth}
        \centering
        \includegraphics[width=\textwidth]{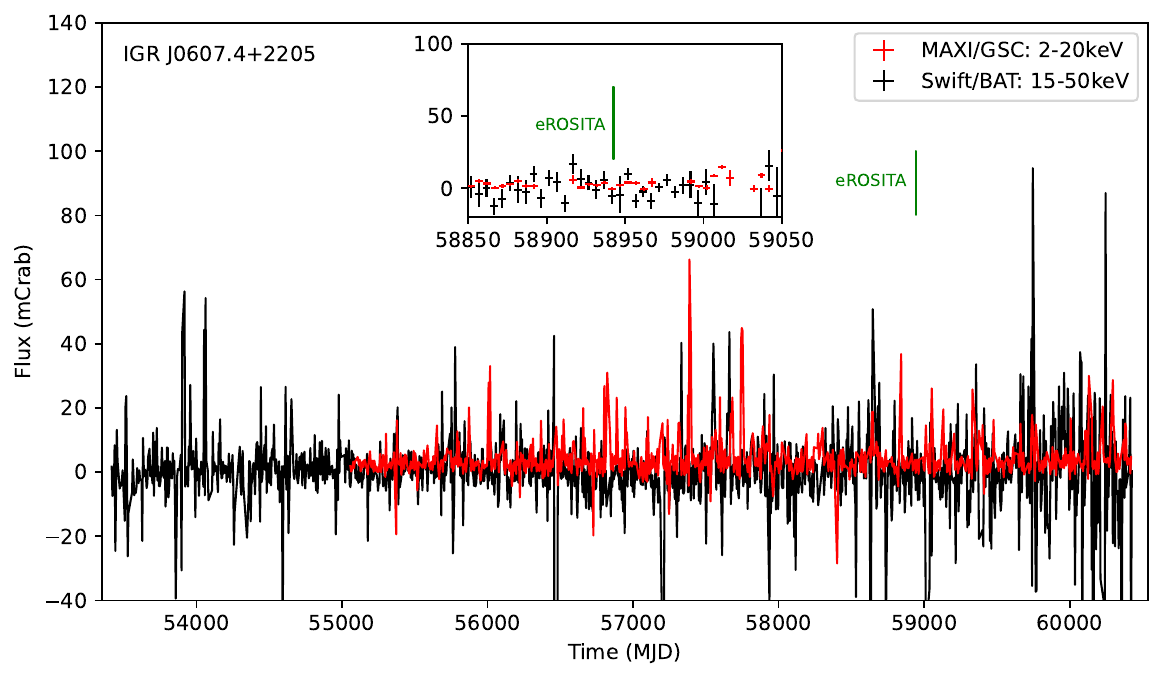}
        \caption{IGR J06074$+$2205}
        \label{fig:IGR_J06074p2205}
    \end{subfigure}
    \begin{subfigure}[b]{0.33\textwidth}
        \centering
        \includegraphics[width=\textwidth]{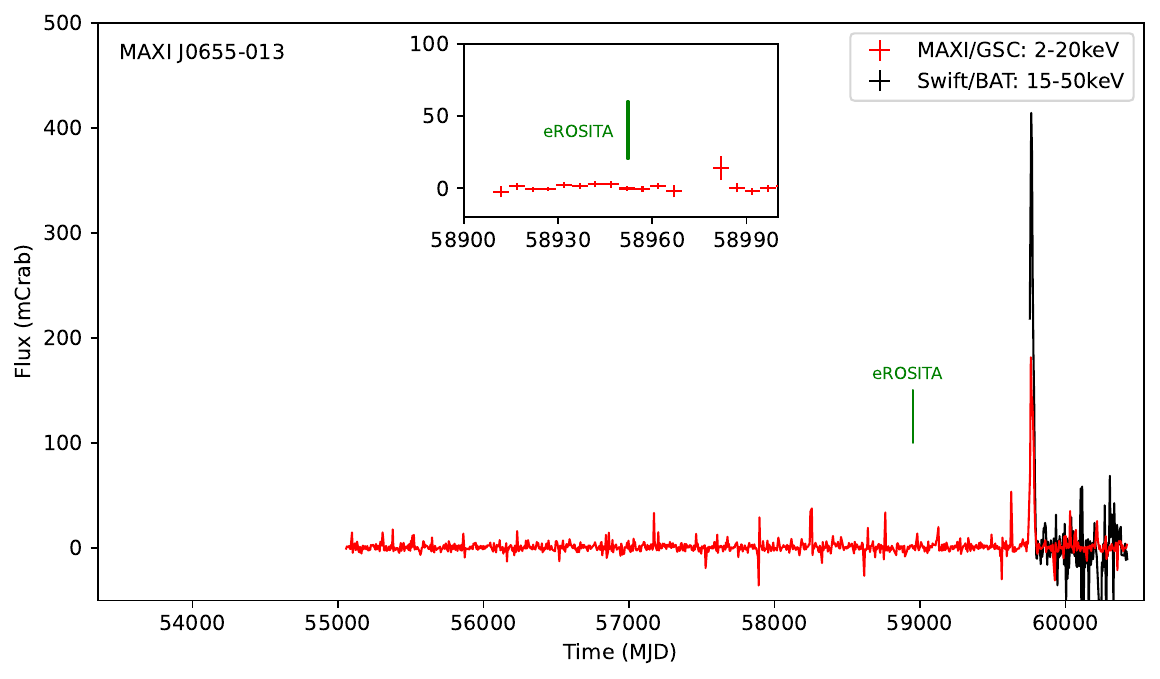}
        \caption{MAXI J0655$-$013}
        \label{fig:MAXI_J0655m013}
    \end{subfigure}
    \begin{subfigure}[b]{0.33\textwidth}
        \centering
        \includegraphics[width=\textwidth]{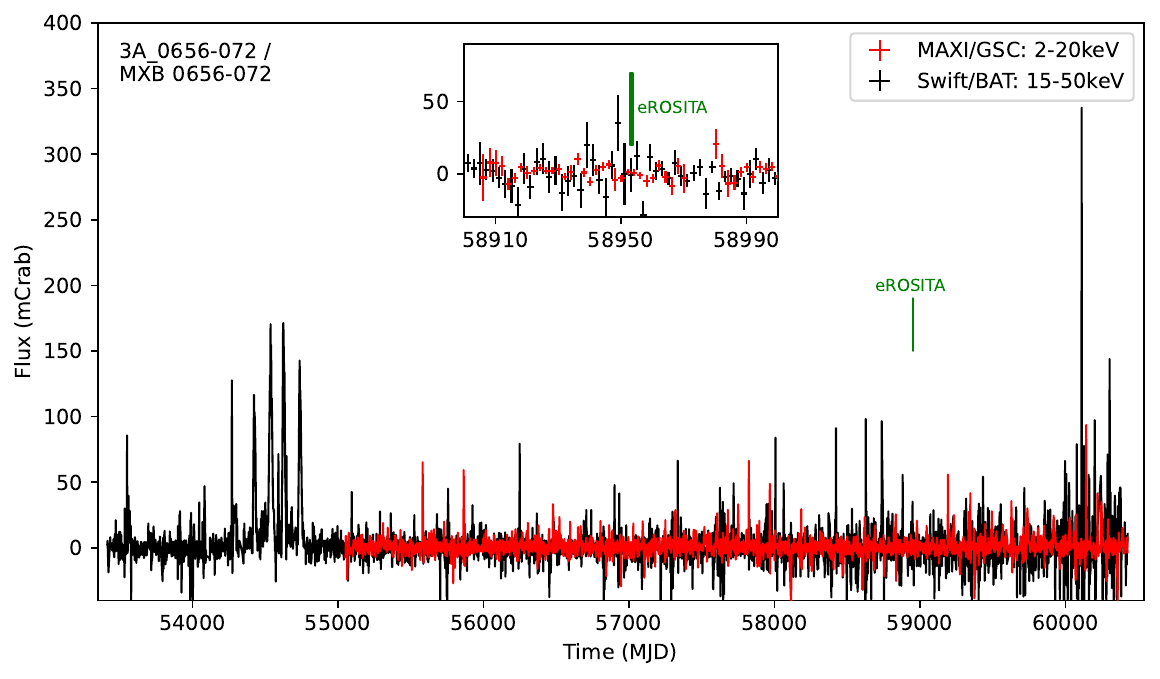}
        \caption{MXB 0656$-$072}
        \label{fig:MXB_0656m072}
    \end{subfigure}
    \begin{subfigure}[b]{0.33\textwidth}
        \centering
        \includegraphics[width=\textwidth]{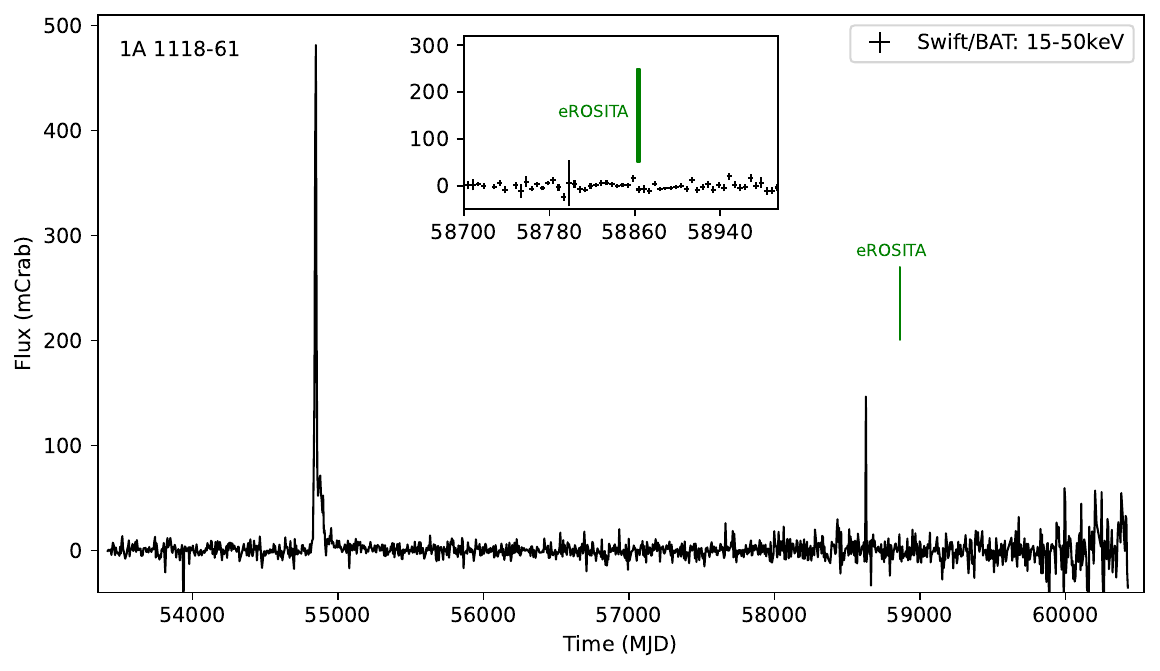}
        \caption{1A 1118$-$61}
        \label{fig:1A_1118m61}
    \end{subfigure}
    \begin{subfigure}[b]{0.33\textwidth}
        \centering
        \includegraphics[width=\textwidth]{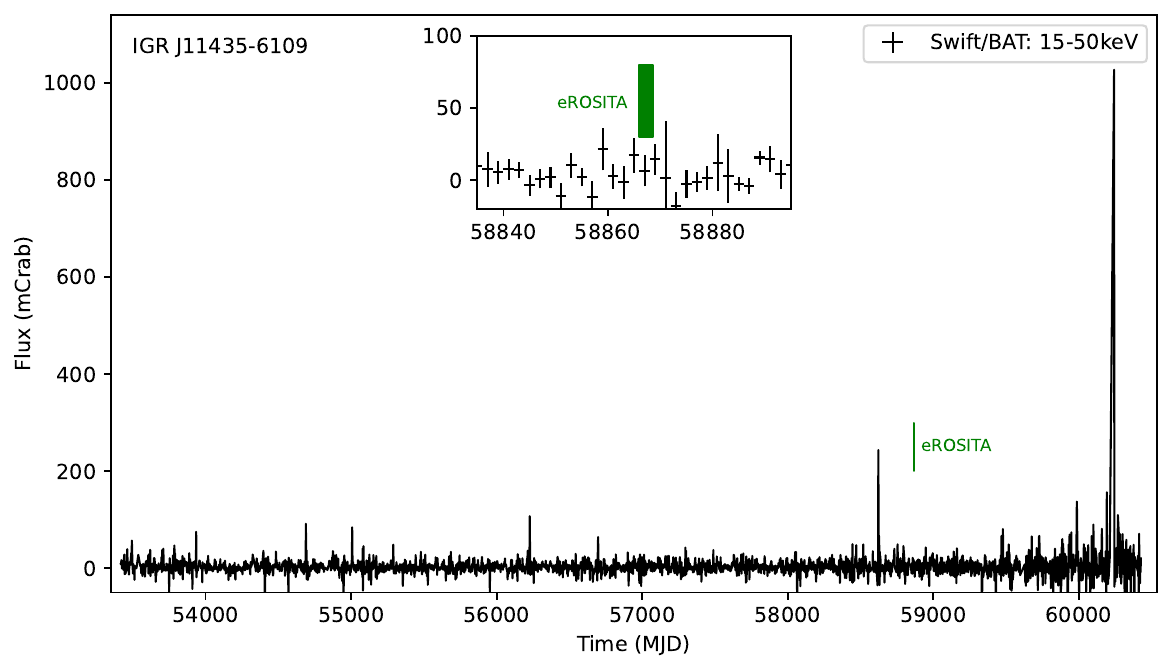}
        \caption{IGR J11435$-$6109}
        \label{fig:IGR_J11435m6109}
    \end{subfigure}
    \begin{subfigure}[b]{0.33\textwidth}
        \centering
        \includegraphics[width=\textwidth]{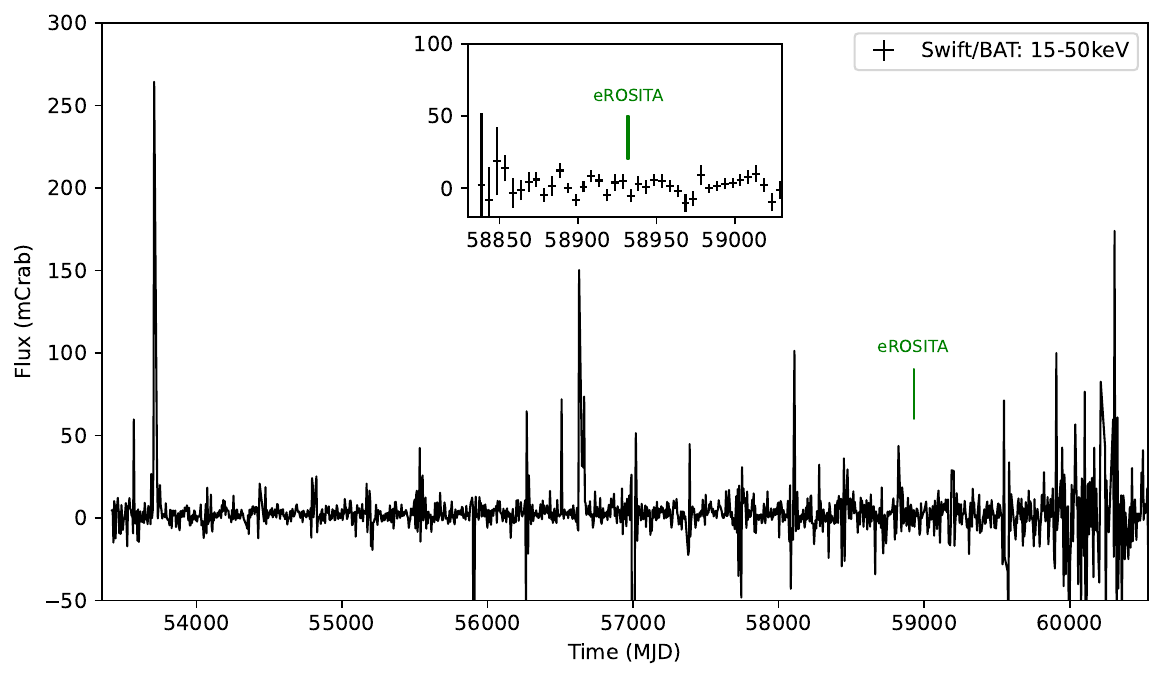}
        \caption{IGR J17200$-$3116}
        \label{fig:IGR_J17200m3116_outburst}
    \end{subfigure}
    \caption{Long-term light-curve of X-ray sources for the 2$-$20 keV MAXI/GSC and 15$-$50 keV Swift/BAT data with a time bin of 1$-$5 days plotted in red and black, respectively. The time interval within which the region around any source was observed in eRASS1 survey is marked by a green vertical line.}
    \label{fig:Outbursts}
\end{figure*}

\clearpage

\section{\text{\texorpdfstring{$\gamma$}{gamma} Cas}-type sources or gamma-ray binaries.}

\texorpdfstring{$\gamma$}{gamma}

\begin{figure*}
    \centering
    \begin{subfigure}[b]{0.33\textwidth}
        \centering
        \includegraphics[width=\textwidth]{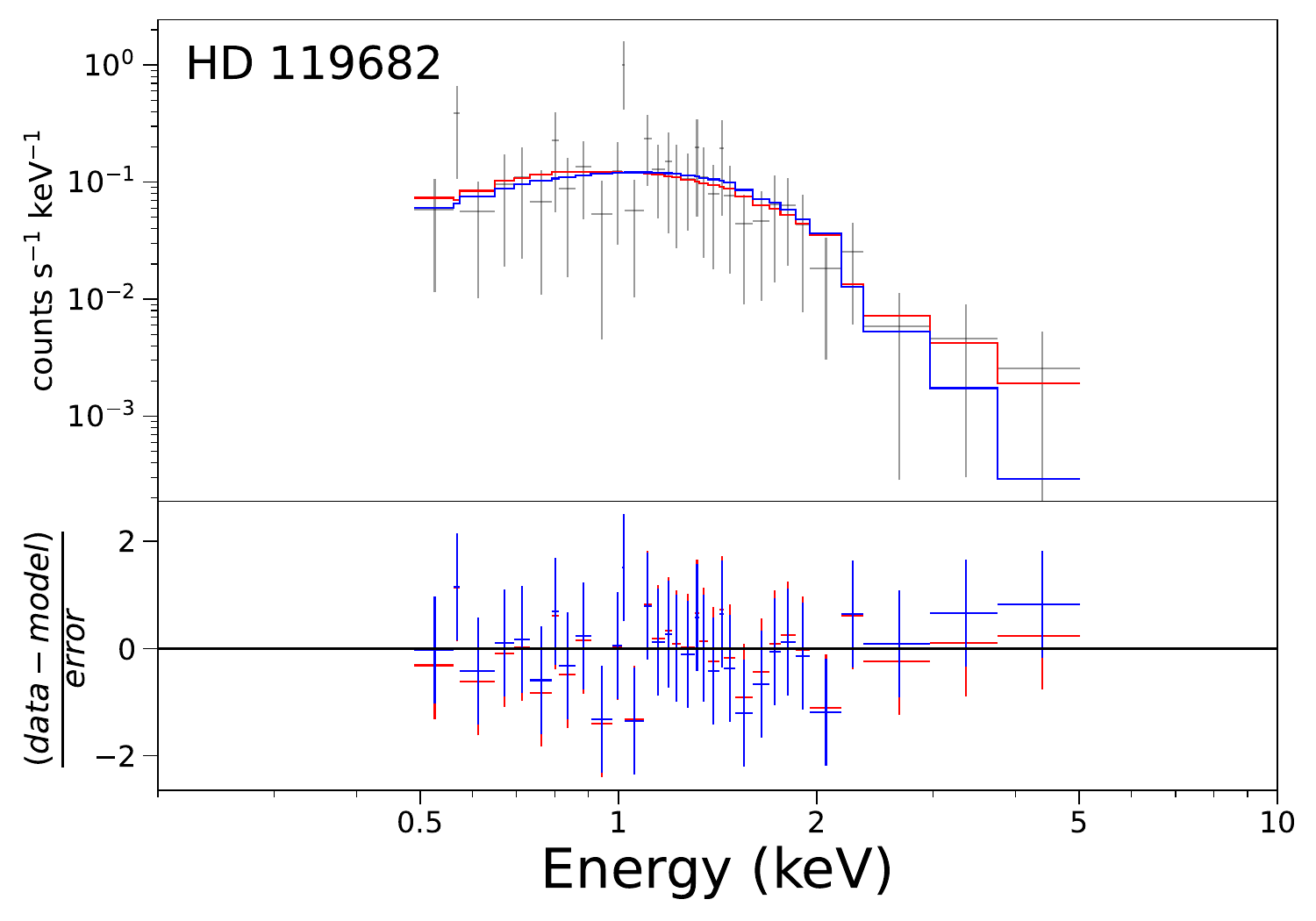}
        \caption{HD 119682}
    \end{subfigure}
    \begin{subfigure}[b]{0.33\textwidth}
        \centering
        \includegraphics[width=\textwidth]{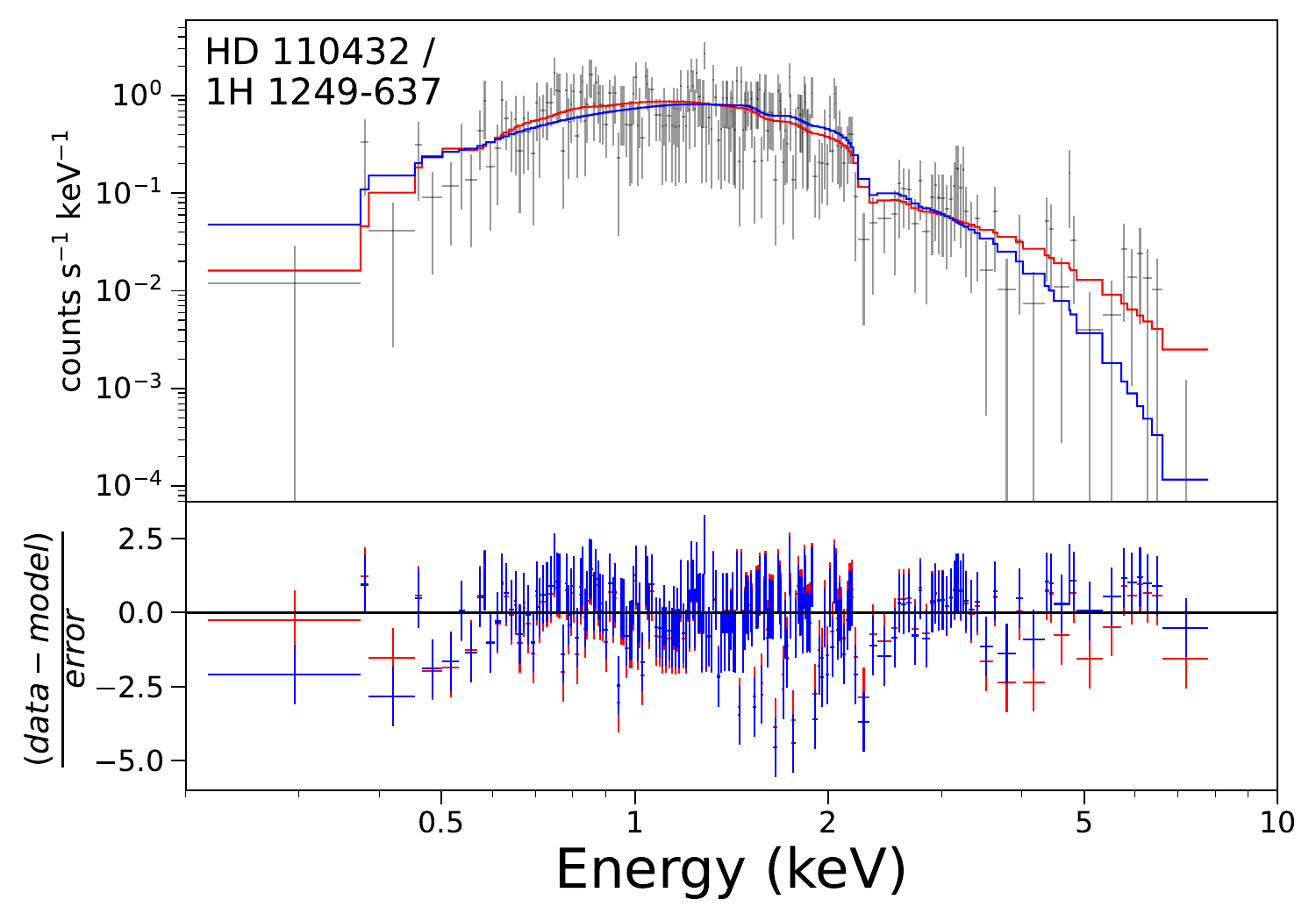}
        \caption{HD 110432 $/$ 1H 1249$-$637}
    \end{subfigure}
    \begin{subfigure}[b]{0.33\textwidth}
        \centering
        \includegraphics[width=\textwidth]{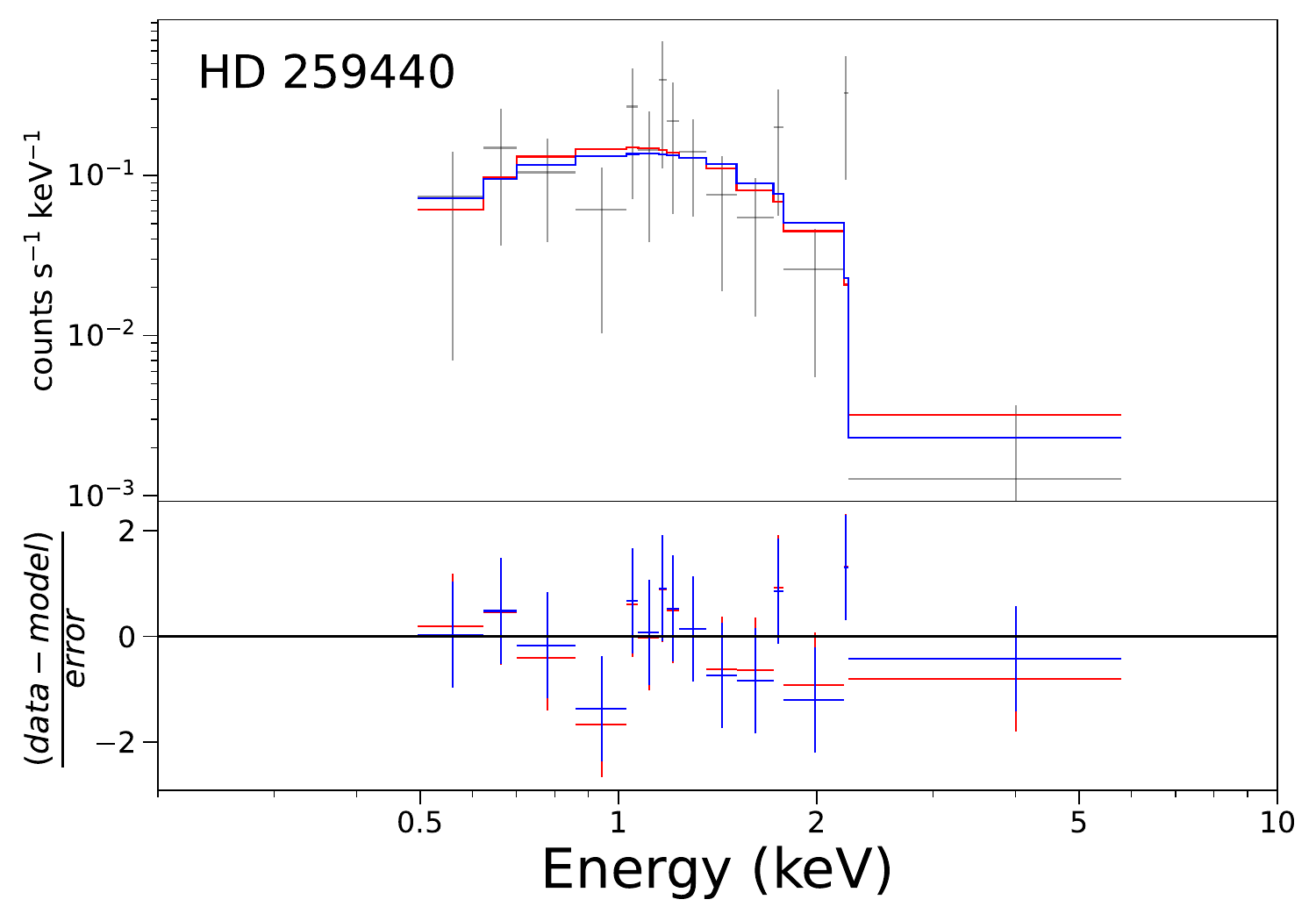}
        \caption{HD 259440}
    \end{subfigure}
    \begin{subfigure}[b]{0.33\textwidth}
        \centering
        \includegraphics[width=\textwidth]{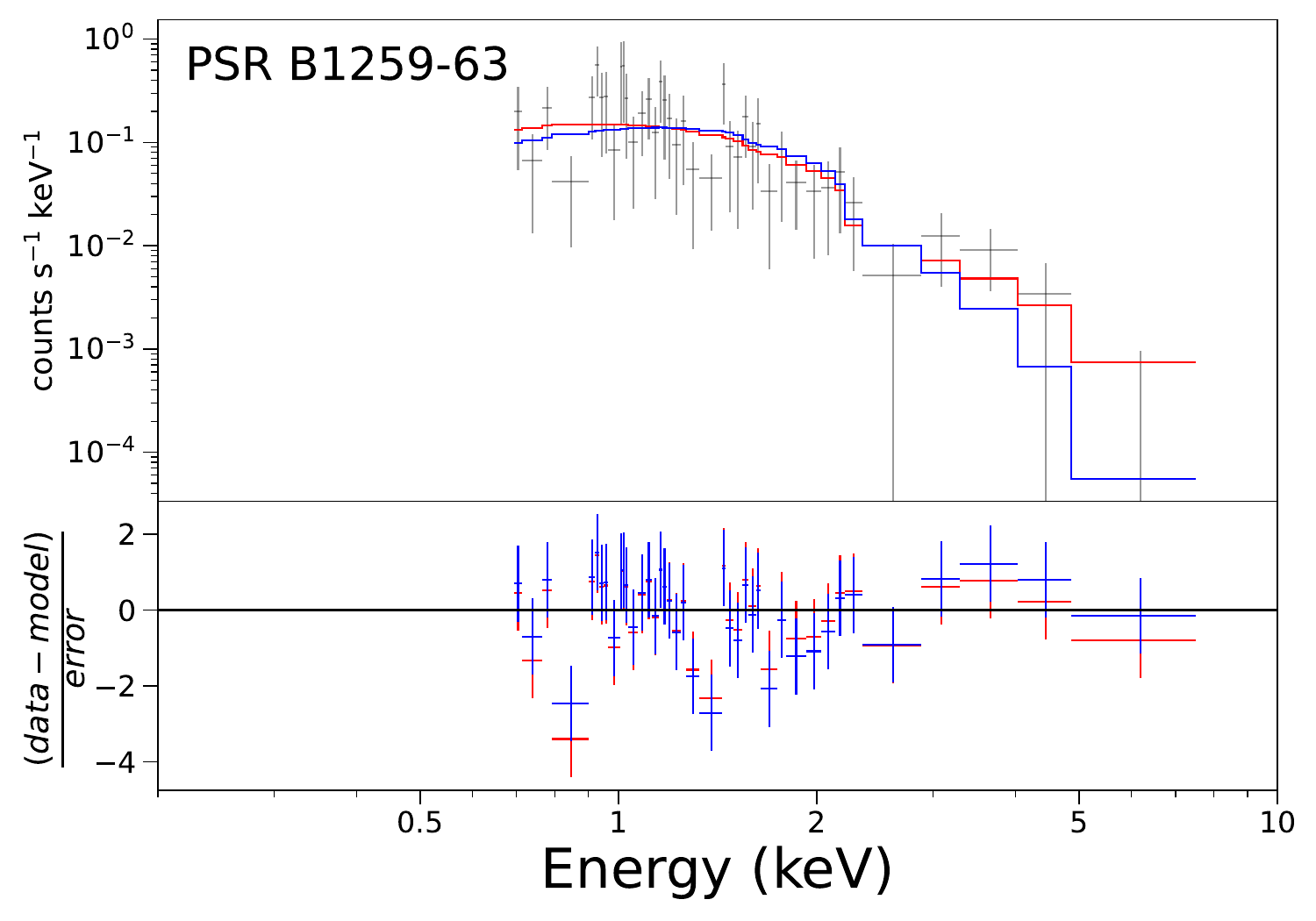}
        \caption{PSR B1259$-$63}
    \end{subfigure}
    \caption{Same as \text{Figure \ref{fig:spectral_best_fit}}.}
    \label{fig:appendix_spectral_best_fit_plot}
\end{figure*}

In Table \ref{tab:appendix_spectral_fit}, we have sources that are not typical BeXRBs in which a NS is the compact star. \text{HD 119682} \citep{ Rakowski_2006ApJ_HD_119682, Torrej_2013ApJ_HD_119682} and \text{HD 110432} $/$ \text{1H 1249$-$637} \citep{Lopes2007AnA_HD_110432_gamma_cas} are $\gamma\text{ Cas}$-type source, a sub-class of HMXBs. The $\gamma\text{ Cas}$-type source may emit X-ray by accreting matter from a Be star onto a white dwarf \citep{Haberl_1995AnA_gammaCas}, or due to interaction between the magnetic field and inner part of circumstellar disc of the Be star \citep{Robinson_2002ApJ_Be_Star_Possible_Magnetic_Dynamo}. Also, the sources \text{HD 259440} $/$ \text{HESS J0632$+$057} \citep{Hinton_2009ApJ_HESS_J0632p057}  and PSR B1259$-$63 \citep{Aharonian_2005AnA_PSR_B1259m63} in Table \ref{tab:appendix_spectral_fit} are another type of HMXBs called the $\gamma$-ray binaries, which emit radiation in the MeV$-$TeV range due to non-thermal emission from the shock region formed between the pulsar wind and the stellar wind from the O$/$Be-type companion star \citep{Dubus_2013AnA_Gamma_Ray_Summary}. The $\gamma$ Cas-type$/$$\gamma-$ray binaries were detected at a luminosity between $L_{x}\sim10^{32}-10^{33}$ erg s$^{-1}$ and at a distance of $\sim0.4-2\text{ kpc}$. Their best-fit spectra are shown in \text{Figure \ref{fig:appendix_spectral_best_fit_plot}}.

\begin{table*}
\caption{List of the X-ray sources and their properties. Best-fit spectral parameters for the \text{0.5$-$8 keV} eRASS1 data fitted for the two spectral models, i.e., i) \texttt{tbabs} $\times$ \texttt{powerlaw}, and ii)  \texttt{tbabs} $\times$ \texttt{bbodyrad}. The upper-limits on flux has a one-sided confidence level of 95$\%$. }
\label{tab:appendix_spectral_fit}
\renewcommand{\arraystretch}{2.5}
\centering
\tiny
\setlength\tabcolsep{5pt}
\begin{tabular}{l|l|cccc|cccccccc}
\hline
   & Source                 & Distance$^{a}$        & Orbital               & Spin Period           & Cyclotron     & Exposure      & N$_H$                         & $\Gamma$                  & kT$_{bb}$                 & R$_{BB}$                  & Observed Flux                         & Luminosity                    & c-stat        \\
   &                        & (pc)                  & Period                & (s)                   & Line          & (seconds)     & ($10^{22}$ cm$^{-2}$)        &                           & (KeV)                     & (km)                      & $(0.5-10\text{ keV})$                  & $(0.5-10\text{ keV})$         & (dof)         \\
   &                        &                       & (d)                   &                       & (keV)         &               &                               &                           &                           &                           & ($10^{-12}$ erg s$^{-1}$ cm$^{-2}$)   & ($10^{34}$ erg s$^{-1}$)    & \\  \hline
1  & HD 110432 $/$          & $439^{+15}_{-15}$     & $-$                   & $-$                   & $-$           & 526.5         & $0.32^{+0.09}_{-0.08}$        & $1.5\pm0.3$               & $-$                       & $-$                       & $9.3^{+2.1}_{-1.8}$                   & $0.021^{+0.007}_{-0.005}$     & 200.86(202)   \\
   & 1H 1249$-$637          &                       &                       &                       &               &               & $<0.07$                       & $-$                       & $0.67^{+0.08}_{-0.07}$    & $0.07\pm0.01$             & $5.2^{+0.5}_{-0.9}$                   & $0.012^{+0.002}_{-0.003}$     & 226.85(202)   \\  \hline
2  & HD 119682              & $1597^{+81}_{-65}$    & $-$                   & $-$                   & $-$           & 413.5         & $<0.6$                        & $1.7^{+1.3}_{-1.0}$       & $-$                       & $-$                       & $1.0^{+1.0}_{-0.6}$                   & $0.03^{+0.04}_{-0.02}$        & 19.24(30)     \\
   &                        &                       &                       &                       &               &               & $<0.2$                        & $-$                       & $0.5\pm0.1$               & $0.15^{+0.15}_{-0.06}$    & $<0.6$                                & $<0.02$                       & 21.19(30)     \\  \hline
3  & HD 259440 $/$          & $1773^{+93}_{-87}$    & $317.3\pm0.7$ [1]    & $-$                   & $-$           & 164.5         & $<1.1$                        & $2.2^{+2.3}_{-1.5}$       & $-$                       & $-$                       & $<1.9$                                & $<0.08$                       & 14.41(12)     \\
   & HESS J0632$+$057       &                       &                       &                       &               &               & $<0.5$                        & $-$                       & $0.5\pm0.2$               & $0.16^{+0.22}_{-0.07}$    & $<0.7$                                & $<0.03$                       & 13.95(12)     \\  \hline
4  & PSR B1259$-$63         & $2171^{+62}_{-59}$    & $1236.724526$         & $47.76\times10^{-3}$ [3]            & $-$           & 500.3         & $<0.7$                        & $1.6^{+1.2}_{-0.7}$       & $-$                       & $-$                       & $1.2^{+0.8}_{-0.7}$                   & $0.07^{+0.05}_{-0.04}$        & 37.58(38)     \\
   &                        &                       & $\pm0.000006$ [2]    &                       &               &               & $<0.2$                        & $-$                       & $0.6^{+0.1}_{-0.2}$       & $0.17^{+0.11}_{-0.04}$    & $<0.7$                                & $<0.04$                       & 43.10(38)     \\  \hline
\end{tabular}

\justifying
\tablecomments{ \scriptsize \\
{$\star$: The parameter was frozen.} \\
{The distance values for sources with known Gaia source identifiers are taken from Gaia early data release 3 (EDR3) distance catalogue $\citep{gaia_edr3_distance}$. For some sources, we use the distance values from the literature.} \\
{
References: [1] \cite{Adams2021ApJ}; [2] \cite{MillerJones2018MNRAS}; [3] \cite{Johnston1994MNRAS_PSR_B1259m63}; \\
}   \\
}
\end{table*}

\clearpage

\bibliographystyle{aasjournalv7}
\bibliography{sample701}

\begin{thebibliography}{}
\expandafter\ifx\csname natexlab\endcsname\relax\def\natexlab#1{#1}\fi
\providecommand{\url}[1]{\href{#1}{#1}}
\providecommand{\dodoi}[1]{doi:~\href{http://doi.org/#1}{\nolinkurl{#1}}}
\providecommand{\doeprint}[1]{\href{http://ascl.net/#1}{\nolinkurl{http://ascl.net/#1}}}
\providecommand{\doarXiv}[1]{\href{https://arxiv.org/abs/#1}{\nolinkurl{https://arxiv.org/abs/#1}}}

\bibitem[{C.~B. {Adams} {et~al.}(2021){Adams}, {Benbow}, {Brill}, {Buckley},
  {Capasso}, {Chromey}, {Errando}, {Falcone}, {A Farrell}, {Feng}, {Finley}, {M
  Foote}, {Fortson}, {Furniss}, {Gent}, {Gillanders}, {Giuri}, {Gueta},
  {Hanna}, {Hassan}, {Hervet}, {Holder}, {Hona}, {Humensky}, {Jin}, {Kaaret},
  {Kertzman}, {Kieda}, {K Kleiner}, {Krennrich}, {Kumar}, {Lang}, {Lundy},
  {Maier}, {E McGrath}, {Moriarty}, {Mukherjee}, {Nieto}, {Nievas-Rosillo},
  {O'Brien}, {Ong}, {Otte}, {Park}, {Patel}, {Pfrang}, {Pichel}, {Pohl},
  {Prado}, {Quinn}, {Ragan}, {Reynolds}, {Ribeiro}, {Roache}, {Rovero}, {Ryan},
  {Santander}, {Schlenstedt}, {Sembroski}, {Shang}, {Tak}, {Vassiliev},
  {Weinstein}, {Williams}, {J Williamson}, {J Williamson}, {Acciari},
  {Ansoldi}, {Antonelli}, {Arbet Engels}, {Artero}, {Asano}, {Baack},
  {Babi{\'c}}, {Baquero}, {Barres de Almeida}, {Barrio}, {Batkovi{\'c}},
  {Becerra Gonz{\'a}lez}, {Bednarek}, {Bellizzi}, {Bernardini}, {Bernardos},
  {Berti}, {Besenrieder}, {Bhattacharyya}, {Bigongiari}, {Biland}, {Blanch},
  {B{\"o}kenkamp}, {Bonnoli}, {Bo{\v{s}}njak}, {Busetto}, {Carosi},
  {Ceribella}, {Cerruti}, {Chai}, {Chilingarian}, {Cikota}, {Colak}, {Colombo},
  {Contreras}, {Cortina}, {Covino}, {D'Amico}, {D'Elia}, {Da Vela}, {Dazzi},
  {De Angelis}, {De Lotto}, {Delfino}, {Delgado}, {Delgado Mendez}, {Depaoli},
  {Di Pierro}, {Di Venere}, {Do Souto Espi{\~n}eira}, {Dominis Prester},
  {Donini}, {Dorner}, {Doro}, {Elsaesser}, {Fallah Ramazani}, {Fattorini},
  {Fonseca}, {Font}, {Fruck}, {Fukami}, {Fukazawa}, {Garc{\'\i}a L{\'o}pez},
  {Garczarczyk}, {Gasparyan}, {Gaug}, {Giglietto}, {Giordano}, {Gliwny},
  {Godinovi{\'c}}, {Green}, {Green}, {Hadasch}, {Hahn}, {Heckmann}, {Herrera},
  {Hoang}, {Hrupec}, {H{\"u}tten}, {Inada}, {Ishio}, {Iwamura}, {Jim{\'e}nez
  Mart{\'\i}nez}, {Jormanainen}, {Jouvin}, {Karjalainen}, {Kerszberg},
  {Kobayashi}, {Kubo}, {Kushida}, {Lamastra}, {Lelas}, {Leone}, {Lindfors},
  {Linhoff}, {Lombardi}, {Longo}, {L{\'o}pez-Coto}, {L{\'o}pez-Moya},
  {L{\'o}pez-Oramas}, {Loporchio}, {Machado de Oliveira Fraga}, {Maggio},
  {Majumdar}, {Makariev}, {Mallamaci}, {Maneva}, {Manganaro}, {Mannheim},
  {Maraschi}, {Mariotti}, {Mart{\'\i}nez}, {Mazin}, {Menchiari}, {Mender},
  {Mi{\'c}anovi{\'c}}, {Miceli}, {Miener}, {Miranda}, {Mirzoyan}, {Molina},
  {Moralejo}, {Morcuende}, {Moreno}, {Moretti}, {Nakamori}, {Nava},
  {Neustroev}, {Nigro}, {Nilsson}, {Nishijima}, {Noda}, {Nozaki}, \&
  {Ohtani}}]{Adams2021ApJ}
{Adams}, C.~B., {Benbow}, W., {Brill}, A., {et~al.} 2021,
  \bibinfo{title}{{Observation of the Gamma-Ray Binary HESS J0632+057 with the
  H.E.S.S., MAGIC, and VERITAS Telescopes},} \apj, 923, 241,
  \dodoi{10.3847/1538-4357/ac29b7}

\bibitem[{F. {Aharonian} {et~al.}(2005){Aharonian}, {Akhperjanian}, {Aye},
  {Bazer-Bachi}, {Beilicke}, {Benbow}, {Berge}, {Berghaus}, {Bernl{\"o}hr},
  {Boisson}, {Bolz}, {Braun}, {Breitling}, {Brown}, {Bussons Gordo},
  {Chadwick}, {Chounet}, {Cornils}, {Costamante}, {Degrange},
  {Djannati-Ata{\"\i}}, {O'C. Drury}, {Dubus}, {Emmanoulopoulos}, {Espigat},
  {Feinstein}, {Fleury}, {Fontaine}, {Fuchs}, {Funk}, {Gallant}, {Giebels},
  {Gillessen}, {Glicenstein}, {Goret}, {Hadjichristidis}, {Hauser},
  {Heinzelmann}, {Henri}, {Hermann}, {Hinton}, {Hofmann}, {Holleran}, {Horns},
  {de Jager}, {Johnston}, {Kh{\'e}lifi}, {Kirk}, {Komin}, {Konopelko},
  {Latham}, {Le Gallou}, {Lemi{\`e}re}, {Lemoine-Goumard}, {Leroy},
  {Martineau-Huynh}, {Lohse}, {Marcowith}, {Masterson}, {McComb}, {de Naurois},
  {Nolan}, {Noutsos}, {Orford}, {Osborne}, {Ouchrif}, {Panter}, {Pelletier},
  {Pita}, {P{\"u}hlhofer}, {Punch}, {Raubenheimer}, {Raue}, {Raux}, {Rayner},
  {Redondo}, {Reimer}, {Reimer}, {Ripken}, {Rob}, {Rolland}, {Rowell},
  {Sahakian}, {Saug{\'e}}, {Schlenker}, {Schlickeiser}, {Schuster}, {Schwanke},
  {Siewert}, {Skj{\ae}raasen}, {Sol}, {Steenkamp}, {Stegmann}, {Tavernet},
  {Terrier}, {Th{\'e}oret}, {Tluczykont}, {Vasileiadis}, {Venter}, {Vincent},
  {V{\"o}lk}, \& {Wagner}}]{Aharonian_2005AnA_PSR_B1259m63}
{Aharonian}, F., {Akhperjanian}, A.~G., {Aye}, K.-M., {et~al.} 2005,
  \bibinfo{title}{{Discovery of the binary pulsar PSR B1259-63 in
  very-high-energy gamma rays around periastron with HESS},} \aap, 442, 1,
  \dodoi{10.1051/0004-6361:20052983}

\bibitem[{H. {Akaike}(1974){Akaike}}]{1974ITAC_Akaike_aic}
{Akaike}, H. 1974, \bibinfo{title}{{A New Look at the Statistical Model
  Identification},} IEEE Transactions on Automatic Control, 19, 716,
  \dodoi{10.1109/TAC.1974.1100705}

\bibitem[{K.~A. {Arnaud}(1996){Arnaud}}]{Arnaud_1996ASPC_XSPEC}
{Arnaud}, K.~A. 1996, \bibinfo{title}{{XSPEC: The First Ten Years},} in
  Astronomical Society of the Pacific Conference Series, Vol. 101, Astronomical
  Data Analysis Software and Systems V, ed. G.~H. {Jacoby} \& J.~{Barnes}, 17

\bibitem[{C.~A.~L. {Bailer-Jones} {et~al.}(2021){Bailer-Jones}, {Rybizki},
  {Fouesneau}, {Demleitner}, \& {Andrae}}]{gaia_edr3_distance}
{Bailer-Jones}, C.~A.~L., {Rybizki}, J., {Fouesneau}, M., {Demleitner}, M., \&
  {Andrae}, R. 2021, \bibinfo{title}{{Estimating Distances from Parallaxes. V.
  Geometric and Photogeometric Distances to 1.47 Billion Stars in Gaia Early
  Data Release 3},} \aj, 161, 147, \dodoi{10.3847/1538-3881/abd806}

\bibitem[{S.~D. {Barthelmy} {et~al.}(2005){Barthelmy}, {Barbier}, {Cummings},
  {Fenimore}, {Gehrels}, {Hullinger}, {Krimm}, {Markwardt}, {Palmer},
  {Parsons}, {Sato}, {Suzuki}, {Takahashi}, {Tashiro}, \&
  {Tueller}}]{Barthelmy2005SSRv_SWIFT_BAT}
{Barthelmy}, S.~D., {Barbier}, L.~M., {Cummings}, J.~R., {et~al.} 2005,
  \bibinfo{title}{{The Burst Alert Telescope (BAT) on the SWIFT Midex
  Mission},} \ssr, 120, 143, \dodoi{10.1007/s11214-005-5096-3}

\bibitem[{A. {Baykal} {et~al.}(2010){Baykal}, {G{\"o}{\v{g}}{\"u}{\c{s}}},
  {{\c{C}}a{\v{g}}da{\c{s}} {\.I}nam}, \&
  {Belloni}}]{Baykal_2010ApJ_Swift_J1626m5156}
{Baykal}, A., {G{\"o}{\v{g}}{\"u}{\c{s}}}, E., {{\c{C}}a{\v{g}}da{\c{s}}
  {\.I}nam}, S., \& {Belloni}, T. 2010, \bibinfo{title}{{The Orbital Period of
  Swift J1626.6-5156},} \apj, 711, 1306, \dodoi{10.1088/0004-637X/711/2/1306}

\bibitem[{J.~K. {Blackburn} {et~al.}(1999){Blackburn}, {Shaw}, {Payne},
  {Hayes}, \& {Heasarc}}]{Blackburn1999_FTOOLS}
{Blackburn}, J.~K., {Shaw}, R.~A., {Payne}, H.~E., {Hayes}, J.~J.~E., \&
  {Heasarc}. 1999, {FTOOLS: A general package of software to manipulate FITS
  files},, Astrophysics Source Code Library, record ascl:9912.002
  \doeprint{9912.002}

\bibitem[{A. {Bodaghee} {et~al.}(2016){Bodaghee}, {Tomsick}, {Fornasini},
  {Krivonos}, {Stern}, {Mori}, {Rahoui}, {Boggs}, {Christensen}, {Craig},
  {Hailey}, {Harrison}, \& {Zhang}}]{Bodaghee_2016ApJ_IGR_J16393m4643}
{Bodaghee}, A., {Tomsick}, J.~A., {Fornasini}, F.~M., {et~al.} 2016,
  \bibinfo{title}{{NuSTAR Discovery of a Cyclotron Line in the Accreting X-Ray
  Pulsar IGR J16393-4643},} \apj, 823, 146, \dodoi{10.3847/0004-637X/823/2/146}

\bibitem[{G. {Boella} {et~al.}(1997){Boella}, {Butler}, {Perola}, {Piro},
  {Scarsi}, \& {Bleeker}}]{Boella_1997AnAS_BeppoSAX}
{Boella}, G., {Butler}, R.~C., {Perola}, G.~C., {et~al.} 1997,
  \bibinfo{title}{{BeppoSAX, the wide band mission for X-ray astronomy},}
  \aaps, 122, 299, \dodoi{10.1051/aas:1997136}

\bibitem[{H.~V. {Bradt} {et~al.}(1993){Bradt}, {Rothschild}, \&
  {Swank}}]{Bradt_1993AnAS_RXTE}
{Bradt}, H.~V., {Rothschild}, R.~E., \& {Swank}, J.~H. 1993,
  \bibinfo{title}{{X-ray timing explorer mission},} \aaps, 97, 355

\bibitem[{E.~F. {Brown} {et~al.}(1998){Brown}, {Bildsten}, \&
  {Rutledge}}]{Brown1998ApJ_Quiescent}
{Brown}, E.~F., {Bildsten}, L., \& {Rutledge}, R.~E. 1998,
  \bibinfo{title}{{Crustal Heating and Quiescent Emission from Transiently
  Accreting Neutron Stars},} \apjl, 504, L95, \dodoi{10.1086/311578}

\bibitem[{D.~N. {Burrows} {et~al.}(2005){Burrows}, {Hill}, {Nousek}, {Kennea},
  {Wells}, {Osborne}, {Abbey}, {Beardmore}, {Mukerjee}, {Short}, {Chincarini},
  {Campana}, {Citterio}, {Moretti}, {Pagani}, {Tagliaferri}, {Giommi},
  {Capalbi}, {Tamburelli}, {Angelini}, {Cusumano}, {Br{\"a}uninger}, {Burkert},
  \& {Hartner}}]{Burrows_2005SSRv_SWIFT_XRT}
{Burrows}, D.~N., {Hill}, J.~E., {Nousek}, J.~A., {et~al.} 2005,
  \bibinfo{title}{{The Swift X-Ray Telescope},} \ssr, 120, 165,
  \dodoi{10.1007/s11214-005-5097-2}

\bibitem[{E.~M. {Cackett} {et~al.}(2010){Cackett}, {Brown}, {Cumming},
  {Degenaar}, {Miller}, \&
  {Wijnands}}]{Cackett_2010ApJ_LMXB_KS1731m260_crustal_cooling}
{Cackett}, E.~M., {Brown}, E.~F., {Cumming}, A., {et~al.} 2010,
  \bibinfo{title}{{Continued Cooling of the Crust in the Neutron Star Low-mass
  X-ray Binary KS 1731-260},} \apjl, 722, L137,
  \dodoi{10.1088/2041-8205/722/2/L137}

\bibitem[{E.~M. {Cackett} {et~al.}(2008){Cackett}, {Wijnands}, {Miller},
  {Brown}, \& {Degenaar}}]{Cackett_2008ApJ_MXB1659m29}
{Cackett}, E.~M., {Wijnands}, R., {Miller}, J.~M., {Brown}, E.~F., \&
  {Degenaar}, N. 2008, \bibinfo{title}{{Cooling of the Crust in the Neutron
  Star Low-Mass X-Ray Binary MXB 1659-29},} \apjl, 687, L87,
  \dodoi{10.1086/593703}

\bibitem[{S. {Campana} {et~al.}(2002){Campana}, {Stella}, {Israel}, {Moretti},
  {Parmar}, \& {Orlandini}}]{Campana2002ApJ_3_sources}
{Campana}, S., {Stella}, L., {Israel}, G.~L., {et~al.} 2002,
  \bibinfo{title}{{The Quiescent X-Ray Emission of Three Transient X-Ray
  Pulsars},} \apj, 580, 389, \dodoi{10.1086/343074}

\bibitem[{J. {Casares} {et~al.}(2014){Casares}, {Negueruela}, {Rib{\'o}},
  {Ribas}, {Paredes}, {Herrero}, \&
  {Sim{\'o}n-D{\'\i}az}}]{Casares_2014Nature_MWC_656}
{Casares}, J., {Negueruela}, I., {Rib{\'o}}, M., {et~al.} 2014,
  \bibinfo{title}{{A Be-type star with a black-hole companion},} \nat, 505,
  378, \dodoi{10.1038/nature12916}

\bibitem[{W. {Cash}(1979){Cash}}]{Cash_1979ApJ_fitting}
{Cash}, W. 1979, \bibinfo{title}{{Parameter estimation in astronomy through
  application of the likelihood ratio.},} \apj, 228, 939,
  \dodoi{10.1086/156922}

\bibitem[{A.~D.
  {Chandra}(2025){Chandra}}]{GX304m1_2025PASA_cold_disc_regime_5yrs}
{Chandra}, A.~D. 2025, \bibinfo{title}{{Long-term spin-down and low luminosity
  regime in the Be/X-ray binary pulsar GX 304-1},} \pasa, 42, e037,
  \dodoi{10.1017/pasa.2025.12}

\bibitem[{S. {Chaty} {et~al.}(2008){Chaty}, {Rahoui}, {Foellmi}, {Tomsick},
  {Rodriguez}, \& {Walter}}]{Chaty2008AnA}
{Chaty}, S., {Rahoui}, F., {Foellmi}, C., {et~al.} 2008,
  \bibinfo{title}{{Multi-wavelength observations of Galactic hard X-ray sources
  discovered by INTEGRAL. I. The nature of the companion star},} \aap, 484,
  783, \dodoi{10.1051/0004-6361:20078768}

\bibitem[{M. {Chernyakova} {et~al.}(2009){Chernyakova}, {Neronov}, {Aharonian},
  {Uchiyama}, \& {Takahashi}}]{Chernyakova2009MNRAS}
{Chernyakova}, M., {Neronov}, A., {Aharonian}, F., {Uchiyama}, Y., \&
  {Takahashi}, T. 2009, \bibinfo{title}{{X-ray observations of PSR B1259-63
  near the 2007 periastron passage},} \mnras, 397, 2123,
  \dodoi{10.1111/j.1365-2966.2009.15116.x}

\bibitem[{W. {Coburn} {et~al.}(2001){Coburn}, {Heindl}, {Gruber}, {Rothschild},
  {Staubert}, {Wilms}, \& {Kreykenbohm}}]{Coburn_2001ApJ_X_Persei}
{Coburn}, W., {Heindl}, W.~A., {Gruber}, D.~E., {et~al.} 2001,
  \bibinfo{title}{{Discovery of a Cyclotron Resonant Scattering Feature in the
  Rossi X-Ray Timing Explorer Spectrum of 4U 0352+309 (X Persei)},} \apj, 552,
  738, \dodoi{10.1086/320565}

\bibitem[{R.~H.~D. {Corbet} {et~al.}(2016){Corbet}, {Coley}, \&
  {Krimm}}]{Corbet_2016ATel}
{Corbet}, R. H.~D., {Coley}, J.~B., \& {Krimm}, H.~A. 2016,
  \bibinfo{title}{{Swift BAT Observations of the Be/X-ray Binary 4U 0728-25 (X
  0726-260)},} The Astronomer's Telegram, 9823, 1

\bibitem[{R.~H.~D. {Corbet} {et~al.}(2017){Corbet}, {Coley}, \&
  {Krimm}}]{Corbet_2017ApJ_pulse_period}
{Corbet}, R. H.~D., {Coley}, J.~B., \& {Krimm}, H.~A. 2017,
  \bibinfo{title}{{Diverse Long-term Variability of Five Candidate High-mass
  X-Ray Binaries from Swift Burst Alert Telescope Observations},} \apj, 846,
  161, \dodoi{10.3847/1538-4357/aa8638}

\bibitem[{R.~H.~D. {Corbet} {et~al.}(2010){Corbet}, {Krimm}, {Barthelmy},
  {Baumgartner}, {Markwardt}, {Skinner}, \&
  {Tueller}}]{Corbet_2010ATel_IGR_J16393m4643}
{Corbet}, R.~H.~D., {Krimm}, H.~A., {Barthelmy}, S.~D., {et~al.} 2010,
  \bibinfo{title}{{A 4.2 Day Period in the X-ray Pulsar IGR J16393-4643 from
  Swift/BAT and RXTE/PCA Observations},} The Astronomer's Telegram, 2570, 1

\bibitem[{R.~H.~D. {Corbet} \& A.~G. {Peele}(2000){Corbet} \&
  {Peele}}]{Corbet_2000ApJ_RX_J08124m3114}
{Corbet}, R. H.~D., \& {Peele}, A.~G. 2000, \bibinfo{title}{{The Orbital Period
  of the Be/Neutron Star Binary RX J0812.4-3114},} \apjl, 530, L33,
  \dodoi{10.1086/312485}

\bibitem[{R.~H.~D. {Corbet} \& R. {Remillard}(2005){Corbet} \&
  {Remillard}}]{Corbet2005ATel}
{Corbet}, R.~H.~D., \& {Remillard}, R. 2005, \bibinfo{title}{{The Orbital
  Period of IGR J11435-6109},} The Astronomer's Telegram, 377, 1

\bibitem[{G. {Cusumano} {et~al.}(2000){Cusumano}, {Maccarone}, {Nicastro},
  {Sacco}, \& {Kaaret}}]{Cusumano_2000ApJ_SAX_J0635p0533}
{Cusumano}, G., {Maccarone}, M.~C., {Nicastro}, L., {Sacco}, B., \& {Kaaret},
  P. 2000, \bibinfo{title}{{Detection of 33.8 Millisecond X-Ray Pulsations in
  SAX J0635+0533},} \apjl, 528, L25, \dodoi{10.1086/312413}

\bibitem[{G. {Cusumano} {et~al.}(2013{\natexlab{a}}){Cusumano}, {Segreto}, {La
  Parola}, {D'A{\`\i}}, {Masetti}, \&
  {Tagliaferri}}]{Cusumano_2013ApJ_IGR_J16283m4838}
{Cusumano}, G., {Segreto}, A., {La Parola}, V., {et~al.} 2013{\natexlab{a}},
  \bibinfo{title}{{Swift Observations of the High-mass X-Ray Binary IGR
  J16283-4838 Unveil a 288 Day Orbital Period},} \apjl, 775, L25,
  \dodoi{10.1088/2041-8205/775/1/L25}

\bibitem[{G. {Cusumano} {et~al.}(2013{\natexlab{b}}){Cusumano}, {Segreto}, {La
  Parola}, {Masetti}, {D'Ai}, \& {Tagliaferri}}]{Cusumano2013MNRAS}
{Cusumano}, G., {Segreto}, A., {La Parola}, V., {et~al.} 2013{\natexlab{b}},
  \bibinfo{title}{{Finding a 61.0d orbital period for the HMXB 4U 1036-56 with
  the Swift-BAT monitoring.},} \mnras, 436, L74, \dodoi{10.1093/mnrasl/slt116}

\bibitem[{A. {D'A{\`\i}} {et~al.}(2011){D'A{\`\i}}, {La Parola}, {Cusumano},
  {Segreto}, {Romano}, {Vercellone}, \& {Robba}}]{DAI_2011AnA_J13186_6257}
{D'A{\`\i}}, A., {La Parola}, V., {Cusumano}, G., {et~al.} 2011,
  \bibinfo{title}{{The Swift-BAT survey reveals the orbital period of three
  high-mass X-ray binaries},} \aap, 529, A30,
  \dodoi{10.1051/0004-6361/201016401}

\bibitem[{A. {Deibel} {et~al.}(2017){Deibel}, {Cumming}, {Brown}, \&
  {Reddy}}]{Deibel_2017ApJ_MXB1659m29_n_KS1731m260}
{Deibel}, A., {Cumming}, A., {Brown}, E.~F., \& {Reddy}, S. 2017,
  \bibinfo{title}{{Late-time Cooling of Neutron Star Transients and the Physics
  of the Inner Crust},} \apj, 839, 95, \dodoi{10.3847/1538-4357/aa6a19}

\bibitem[{A. {Deo Chandra}(2023){Deo Chandra}}]{Chandra_2023arXiv_GX_304m1}
{Deo Chandra}, A. 2023, \bibinfo{title}{{Long-term spin-down and low luminosity
  regime in the Be/X-ray binary pulsar GX 304-1},} arXiv e-prints,
  arXiv:2311.13303, \dodoi{10.48550/arXiv.2311.13303}

\bibitem[{{\c{C}}.~K. {D{\"o}nmez} {et~al.}(2020){D{\"o}nmez}, {Serim},
  {{\.I}nam}, {{\c{S}}ahiner}, {Serim}, \&
  {Baykal}}]{Donmez2020MNRAS_MAXI_J1409m619}
{D{\"o}nmez}, {\c{C}}.~K., {Serim}, M.~M., {{\.I}nam}, S.~{\c{C}}., {et~al.}
  2020, \bibinfo{title}{{Comprehensive analysis of the transient X-ray pulsar
  MAXI J1409-619},} \mnras, 496, 1768, \dodoi{10.1093/mnras/staa1562}

\bibitem[{V. {Doroshenko} {et~al.}(2014){Doroshenko}, {Santangelo},
  {Doroshenko}, {Caballero}, {Tsygankov}, \&
  {Rothschild}}]{Doroshenko2014AnA_1A0535p262}
{Doroshenko}, V., {Santangelo}, A., {Doroshenko}, R., {et~al.} 2014,
  \bibinfo{title}{{XMM-Newton observations of <ASTROBJ>1A 0535+262</ASTROBJ> in
  quiescence},} \aap, 561, A96, \dodoi{10.1051/0004-6361/201322472}

\bibitem[{V. {Doroshenko} {et~al.}(2010){Doroshenko}, {Suchy}, {Santangelo},
  {Staubert}, {Kreykenbohm}, {Rothschild}, {Pottschmidt}, \&
  {Wilms}}]{Doroshenko2010aanda_1A_1118m61}
{Doroshenko}, V., {Suchy}, S., {Santangelo}, A., {et~al.} 2010,
  \bibinfo{title}{{RXTE observations of the 1A 1118-61 in an outburst, and the
  discovery of a cyclotron line},} \aap, 515, L1,
  \dodoi{10.1051/0004-6361/201014858}

\bibitem[{V. {Doroshenko} {et~al.}(2022){Doroshenko}, {Staubert}, {Maitra},
  {Rau}, {Haberl}, {Santangelo}, {Schwope}, {Wilms}, {Buckley}, {Semena},
  {Mereminskiy}, {Lutovinov}, {Gromadzki}, {Townsend}, \&
  {Monageng}}]{Doroshenko2022AnA_SRGA_J124404m632232}
{Doroshenko}, V., {Staubert}, R., {Maitra}, C., {et~al.} 2022,
  \bibinfo{title}{{SRGA J124404.1-632232/SRGU J124403.8-632231: New X-ray
  pulsar discovered in the all-sky survey by the SRG},} \aap, 661, A21,
  \dodoi{10.1051/0004-6361/202141147}

\bibitem[{G. {Dubus}(2013){Dubus}}]{Dubus_2013AnA_Gamma_Ray_Summary}
{Dubus}, G. 2013, \bibinfo{title}{{Gamma-ray binaries and related systems},}
  \aapr, 21, 64, \dodoi{10.1007/s00159-013-0064-5}

\bibitem[{P. {Esposito} {et~al.}(2014){Esposito}, {Israel}, {Sidoli}, {Tiengo},
  {Campana}, \& {Moretti}}]{Esposito2014MNRAS_IGR_J17200m3116}
{Esposito}, P., {Israel}, G.~L., {Sidoli}, L., {et~al.} 2014,
  \bibinfo{title}{{The X-ray emission of the high-mass X-ray binary IGR
  J17200-3116},} \mnras, 441, 1126, \dodoi{10.1093/mnras/stu659}

\bibitem[{M.~H. {Finger} {et~al.}(1994){Finger}, {Cominsky}, {Wilson},
  {Harmon}, \& {Fishman}}]{Finger_1994AIPC}
{Finger}, M.~H., {Cominsky}, L.~R., {Wilson}, R.~B., {Harmon}, B.~A., \&
  {Fishman}, G.~J. 1994, \bibinfo{title}{{Hard X-Ray Observations of A
  0535+262},} in American Institute of Physics Conference Series, Vol. 308, The
  Evolution of X-ray Binariese, ed. S.~{Holt} \& C.~S. {Day} (AIP), 459,
  \dodoi{10.1063/1.46032}

\bibitem[{M.~H. {Finger} {et~al.}(1996){Finger}, {Wilson}, \&
  {Chakrabarty}}]{Finger_1996AandAS}
{Finger}, M.~H., {Wilson}, R.~B., \& {Chakrabarty}, D. 1996,
  \bibinfo{title}{{Reappearance of the X-ray binary pulsar 2S 1417-624.},}
  \aaps, 120, 209

\bibitem[{W. {Forman} {et~al.}(1978){Forman}, {Jones}, {Cominsky}, {Julien},
  {Murray}, {Peters}, {Tananbaum}, \& {Giacconi}}]{Forman_1978ApJS_UHURU}
{Forman}, W., {Jones}, C., {Cominsky}, L., {et~al.} 1978, \bibinfo{title}{{The
  fourth Uhuru catalog of X-ray sources.},} \apjs, 38, 357,
  \dodoi{10.1086/190561}

\bibitem[{F. {Fortin} {et~al.}(2018){Fortin}, {Chaty}, {Coleiro}, {Tomsick}, \&
  {Nitschelm}}]{Fortin_2018AnA_IGR_J13186m6257}
{Fortin}, F., {Chaty}, S., {Coleiro}, A., {Tomsick}, J.~A., \& {Nitschelm},
  C.~H.~R. 2018, \bibinfo{title}{{Spectroscopic identification of INTEGRAL
  high-energy sources with VLT/ISAAC},} \aap, 618, A150,
  \dodoi{10.1051/0004-6361/201731265}

\bibitem[{F. {Fortin} {et~al.}(2023){Fortin}, {Garc{\'\i}a}, {Simaz Bunzel}, \&
  {Chaty}}]{Fortin_2023AnA__catalogue}
{Fortin}, F., {Garc{\'\i}a}, F., {Simaz Bunzel}, A., \& {Chaty}, S. 2023,
  \bibinfo{title}{{A catalogue of high-mass X-ray binaries in the Galaxy: from
  the INTEGRAL to the Gaia era},} \aap, 671, A149,
  \dodoi{10.1051/0004-6361/202245236}

\bibitem[{M.~Y. {Ge} {et~al.}(2020){Ge}, {Ji}, {Zhang}, {Santangelo}, {Liu},
  {Doroshenko}, {Staubert}, {Qu}, {Zhang}, {Lu}, {Song}, {Li}, {Tao}, {Xu},
  {Cao}, {Chen}, {Bu}, {Cai}, {Chang}, {Chen}, {Chen}, {Chen}, {Chen}, {Chen},
  {Cui}, {Cui}, {Deng}, {Dong}, {Du}, {Fu}, {Gao}, {Gao}, {Gao}, {Gu}, {Guan},
  {Guo}, {Han}, {Huang}, {Huo}, {Jia}, {Jiang}, {Jiang}, {Jin}, {Jin}, {Kong},
  {Li}, {Li}, {Li}, {Li}, {Li}, {Li}, {Li}, {Li}, {Li}, {Li}, {Liang}, {Liao},
  {Liu}, {Liu}, {Liu}, {Liu}, {Liu}, {Lu}, {Lu}, {Luo}, {Luo}, {Ma}, {Meng},
  {Nang}, {Nie}, {Ou}, {Sai}, {Shang}, {Song}, {Sun}, {Tan}, {Tuo}, {Wang},
  {Wang}, {Wang}, {Wang}, {Wang}, {Wang}, {Wang}, {Wen}, {Wu}, {Wu}, {Wu},
  {Xiao}, {Xiao}, {Xiong}, {Xu}, {Yang}, {Yang}, {Yang}, {Yang}, {Yi}, {Yin},
  {You}, {Zhang}, {Zhang}, {Zhang}, {Zhang}, {Zhang}, {Zhang}, {Zhang},
  {Zhang}, {Zhang}, {Zhang}, {Zhang}, {Zhang}, {Zhang}, {Zhang}, {Zhang},
  {Zhang}, {Zhao}, {Zhao}, {Zheng}, {Zheng}, {Zhou}, {Zhou}, {Zhuang}, {Zhu},
  \& {Zhu}}]{Ge2020ApJ_GRO_J1008m57}
{Ge}, M.~Y., {Ji}, L., {Zhang}, S.~N., {et~al.} 2020,
  \bibinfo{title}{{Insight-HXMT Firm Detection of the Highest-energy
  Fundamental Cyclotron Resonance Scattering Feature in the Spectrum of GRO
  J1008-57},} \apjl, 899, L19, \dodoi{10.3847/2041-8213/abac05}

\bibitem[{M. {Ghimiray} {et~al.}(2024){Ghimiray}, {Sharma}, \&
  {Subba}}]{Ghimiray2024MNRAS}
{Ghimiray}, M., {Sharma}, P., \& {Subba}, N. 2024,
  \bibinfo{title}{{Low-luminosity accretion of Be/X-ray pulsar MAXI
  J1409-619},} \mnras, 531, 3386, \dodoi{10.1093/mnras/stae1369}

\bibitem[{P. {Ghosh} \& F.~K. {Lamb}(1978){Ghosh} \&
  {Lamb}}]{Ghosh1978ApJ_accretion_flow_factor}
{Ghosh}, P., \& {Lamb}, F.~K. 1978, \bibinfo{title}{{Disk accretion by magnetic
  neutron stars.},} \apjl, 223, L83, \dodoi{10.1086/182734}

\bibitem[{E. {G{\"o}{\v{g}}{\"u}{\c{s}}}
  {et~al.}(2005){G{\"o}{\v{g}}{\"u}{\c{s}}}, {Patel}, {Wilson}, {Woods},
  {Finger}, \& {Kouveliotou}}]{Ersin2005ApJ_XTE_J1906p090}
{G{\"o}{\v{g}}{\"u}{\c{s}}}, E., {Patel}, S.~K., {Wilson}, C.~A., {et~al.}
  2005, \bibinfo{title}{{Chandra Localization of XTE J1906+090 and Discovery of
  its Optical and Infrared Counterparts},} \apj, 632, 1069,
  \dodoi{10.1086/444373}

\bibitem[{F. {Haberl}(1995){Haberl}}]{Haberl_1995AnA_gammaCas}
{Haberl}, F. 1995, \bibinfo{title}{{{\ensuremath{\gamma}} Cassiopeiae: evidence
  for a Be star/white dwarf X-ray binary?},} \aap, 296, 685

\bibitem[{R.~W. {Hanuschik}(1996){Hanuschik}}]{Hanuschik1996AnA}
{Hanuschik}, R.~W. 1996, \bibinfo{title}{{On the structure of Be star disks.},}
  \aap, 308, 170

\bibitem[{F.~A. {Harrison} {et~al.}(2013){Harrison}, {Craig}, {Christensen},
  {Hailey}, {Zhang}, {Boggs}, {Stern}, {Cook}, {Forster}, {Giommi},
  {Grefenstette}, {Kim}, {Kitaguchi}, {Koglin}, {Madsen}, {Mao}, {Miyasaka},
  {Mori}, {Perri}, {Pivovaroff}, {Puccetti}, {Rana}, {Westergaard}, {Willis},
  {Zoglauer}, {An}, {Bachetti}, {Barri{\`e}re}, {Bellm}, {Bhalerao},
  {Brejnholt}, {Fuerst}, {Liebe}, {Markwardt}, {Nynka}, {Vogel}, {Walton},
  {Wik}, {Alexander}, {Cominsky}, {Hornschemeier}, {Hornstrup}, {Kaspi},
  {Madejski}, {Matt}, {Molendi}, {Smith}, {Tomsick}, {Ajello}, {Ballantyne},
  {Balokovi{\'c}}, {Barret}, {Bauer}, {Blandford}, {Brandt}, {Brenneman},
  {Chiang}, {Chakrabarty}, {Chenevez}, {Comastri}, {Dufour}, {Elvis}, {Fabian},
  {Farrah}, {Fryer}, {Gotthelf}, {Grindlay}, {Helfand}, {Krivonos}, {Meier},
  {Miller}, {Natalucci}, {Ogle}, {Ofek}, {Ptak}, {Reynolds}, {Rigby},
  {Tagliaferri}, {Thorsett}, {Treister}, \& {Urry}}]{NuSTAR_paper_2013ApJ}
{Harrison}, F.~A., {Craig}, W.~W., {Christensen}, F.~E., {et~al.} 2013,
  \bibinfo{title}{{The Nuclear Spectroscopic Telescope Array (NuSTAR)
  High-energy X-Ray Mission},} \apj, 770, 103,
  \dodoi{10.1088/0004-637X/770/2/103}

\bibitem[{ {HI4PI Collaboration} {et~al.}(2016){HI4PI Collaboration}, {Ben
  Bekhti}, {Fl{\"o}er}, {Keller}, {Kerp}, {Lenz}, {Winkel}, {Bailin},
  {Calabretta}, {Dedes}, {Ford}, {Gibson}, {Haud}, {Janowiecki}, {Kalberla},
  {Lockman}, {McClure-Griffiths}, {Murphy}, {Nakanishi}, {Pisano}, \&
  {Staveley-Smith}}]{HI4PI_2016AmA_NH}
{HI4PI Collaboration}, {Ben Bekhti}, N., {Fl{\"o}er}, L., {et~al.} 2016,
  \bibinfo{title}{{HI4PI: A full-sky H I survey based on EBHIS and GASS},}
  \aap, 594, A116, \dodoi{10.1051/0004-6361/201629178}

\bibitem[{J.~A. {Hinton} {et~al.}(2009){Hinton}, {Skilton}, {Funk}, {Brucker},
  {Aharonian}, {Dubus}, {Fiasson}, {Gallant}, {Hofmann}, {Marcowith}, \&
  {Reimer}}]{Hinton_2009ApJ_HESS_J0632p057}
{Hinton}, J.~A., {Skilton}, J.~L., {Funk}, S., {et~al.} 2009,
  \bibinfo{title}{{HESS J0632+057: A New Gamma-Ray Binary?},} \apjl, 690, L101,
  \dodoi{10.1088/0004-637X/690/2/L101}

\bibitem[{J. {Homan} {et~al.}(2014){Homan}, {Fridriksson}, {Wijnands},
  {Cackett}, {Degenaar}, {Linares}, {Lin}, \&
  {Remillard}}]{Homan_2014ApJ_decay_timescales_low_mag_field_LMXBs}
{Homan}, J., {Fridriksson}, J.~K., {Wijnands}, R., {et~al.} 2014,
  \bibinfo{title}{{A Strongly Heated Neutron Star in the Transient Z Source
  MAXI J0556-332},} \apj, 795, 131, \dodoi{10.1088/0004-637X/795/2/131}

\bibitem[{A.~F. {Illarionov} \& R.~A. {Sunyaev}(1975){Illarionov} \&
  {Sunyaev}}]{Illarionov_1975AnA_propeller_regime}
{Illarionov}, A.~F., \& {Sunyaev}, R.~A. 1975, \bibinfo{title}{{Why the Number
  of Galactic X-ray Stars Is so Small?},} \aap, 39, 185

\bibitem[{J. {in't Zand} \& J. {Heise}(2004){in't Zand} \&
  {Heise}}]{2004ATel_IGR_J11435M6109}
{in't Zand}, J., \& {Heise}, J. 2004, \bibinfo{title}{{BeppoSAX-WFC
  observations of IGR J11435-6109 and J00291+5934},} The Astronomer's Telegram,
  362, 1

\bibitem[{J.~J.~M. {in't Zand} {et~al.}(2001){in't Zand}, {Corbet}, \&
  {Marshall}}]{intZand_2001ApJ_XTE_J1543m568_dist}
{in't Zand}, J.~J.~M., {Corbet}, R.~H.~D., \& {Marshall}, F.~E. 2001,
  \bibinfo{title}{{Discovery of a 75 Day Orbit in XTE J1543-568},} \apjl, 553,
  L165, \dodoi{10.1086/320688}

\bibitem[{N. {Islam} {et~al.}(2015){Islam}, {Maitra}, {Pradhan}, \&
  {Paul}}]{Islam_2015MNRAS_IGR_J16393m4643}
{Islam}, N., {Maitra}, C., {Pradhan}, P., \& {Paul}, B. 2015,
  \bibinfo{title}{{A Suzaku view of IGR J16393-4643},} \mnras, 446, 4148,
  \dodoi{10.1093/mnras/stu2395}

\bibitem[{J.~C. {Ives} {et~al.}(1975){Ives}, {Sanford}, \& {Bell
  Burnell}}]{Ives1975Natur_1A_1118m615}
{Ives}, J.~C., {Sanford}, P.~W., \& {Bell Burnell}, S.~J. 1975,
  \bibinfo{title}{{Observations of a transient X-ray source with regular
  periodicity of 6.75 min},} \nat, 254, 578, \dodoi{10.1038/254578a0}

\bibitem[{N. {Iyer} \& B. {Paul}(2017){Iyer} \&
  {Paul}}]{Iyer_2017MNRAS_IGR_J16318m4848}
{Iyer}, N., \& {Paul}, B. 2017, \bibinfo{title}{{Orbital variations in
  intensity and spectral properties of the highly obscured sgHMXB IGR
  J16318-4848},} \mnras, 471, 355, \dodoi{10.1093/mnras/stx1575}

\bibitem[{C. {Jain} {et~al.}(2009){Jain}, {Paul}, \&
  {Dutta}}]{Jain_2009RAA_IGR_J16318m4848}
{Jain}, C., {Paul}, B., \& {Dutta}, A. 2009, \bibinfo{title}{{Orbital X-ray
  modulation study of three supergiant HMXBs},} Research in Astronomy and
  Astrophysics, 9, 1303, \dodoi{10.1088/1674-4527/9/12/002}

\bibitem[{F. {Jansen} {et~al.}(2001){Jansen}, {Lumb}, {Altieri}, {Clavel},
  {Ehle}, {Erd}, {Gabriel}, {Guainazzi}, {Gondoin}, {Much}, {Munoz}, {Santos},
  {Schartel}, {Texier}, \& {Vacanti}}]{Jansen_2001AnA_XMM_newton}
{Jansen}, F., {Lumb}, D., {Altieri}, B., {et~al.} 2001,
  \bibinfo{title}{{XMM-Newton observatory. I. The spacecraft and operations},}
  \aap, 365, L1, \dodoi{10.1051/0004-6361:20000036}

\bibitem[{S. {Janssens} {et~al.}(2023){Janssens}, {Shenar}, {Degenaar},
  {Bodensteiner}, {Sana}, {Audenaert}, \&
  {Frost}}]{Janssens_2023AnA_MWC656_BeBH_or_not}
{Janssens}, S., {Shenar}, T., {Degenaar}, N., {et~al.} 2023,
  \bibinfo{title}{{MWC 656 is unlikely to contain a black hole},} \aap, 677,
  L9, \dodoi{10.1051/0004-6361/202347318}

\bibitem[{S. {Johnston} {et~al.}(1994){Johnston}, {Manchester}, {Lyne},
  {Nicastro}, \& {Spyromilio}}]{Johnston1994MNRAS_PSR_B1259m63}
{Johnston}, S., {Manchester}, R.~N., {Lyne}, A.~G., {Nicastro}, L., \&
  {Spyromilio}, J. 1994, \bibinfo{title}{{Radio and Optical Observations of the
  PSR:B1259-63 / SS:2883 Be-Star Binary System},} \mnras, 268, 430,
  \dodoi{10.1093/mnras/268.2.430}

\bibitem[{P. {Kaaret} {et~al.}(2000){Kaaret}, {Cusumano}, \&
  {Sacco}}]{Kaaret_2000ApJ_SAX_J0635p0533}
{Kaaret}, P., {Cusumano}, G., \& {Sacco}, B. 2000, \bibinfo{title}{{X-Ray
  Timing of the 34 Millisecond Binary Pulsar SAX J0635+0533},} \apjl, 542, L41,
  \dodoi{10.1086/312918}

\bibitem[{R. {Kaur} {et~al.}(2009){Kaur}, {Wijnands}, {Patruno}, {Testa},
  {Israel}, {Degenaar}, {Paul}, \& {Kumar}}]{Kaur2009MNRAS_SAX_J13244m6200}
{Kaur}, R., {Wijnands}, R., {Patruno}, A., {et~al.} 2009,
  \bibinfo{title}{{Chandra and XMM-Newton observations of the low-luminosity
  X-ray pulsators SAX J1324.4-6200 and SAX J1452.8-5949},} \mnras, 394, 1597,
  \dodoi{10.1111/j.1365-2966.2009.14438.x}

\bibitem[{R.~L. {Kelley} {et~al.}(1981){Kelley}, {Apparao}, {Doxsey},
  {Jernigan}, {Naranan}, \& {Rappaport}}]{Kelley1981ApJ_2S_1417m624}
{Kelley}, R.~L., {Apparao}, K.~M.~V., {Doxsey}, R.~E., {et~al.} 1981,
  \bibinfo{title}{{Discovery of X-ray pulsations from 2S 1417-624.},} \apj,
  243, 251, \dodoi{10.1086/158591}

\bibitem[{J.~A. {Kennea} {et~al.}(2010){Kennea}, {Curran}, {Krimm}, {Romano},
  {Mangano}, {Evans}, {Yamaoka}, \& {Burrows}}]{Kennea_2010ATel_MAXI_J1409m619}
{Kennea}, J.~A., {Curran}, P., {Krimm}, H., {et~al.} 2010,
  \bibinfo{title}{{Swift detection of an \raisebox{-0.5ex}\textasciitilde500s
  period in MAXI J1409-619},} The Astronomer's Telegram, 3060, 1

\bibitem[{P. {Kretschmar} {et~al.}(2019){Kretschmar}, {F{\"u}rst}, {Sidoli},
  {Bozzo}, {Alfonso-Garz{\'o}n}, {Bodaghee}, {Chaty}, {Chernyakova},
  {Ferrigno}, {Manousakis}, {Negueruela}, {Postnov}, {Paizis}, {Reig},
  {Rodes-Roca}, {Tsygankov}, {Bird}, {Bissinger n{\'e} K{\"u}hnel}, {Blay},
  {Caballero}, {Coe}, {Domingo}, {Doroshenko}, {Ducci}, {Falanga}, {Grebenev},
  {Grinberg}, {Hemphill}, {Kreykenbohm}, {Kreykenbohm n{\'e} Fritz}, {Li},
  {Lutovinov}, {Mart{\'\i}nez-N{\'u}{\~n}ez}, {Mas-Hesse}, {Masetti},
  {McBride}, {Neronov}, {Pottschmidt}, {Rodriguez}, {Romano}, {Rothschild},
  {Santangelo}, {Sguera}, {Staubert}, {Tomsick}, {Torrej{\'o}n}, {Torres},
  {Walter}, {Wilms}, {Wilson-Hodge}, \&
  {Zhang}}]{Kretschmar_2019NewAR_sgBe_2_sources}
{Kretschmar}, P., {F{\"u}rst}, F., {Sidoli}, L., {et~al.} 2019,
  \bibinfo{title}{{Advances in Understanding High-Mass X-ray Binaries with
  INTEGRALand Future Directions},} \nar, 86, 101546,
  \dodoi{10.1016/j.newar.2020.101546}

\bibitem[{H.~A. {Krimm} {et~al.}(2013){Krimm}, {Holland}, {Corbet}, {Pearlman},
  {Romano}, {Kennea}, {Bloom}, {Barthelmy}, {Baumgartner}, {Cummings},
  {Gehrels}, {Lien}, {Markwardt}, {Palmer}, {Sakamoto}, {Stamatikos}, \&
  {Ukwatta}}]{Krimm2013ApJS_SWIFT_BAT}
{Krimm}, H.~A., {Holland}, S.~T., {Corbet}, R.~H.~D., {et~al.} 2013,
  \bibinfo{title}{{The Swift/BAT Hard X-Ray Transient Monitor},} \apjs, 209,
  14, \dodoi{10.1088/0067-0049/209/1/14}

\bibitem[{R.~A. {Krivonos} {et~al.}(2015){Krivonos}, {Tsygankov}, {Lutovinov},
  {Tomsick}, {Chakrabarty}, {Bachetti}, {Boggs}, {Chernyakova}, {Christensen},
  {Craig}, {F{\"u}rst}, {Hailey}, {Harrison}, {Lansbury}, {Rahoui}, {Stern}, \&
  {Zhang}}]{Krivonos_2015ApJ_2RXPJ1301596m635806}
{Krivonos}, R.~A., {Tsygankov}, S.~S., {Lutovinov}, A.~A., {et~al.} 2015,
  \bibinfo{title}{{NuSTAR Discovery of an Unusually Steady Long-term Spin-up of
  the Be Binary 2RXP J130159.6-635806},} \apj, 809, 140,
  \dodoi{10.1088/0004-637X/809/2/140}

\bibitem[{N. {La Palombara} \& S. {Mereghetti}(2017){La Palombara} \&
  {Mereghetti}}]{La_Palombara_2017AnA_SAX_J0635p0533}
{La Palombara}, N., \& {Mereghetti}, S. 2017, \bibinfo{title}{{Swift monitoring
  of the massive X-ray binary SAX J0635.2+0533},} \aap, 602, A114,
  \dodoi{10.1051/0004-6361/201730887}

\bibitem[{N. {La Palombara} {et~al.}(2021){La Palombara}, {Sidoli}, {Esposito},
  {Israel}, \& {Rodr{\'\i}guez
  Castillo}}]{Palombara2021AnA_CXOU_J2253551p624336}
{La Palombara}, N., {Sidoli}, L., {Esposito}, P., {Israel}, G.~L., \&
  {Rodr{\'\i}guez Castillo}, G.~A. 2021, \bibinfo{title}{{A deep XMM-Newton
  observation of the X-Persei-like binary system CXOU J225355.1+624336},} \aap,
  649, A118, \dodoi{10.1051/0004-6361/202140760}

\bibitem[{N. {La Palombara} {et~al.}(2012){La Palombara}, {Sidoli}, {Esposito},
  {Tiengo}, \& {Mereghetti}}]{Palombara_RX_J04409p4431}
{La Palombara}, N., {Sidoli}, L., {Esposito}, P., {Tiengo}, A., \&
  {Mereghetti}, S. 2012, \bibinfo{title}{{XMM-Newton observation of the
  persistent Be/NS X-ray binary pulsar RX J0440.9+4431},} \aap, 539, A82,
  \dodoi{10.1051/0004-6361/201118221}

\bibitem[{N. {La Palombara} {et~al.}(2025{\natexlab{a}}){La Palombara},
  {Sidoli}, {Mereghetti}, {Esposito}, \&
  {Israel}}]{Palombara_2025AnA_4U_0728m25}
{La Palombara}, N., {Sidoli}, L., {Mereghetti}, S., {Esposito}, P., \&
  {Israel}, G.~L. 2025{\natexlab{a}}, \bibinfo{title}{{The persistent nature of
  the Be X-ray binary pulsar 4U 0728-25},} \aap, 696, A197,
  \dodoi{10.1051/0004-6361/202554103}

\bibitem[{N. {La Palombara} {et~al.}(2025{\natexlab{b}}){La Palombara},
  {Sidoli}, {Mereghetti}, {Israel}, \&
  {Esposito}}]{La_Palombara_2025AN_persistent_BeXRB}
{La Palombara}, N., {Sidoli}, L., {Mereghetti}, S., {Israel}, G.~L., \&
  {Esposito}, P. 2025{\natexlab{b}}, \bibinfo{title}{{The Role of
  XMM{\textendash}Newton in the Investigation of Persistent BeXRBs},}
  Astronomische Nachrichten, 346, e20240100, \dodoi{10.1002/asna.20240100}

\bibitem[{V. {La Parola} {et~al.}(2013){La Parola}, {D'A{\`\i}}, {Cusumano},
  {Segreto}, {Masetti}, \& {Melandri}}]{LaParola_2013arXiv_IGR_J11305m6256}
{La Parola}, V., {D'A{\`\i}}, A., {Cusumano}, G., {et~al.} 2013,
  \bibinfo{title}{{Timing and spectral study of the Be XRB IGR J11305-6256:
  Swift discovers the orbital period and a soft X-ray excess},} arXiv e-prints,
  arXiv:1305.3916, \dodoi{10.48550/arXiv.1305.3916}

\bibitem[{J.~P. {Lasota}(1997){Lasota}}]{Lasota_1997ASPC_cold_disc_temp}
{Lasota}, J.~P. 1997, \bibinfo{title}{{Disc Instabilities and ``Soft'' X-Ray
  Transients},} in Astronomical Society of the Pacific Conference Series, Vol.
  121, IAU Colloquium 163: Accretion Phenomena and Related Outflows, ed. D.~T.
  {Wickramasinghe}, G.~V. {Bicknell}, \& L.~{Ferrario}, 351,
  \dodoi{10.48550/arXiv.astro-ph/9610068}

\bibitem[{A.~M. {Levine} {et~al.}(1996){Levine}, {Bradt}, {Cui}, {Jernigan},
  {Morgan}, {Remillard}, {Shirey}, \& {Smith}}]{Levine_1996ApJ_RXTE_ASM}
{Levine}, A.~M., {Bradt}, H., {Cui}, W., {et~al.} 1996, \bibinfo{title}{{First
  Results from the All-Sky Monitor on the Rossi X-Ray Timing Explorer},} \apjl,
  469, L33, \dodoi{10.1086/310260}

\bibitem[{A.~M. {Levine} \& R. {Corbet}(2006){Levine} \&
  {Corbet}}]{Levine_n_Corbet2006}
{Levine}, A.~M., \& {Corbet}, R. 2006, \bibinfo{title}{{Detection of Additional
  Periodicities in RXTE ASM Light Curves},} The Astronomer's Telegram, 940, 1

\bibitem[{J. {Li} {et~al.}(2012){Li}, {Torres}, {Zhang}, {Papitto}, {Chen}, \&
  {Wang}}]{Li2012ApJ_4U_1036m56}
{Li}, J., {Torres}, D.~F., {Zhang}, S., {et~al.} 2012,
  \bibinfo{title}{{INTEGRAL and Swift Observations of the Be X-Ray Binary 4U
  1036-56 (RX J1037.5-5647) and its Possible Relation with
  {\ensuremath{\gamma}}-Ray Transients},} \apj, 761, 49,
  \dodoi{10.1088/0004-637X/761/1/49}

\bibitem[{Q. {Liu} {et~al.}(2024){Liu}, {Santangelo}, {Kong}, {Ducci}, {Ji},
  {Wang}, {Serim}, {G{\"u}ng{\"o}r}, {Tuo}, \&
  {Serim}}]{Liu2024aanda_2S_1417m624}
{Liu}, Q., {Santangelo}, A., {Kong}, L.~D., {et~al.} 2024, \bibinfo{title}{{The
  high-energy cyclotron line in 2S 1417-624 discovered with Insight-HXMT during
  the 2018 outburst},} \aap, 691, A215, \dodoi{10.1051/0004-6361/202451062}

\bibitem[{Q.~Z. {Liu} {et~al.}(2006){Liu}, {van Paradijs}, \& {van den
  Heuvel}}]{Liu_2006AnA_catalogue}
{Liu}, Q.~Z., {van Paradijs}, J., \& {van den Heuvel}, E.~P.~J. 2006,
  \bibinfo{title}{{Catalogue of high-mass X-ray binaries in the Galaxy (4th
  edition)},} \aap, 455, 1165, \dodoi{10.1051/0004-6361:20064987}

\bibitem[{R. {Lopes de Oliveira} {et~al.}(2007){Lopes de Oliveira}, {Motch},
  {Smith}, {Negueruela}, \& {Torrej{\'o}n}}]{Lopes2007AnA_HD_110432_gamma_cas}
{Lopes de Oliveira}, R., {Motch}, C., {Smith}, M.~A., {Negueruela}, I., \&
  {Torrej{\'o}n}, J.~M. 2007, \bibinfo{title}{{On the X-ray and optical
  properties of the Be star <ASTROBJ>HD 110432</ASTROBJ>: a very hard-thermal
  X-ray emitter},} \aap, 474, 983, \dodoi{10.1051/0004-6361:20077295}

\bibitem[{A.~A. {Lutovinov} {et~al.}(2019){Lutovinov}, {Tsygankov}, {Karasev},
  {Molkov}, \& {Doroshenko}}]{Lutovinov2019MNRAS_GRO_J1750m27}
{Lutovinov}, A.~A., {Tsygankov}, S.~S., {Karasev}, D.~I., {Molkov}, S.~V., \&
  {Doroshenko}, V. 2019, \bibinfo{title}{{GRO J1750-27: A neutron star far
  behind the Galactic Center switching into the propeller regime},} \mnras,
  485, 770, \dodoi{10.1093/mnras/stz437}

\bibitem[{C. {Maitra}(2017){Maitra}}]{Maitra_2017JApA_crsf}
{Maitra}, C. 2017, \bibinfo{title}{{Cyclotron Lines: From Magnetic Field
  Strength Estimators to Geometry Tracers in Neutron Stars},} Journal of
  Astrophysics and Astronomy, 38, 50, \dodoi{10.1007/s12036-017-9476-3}

\bibitem[{C. {Malacaria} {et~al.}(2022){Malacaria}, {Bhargava}, {Coley},
  {Ducci}, {Pradhan}, {Ballhausen}, {Fuerst}, {Islam}, {Jaisawal}, {Jenke},
  {Kretschmar}, {Kreykenbohm}, {Pottschmidt}, {Sokolova-Lapa}, {Staubert},
  {Wilms}, {Wilson-Hodge}, \& {Wolff}}]{Malacaria_2022ApJ_2S_1553m542}
{Malacaria}, C., {Bhargava}, Y., {Coley}, J.~B., {et~al.} 2022,
  \bibinfo{title}{{Accreting on the Edge: A Luminosity-dependent Cyclotron Line
  in the Be/X-Ray Binary 2S 1553-542 Accompanied by Accretion Regimes
  Transition},} \apj, 927, 194, \dodoi{10.3847/1538-4357/ac524f}

\bibitem[{C.~B. {Markwardt} {et~al.}(2010){Markwardt}, {Baumgartner},
  {Skinner}, \& {Corbet}}]{Markwardt2010ATel_AXJ17002m4220}
{Markwardt}, C.~B., {Baumgartner}, W.~H., {Skinner}, G.~K., \& {Corbet}, R.
  H.~D. 2010, \bibinfo{title}{{AX J1700.2-4220 is a 54 second X-ray pulsar},}
  The Astronomer's Telegram, 2564, 1

\bibitem[{C.~B. {Markwardt} {et~al.}(2005){Markwardt}, {Tueller}, {Skinner},
  {Gehrels}, {Barthelmy}, \& {Mushotzky}}]{Markwardt_2005ApJ_Swift_BAT}
{Markwardt}, C.~B., {Tueller}, J., {Skinner}, G.~K., {et~al.} 2005,
  \bibinfo{title}{{The Swift/BAT High-Latitude Survey: First Results},} \apjl,
  633, L77, \dodoi{10.1086/498569}

\bibitem[{A.~B. {Mason} {et~al.}(2024){Mason}, {Norton}, {Clark}, {Farrell}, \&
  {Gosling}}]{Mason2024PASA_4XMM_J1825315m144036}
{Mason}, A.~B., {Norton}, A.~J., {Clark}, J.~S., {Farrell}, S.~A., \&
  {Gosling}, A.~J. 2024, \bibinfo{title}{{4XMM J182531.5{\ensuremath{-}}144036:
  A new persistent Be/X-ray binary found within the XMM-Newton serendipitous
  survey},} \pasa, 41, e008, \dodoi{10.1017/pasa.2024.6}

\bibitem[{M. {Matsuoka} {et~al.}(2009){Matsuoka}, {Kawasaki}, {Ueno}, {Tomida},
  {Kohama}, {Suzuki}, {Adachi}, {Ishikawa}, {Mihara}, {Sugizaki}, {Isobe},
  {Nakagawa}, {Tsunemi}, {Miyata}, {Kawai}, {Kataoka}, {Morii}, {Yoshida},
  {Negoro}, {Nakajima}, {Ueda}, {Chujo}, {Yamaoka}, {Yamazaki}, {Nakahira},
  {You}, {Ishiwata}, {Miyoshi}, {Eguchi}, {Hiroi}, {Katayama}, \&
  {Ebisawa}}]{Matsuoka2009PASJ_MAXI}
{Matsuoka}, M., {Kawasaki}, K., {Ueno}, S., {et~al.} 2009, \bibinfo{title}{{The
  MAXI Mission on the ISS: Science and Instruments for Monitoring All-Sky X-Ray
  Images},} \pasj, 61, 999, \dodoi{10.1093/pasj/61.5.999}

\bibitem[{V.~A. {McBride} {et~al.}(2006){McBride}, {Wilms}, {Coe},
  {Kreykenbohm}, {Rothschild}, {Coburn}, {Galache}, {Kretschmar}, {Edge}, \&
  {Staubert}}]{McBride2006aanda_MXB_0656m072}
{McBride}, V.~A., {Wilms}, J., {Coe}, M.~J., {et~al.} 2006,
  \bibinfo{title}{{Study of the cyclotron feature in <ASTROBJ>MXB
  0656-072</ASTROBJ>},} \aap, 451, 267, \dodoi{10.1051/0004-6361:20054239}

\bibitem[{J.~E. {McClintock} {et~al.}(1977){McClintock}, {Rappaport}, {Nugent},
  \& {Li}}]{McClintock1977ApJ_gx_304m1}
{McClintock}, J.~E., {Rappaport}, S.~A., {Nugent}, J.~J., \& {Li}, F.~K. 1977,
  \bibinfo{title}{{Discovery of a 272 second periodic variation in the X-ray
  source GX 304-1.},} \apjl, 216, L15, \dodoi{10.1086/182499}

\bibitem[{A. {Merloni} {et~al.}(2024){Merloni}, {Lamer}, {Liu}, {Ramos-Ceja},
  {Brunner}, {Bulbul}, {Dennerl}, {Doroshenko}, {Freyberg}, {Friedrich},
  {Gatuzz}, {Georgakakis}, {Haberl}, {Igo}, {Kreykenbohm}, {Liu}, {Maitra},
  {Malyali}, {Mayer}, {Nandra}, {Predehl}, {Robrade}, {Salvato}, {Sanders},
  {Stewart}, {Tub{\'\i}n-Arenas}, {Weber}, {Wilms}, {Arcodia}, {Artis},
  {Aschersleben}, {Avakyan}, {Aydar}, {Bahar}, {Balzer}, {Becker}, {Berger},
  {Boller}, {Bornemann}, {Br{\"u}ggen}, {Brusa}, {Buchner}, {Burwitz},
  {Camilloni}, {Clerc}, {Comparat}, {Coutinho}, {Czesla}, {Dannhauer},
  {Dauner}, {Dauser}, {Dietl}, {Dolag}, {Dwelly}, {Egg}, {Ehl}, {Freund},
  {Friedrich}, {Gaida}, {Garrel}, {Ghirardini}, {Gokus}, {Gr{\"u}nwald},
  {Grandis}, {Grotova}, {Gruen}, {Gueguen}, {H{\"a}mmerich}, {Hamaus},
  {Hasinger}, {Haubner}, {Homan}, {Ider Chitham}, {Joseph}, {Joyce},
  {K{\"o}nig}, {Kaltenbrunner}, {Khokhriakova}, {Kink}, {Kirsch}, {Kluge},
  {Knies}, {Krippendorf}, {Krumpe}, {Kurpas}, {Li}, {Liu}, {Locatelli},
  {Lorenz}, {M{\"u}ller}, {Magaudda}, {Mannes}, {McCall}, {Meidinger},
  {Michailidis}, {Migkas}, {Mu{\~n}oz-Giraldo}, {Musiimenta}, {Nguyen-Dang},
  {Ni}, {Olechowska}, {Ota}, {Pacaud}, {Pasini}, {Perinati}, {Pires},
  {Pommranz}, {Ponti}, {Poppenhaeger}, {P{\"u}hlhofer}, {Rau}, {Reh},
  {Reiprich}, {Roster}, {Saeedi}, {Santangelo}, {Sasaki}, {Schmitt},
  {Schneider}, {Schrabback}, {Schuster}, {Schwope}, {Seppi}, {Serim},
  {Shreeram}, {Sokolova-Lapa}, {Starck}, {Stelzer}, {Stierhof}, {Suleimanov},
  {Tenzer}, {Traulsen}, {Tr{\"u}mper}, {Tsuge}, {Urrutia}, {Veronica},
  {Waddell}, {Willer}, {Wolf}, {Yeung}, {Zainab}, {Zangrandi}, {Zhang},
  {Zhang}, \& {Zheng}}]{Merloni2024AnA}
{Merloni}, A., {Lamer}, G., {Liu}, T., {et~al.} 2024, \bibinfo{title}{{The
  SRG/eROSITA all-sky survey. First X-ray catalogues and data release of the
  western Galactic hemisphere},} \aap, 682, A34,
  \dodoi{10.1051/0004-6361/202347165}

\bibitem[{T. {Mihara} {et~al.}(2011){Mihara}, {Nakajima}, {Sugizaki}, {Serino},
  {Matsuoka}, {Kohama}, {Kawasaki}, {Tomida}, {Ueno}, {Kawai}, {Kataoka},
  {Morii}, {Yoshida}, {Yamaoka}, {Nakahira}, {Negoro}, {Isobe}, {Yamauchi}, \&
  {Sakurai}}]{Mihara_2011PASJ_MAXI_GSC}
{Mihara}, T., {Nakajima}, M., {Sugizaki}, M., {et~al.} 2011,
  \bibinfo{title}{{Gas Slit Camera (GSC) onboard MAXI on ISS},} \pasj, 63,
  S623, \dodoi{10.1093/pasj/63.sp3.S623}

\bibitem[{J.~C.~A. {Miller-Jones} {et~al.}(2018){Miller-Jones}, {Deller},
  {Shannon}, {Dodson}, {Mold{\'o}n}, {Rib{\'o}}, {Dubus}, {Johnston},
  {Paredes}, {Ransom}, \& {Tomsick}}]{MillerJones2018MNRAS}
{Miller-Jones}, J.~C.~A., {Deller}, A.~T., {Shannon}, R.~M., {et~al.} 2018,
  \bibinfo{title}{{The geometric distance and binary orbit of PSR B1259-63},}
  \mnras, 479, 4849, \dodoi{10.1093/mnras/sty1775}

\bibitem[{K. {Mitsuda} {et~al.}(2007){Mitsuda}, {Bautz}, {Inoue}, {Kelley},
  {Koyama}, {Kunieda}, {Makishima}, {Ogawara}, {Petre}, {Takahashi}, {Tsunemi},
  {White}, {Anabuki}, {Angelini}, {Arnaud}, {Awaki}, {Bamba}, {Boyce}, {Brown},
  {Chan}, {Cottam}, {Dotani}, {Doty}, {Ebisawa}, {Ezoe}, {Fabian}, {Figueroa},
  {Fujimoto}, {Fukazawa}, {Furusho}, {Furuzawa}, {Gendreau}, {Griffiths},
  {Haba}, {Hamaguchi}, {Harrus}, {Hasinger}, {Hatsukade}, {Hayashida}, {Henry},
  {Hiraga}, {Holt}, {Hornschemeier}, {Hughes}, {Hwang}, {Ishida}, {Ishisaki},
  {Isobe}, {Itoh}, {Iyomoto}, {Kahn}, {Kamae}, {Katagiri}, {Kataoka},
  {Katayama}, {Kawai}, {Kilbourne}, {Kinugasa}, {Kissel}, {Kitamoto}, {Kohama},
  {Kohmura}, {Kokubun}, {Kotani}, {Kotoku}, {Kubota}, {Madejski}, {Maeda},
  {Makino}, {Markowitz}, {Matsumoto}, {Matsumoto}, {Matsuoka}, {Matsushita},
  {McCammon}, {Mihara}, {Misaki}, {Miyata}, {Mizuno}, {Mori}, {Mori}, {Morii},
  {Moseley}, {Mukai}, {Murakami}, {Murakami}, {Mushotzky}, {Nagase}, {Namiki},
  {Negoro}, {Nakazawa}, {Nousek}, {Okajima}, {Ogasaka}, {Ohashi}, {Oshima},
  {Ota}, {Ozaki}, {Ozawa}, {Parmar}, {Pence}, {Porter}, {Reeves}, {Ricker},
  {Sakurai}, {Sanders}, {Senda}, {Serlemitsos}, {Shibata}, {Soong}, {Smith},
  {Suzuki}, {Szymkowiak}, {Takahashi}, {Tamagawa}, {Tamura}, {Tamura},
  {Tanaka}, {Tashiro}, {Tawara}, {Terada}, {Terashima}, {Tomida}, {Torii},
  {Tsuboi}, {Tsujimoto}, {Tsuru}, {Turner}, {Ueda}, {Ueno}, {Ueno}, {Uno},
  {Urata}, {Watanabe}, {Yamamoto}, {Yamaoka}, {Yamasaki}, {Yamashita},
  {Yamauchi}, {Yamauchi}, {Yaqoob}, {Yonetoku}, \&
  {Yoshida}}]{Mitsuda_2007PASJ_suzaku}
{Mitsuda}, K., {Bautz}, M., {Inoue}, H., {et~al.} 2007, \bibinfo{title}{{The
  X-Ray Observatory Suzaku},} \pasj, 59, S1, \dodoi{10.1093/pasj/59.sp1.S1}

\bibitem[{H. {Miyasaka} {et~al.}(2013){Miyasaka}, {Bachetti}, {Harrison},
  {F{\"u}rst}, {Barret}, {Bellm}, {Boggs}, {Chakrabarty}, {Chenevez},
  {Christensen}, {Craig}, {Grefenstette}, {Hailey}, {Madsen}, {Natalucci},
  {Pottschmidt}, {Stern}, {Tomsick}, {Walton}, {Wilms}, \&
  {Zhang}}]{Miyasaka_2013ApJ_GS_0834m430}
{Miyasaka}, H., {Bachetti}, M., {Harrison}, F.~A., {et~al.} 2013,
  \bibinfo{title}{{NuSTAR Detection of Hard X-Ray Phase Lags from the Accreting
  Pulsar GS 0834-430},} \apj, 775, 65, \dodoi{10.1088/0004-637X/775/1/65}

\bibitem[{S. {Molkov} {et~al.}(2021){Molkov}, {Doroshenko}, {Lutovinov},
  {Tsygankov}, {Santangelo}, {Mereminskiy}, \&
  {Semena}}]{Molkov_2021ApJ_Swift_J16266m5156_crsf}
{Molkov}, S., {Doroshenko}, V., {Lutovinov}, A., {et~al.} 2021,
  \bibinfo{title}{{Discovery of the 5 keV Cyclotron Line Followed by Three
  Harmonics in Swift J1626.6-5156},} \apjl, 915, L27,
  \dodoi{10.3847/2041-8213/ac0c15}

\bibitem[{E. {Morgan} {et~al.}(2003){Morgan}, {Remillard}, \&
  {Swank}}]{Morgan2003ATel_MXB_0656m072}
{Morgan}, E., {Remillard}, R., \& {Swank}, J. 2003,
  \bibinfo{title}{{XTEJ0658-073 (=MX0656-072) is a Pulsar},} The Astronomer's
  Telegram, 199, 1

\bibitem[{P. {Mroz} \& A. {Udalski}(2021){Mroz} \&
  {Udalski}}]{Mroz_2021ATel14361_SRGA_J1244041m632232}
{Mroz}, P., \& {Udalski}, A. 2021, \bibinfo{title}{{Periodic variability in the
  archival OGLE light curve of X-ray transient SRGA J124404.1-632232 (SRGE
  J124403.8-632231)},} The Astronomer's Telegram, 14361, 1

\bibitem[{U. {Mukherjee} \& B. {Paul}(2005){Mukherjee} \&
  {Paul}}]{Mukherjee2005aanda_3a0535p262}
{Mukherjee}, U., \& {Paul}, B. 2005, \bibinfo{title}{{X-ray properties of the
  transient pulsar 3A 0535+262 in quiescence},} \aap, 431, 667,
  \dodoi{10.1051/0004-6361:20041665}

\bibitem[{A.~A. {Mushtukov} {et~al.}(2015){Mushtukov}, {Suleimanov},
  {Tsygankov}, \& {Poutanen}}]{Mushtukov_2015MNRAS_optical_thickness}
{Mushtukov}, A.~A., {Suleimanov}, V.~F., {Tsygankov}, S.~S., \& {Poutanen}, J.
  2015, \bibinfo{title}{{The critical accretion luminosity for magnetized
  neutron stars},} \mnras, 447, 1847, \dodoi{10.1093/mnras/stu2484}

\bibitem[{F. {Nagase}(1989){Nagase}}]{Nagase1989PASJ}
{Nagase}, F. 1989, \bibinfo{title}{{Accretion-powered X-ray pulsars.},} \pasj,
  41, 1

\bibitem[{M. {Neumann} {et~al.}(2023){Neumann}, {Avakyan}, {Doroshenko}, \&
  {Santangelo}}]{Neumann_2023AnA_catalogue}
{Neumann}, M., {Avakyan}, A., {Doroshenko}, V., \& {Santangelo}, A. 2023,
  \bibinfo{title}{{XRBcats: Galactic High Mass X-ray Binary Catalogue★},}
  \aap, 677, A134, \dodoi{10.1051/0004-6361/202245728}

\bibitem[{E. {Nichelli} {et~al.}(2011){Nichelli}, {Israel}, {Moretti},
  {Campana}, {G{\"o}z}, \& {Stella}}]{Nichelli_2011ATel_IGR_J17200m3116}
{Nichelli}, E., {Israel}, G.~L., {Moretti}, A., {et~al.} 2011,
  \bibinfo{title}{{Swift discovery of 328s coherent pulsations from the HMXB
  IGR J17200-3116},} The Astronomer's Telegram, 3205, 1

\bibitem[{M.~A. {Nowak} {et~al.}(2012){Nowak}, {Paizis}, {Rodriguez}, {Chaty},
  {Del Santo}, {Grinberg}, {Wilms}, {Ubertini}, \& {Chini}}]{Nowak2012ApJ}
{Nowak}, M.~A., {Paizis}, A., {Rodriguez}, J., {et~al.} 2012,
  \bibinfo{title}{{X-Ray and Near-infrared Observations of the Obscured
  Accreting Pulsar IGR J18179-1621},} \apj, 757, 143,
  \dodoi{10.1088/0004-637X/757/2/143}

\bibitem[{L.~S. {Ootes} {et~al.}(2016){Ootes}, {Page}, {Wijnands}, \&
  {Degenaar}}]{Ootes_2016MNRAS_KS1731m260_crustcool}
{Ootes}, L.~S., {Page}, D., {Wijnands}, R., \& {Degenaar}, N. 2016,
  \bibinfo{title}{{Neutron star crust cooling in KS 1731-260: the influence of
  accretion outburst variability on the crustal temperature evolution},}
  \mnras, 461, 4400, \dodoi{10.1093/mnras/stw1799}

\bibitem[{M. {Orlandini} {et~al.}(2012){Orlandini}, {Frontera}, {Masetti},
  {Sguera}, \& {Sidoli}}]{Orlandini2012ApJ}
{Orlandini}, M., {Frontera}, F., {Masetti}, N., {Sguera}, V., \& {Sidoli}, L.
  2012, \bibinfo{title}{{BeppoSAX Observations of the X-Ray Pulsar MAXI
  J1409-619 in Low State: Discovery of Cyclotron Resonance Features},} \apj,
  748, 86, \dodoi{10.1088/0004-637X/748/2/86}

\bibitem[{M. {{\"O}zbey Arabac{\i}} {et~al.}(2015){{\"O}zbey Arabac{\i}},
  {Camero-Arranz}, {Zurita}, {Guti{\'e}rrez-Soto}, {Nespoli}, {Suso},
  {Kiaeerad}, {Garc{\'\i}a-Rojas}, \&
  {K{\i}z{\i}lo{\v{g}}lu}}]{Arabac_2015AnA_XTE_J1946p274}
{{\"O}zbey Arabac{\i}}, M., {Camero-Arranz}, A., {Zurita}, C., {et~al.} 2015,
  \bibinfo{title}{{Detection of a large Be circumstellar disk during X-ray
  quiescence of XTE J1946+274},} \aap, 582, A53,
  \dodoi{10.1051/0004-6361/201425488}

\bibitem[{J.~M. {Paredes} {et~al.}(2007){Paredes}, {Rib{\'o}}, {Bosch-Ramon},
  {West}, {Butt}, {Torres}, \& {Mart{\'\i}}}]{Paredes2007ApJ}
{Paredes}, J.~M., {Rib{\'o}}, M., {Bosch-Ramon}, V., {et~al.} 2007,
  \bibinfo{title}{{Chandra Observations of the Gamma-Ray Binary LS I +61 303:
  Extended X-Ray Structure?},} \apjl, 664, L39, \dodoi{10.1086/520674}

\bibitem[{A.~S. {Parikh} {et~al.}(2019){Parikh}, {Wijnands}, {Ootes}, {Page},
  {Degenaar}, {Bahramian}, {Brown}, {Cackett}, {Cumming}, {Heinke}, {Homan},
  {Rouco Escorial}, \&
  {Wijngaarden}}]{Parikh_2019AnA_LMXB_MXB1659m29_crustcool}
{Parikh}, A.~S., {Wijnands}, R., {Ootes}, L.~S., {et~al.} 2019,
  \bibinfo{title}{{Consistent accretion-induced heating of the neutron-star
  crust in MXB 1659-29 during two different outbursts},} \aap, 624, A84,
  \dodoi{10.1051/0004-6361/201834412}

\bibitem[{B. {Paul} \& S. {Naik}(2011){Paul} \&
  {Naik}}]{Paul_Naik_2011BASI_transient_hmxb}
{Paul}, B., \& {Naik}, S. 2011, \bibinfo{title}{{Transient High Mass X-ray
  Binaries},} Bulletin of the Astronomical Society of India, 39, 429,
  \dodoi{10.48550/arXiv.1110.4446}

\bibitem[{M. {Pavlinsky} {et~al.}(2021){Pavlinsky}, {Tkachenko}, {Levin},
  {Alexandrovich}, {Arefiev}, {Babyshkin}, {Batanov}, {Bodnar}, {Bogomolov},
  {Bubnov}, {Buntov}, {Burenin}, {Chelovekov}, {Chen}, {Drozdova}, {Ehlert},
  {Filippova}, {Frolov}, {Gamkov}, {Garanin}, {Garin}, {Glushenko}, {Gorelov},
  {Grebenev}, {Grigorovich}, {Gureev}, {Gurova}, {Ilkaev}, {Katasonov},
  {Krivchenko}, {Krivonos}, {Korotkov}, {Kudelin}, {Kuznetsova}, {Lazarchuk},
  {Lomakin}, {Lapshov}, {Lipilin}, {Lutovinov}, {Mereminskiy}, {Molkov},
  {Nazarov}, {Oleinikov}, {Pikalov}, {Ramsey}, {Roiz}, {Rotin}, {Ryadov},
  {Sankin}, {Sazonov}, {Sedov}, {Semena}, {Semena}, {Serbinov}, {Shirshakov},
  {Shtykovsky}, {Shvetsov}, {Sunyaev}, {Swartz}, {Tambov}, {Voron}, \&
  {Yaskovich}}]{Pavlinsky_2021AnA_ARTmXC_paper}
{Pavlinsky}, M., {Tkachenko}, A., {Levin}, V., {et~al.} 2021,
  \bibinfo{title}{{The ART-XC telescope on board the SRG observatory},} \aap,
  650, A42, \dodoi{10.1051/0004-6361/202040265}

\bibitem[{L.~J. {Pellizza} {et~al.}(2011){Pellizza}, {Chaty}, \&
  {Chisari}}]{Pellizza_2011AnA_IGR_J16283m4838}
{Pellizza}, L.~J., {Chaty}, S., \& {Chisari}, N.~E. 2011,
  \bibinfo{title}{{Unveiling the nature of IGR J16283-4838},} \aap, 526, A15,
  \dodoi{10.1051/0004-6361/201014954}

\bibitem[{V.~F. {Polcaro} \& L. {Norci}(2001){Polcaro} \&
  {Norci}}]{Polcaro2001AdSpR}
{Polcaro}, V.~F., \& {Norci}, L. 2001, \bibinfo{title}{{The HMXRB system
  BD+37{\textdegree} 1160/4U 0515+38},} Advances in Space Research, 28, 395,
  \dodoi{10.1016/S0273-1177(01)00437-9}

\bibitem[{P. {Pradhan} {et~al.}(2013){Pradhan}, {Maitra}, {Paul}, \&
  {Paul}}]{Pradhan2013MNRAS_SW_J20006p3210}
{Pradhan}, P., {Maitra}, C., {Paul}, B., \& {Paul}, B.~C. 2013,
  \bibinfo{title}{{Revisiting SW J2000.6+3210: a persistent Be X-ray pulsar?},}
  \mnras, 436, 945, \dodoi{10.1093/mnras/stt1504}

\bibitem[{P. {Predehl} {et~al.}(2021){Predehl}, {Andritschke}, {Arefiev},
  {Babyshkin}, {Batanov}, {Becker}, {B{\"o}hringer}, {Bogomolov}, {Boller},
  {Borm}, {Bornemann}, {Br{\"a}uninger}, {Br{\"u}ggen}, {Brunner}, {Brusa},
  {Bulbul}, {Buntov}, {Burwitz}, {Burkert}, {Clerc}, {Churazov}, {Coutinho},
  {Dauser}, {Dennerl}, {Doroshenko}, {Eder}, {Emberger}, {Eraerds},
  {Finoguenov}, {Freyberg}, {Friedrich}, {Friedrich}, {F{\"u}rmetz},
  {Georgakakis}, {Gilfanov}, {Granato}, {Grossberger}, {Gueguen}, {Gureev},
  {Haberl}, {H{\"a}lker}, {Hartner}, {Hasinger}, {Huber}, {Ji}, {Kienlin},
  {Kink}, {Korotkov}, {Kreykenbohm}, {Lamer}, {Lomakin}, {Lapshov}, {Liu},
  {Maitra}, {Meidinger}, {Menz}, {Merloni}, {Mernik}, {Mican}, {Mohr},
  {M{\"u}ller}, {Nandra}, {Nazarov}, {Pacaud}, {Pavlinsky}, {Perinati},
  {Pfeffermann}, {Pietschner}, {Ramos-Ceja}, {Rau}, {Reiffers}, {Reiprich},
  {Robrade}, {Salvato}, {Sanders}, {Santangelo}, {Sasaki}, {Scheuerle},
  {Schmid}, {Schmitt}, {Schwope}, {Shirshakov}, {Steinmetz}, {Stewart},
  {Str{\"u}der}, {Sunyaev}, {Tenzer}, {Tiedemann}, {Tr{\"u}mper}, {Voron},
  {Weber}, {Wilms}, \& {Yaroshenko}}]{Predehl2021AnA}
{Predehl}, P., {Andritschke}, R., {Arefiev}, V., {et~al.} 2021,
  \bibinfo{title}{{The eROSITA X-ray telescope on SRG},} \aap, 647, A1,
  \dodoi{10.1051/0004-6361/202039313}

\bibitem[{B. {Rai} {et~al.}(2023){Rai}, {Tobrej}, {Ghising}, {Tamang}, \&
  {Paul}}]{Rai2023MNRAS}
{Rai}, B., {Tobrej}, M., {Ghising}, M., {Tamang}, R., \& {Paul}, B.~C. 2023,
  \bibinfo{title}{{Study of recently discovered Be/X-ray pulsar MAXI J0655-013
  using NuSTAR},} \mnras, 524, 1352, \dodoi{10.1093/mnras/stad1944}

\bibitem[{C.~E. {Rakowski} {et~al.}(2006){Rakowski}, {Schulz}, {Wolk}, \&
  {Testa}}]{Rakowski_2006ApJ_HD_119682}
{Rakowski}, C.~E., {Schulz}, N.~S., {Wolk}, S.~J., \& {Testa}, P. 2006,
  \bibinfo{title}{{Extraordinarily Hot X-Ray Emission from the O9 Emission-Line
  Star HD 119682},} \apjl, 649, L111, \dodoi{10.1086/508350}

\bibitem[{G. {Raman} {et~al.}(2023){Raman}, {Varun}, {Pradhan}, \&
  {Kennea}}]{Raman2023MNRAS_quiescent}
{Raman}, G., {Varun}, {Pradhan}, P., \& {Kennea}, J. 2023,
  \bibinfo{title}{{Quiet, but not silent: uncovering quiescent state properties
  of two transient high-mass X-ray binaries},} \mnras, 526, 3262,
  \dodoi{10.1093/mnras/stad2577}

\bibitem[{P. {Reig}(2011){Reig}}]{PabloReig2011Ap}
{Reig}, P. 2011, \bibinfo{title}{{Be/X-ray binaries},} \apss, 332, 1,
  \dodoi{10.1007/s10509-010-0575-8}

\bibitem[{P. {Reig} \& P. {Roche}(1999){Reig} \&
  {Roche}}]{Reig_1999MNRAS_RX_J08124m3114}
{Reig}, P., \& {Roche}, P. 1999, \bibinfo{title}{{Discovery of X-ray pulsations
  in the Be/X-ray binary LS 992/RX J0812.4-3114},} \mnras, 306, 95,
  \dodoi{10.1046/j.1365-8711.1999.02463.x}

\bibitem[{P. {Reig} \& A. {Zezas}(2018){Reig} \&
  {Zezas}}]{Reig2018aanda_IGR_J06074p2205}
{Reig}, P., \& {Zezas}, A. 2018, \bibinfo{title}{{Discovery of X-ray pulsations
  in the Be/X-ray binary IGR J06074+2205},} \aap, 613, A52,
  \dodoi{10.1051/0004-6361/201732533}

\bibitem[{M. {Rib{\'o}} {et~al.}(2017){Rib{\'o}}, {Munar-Adrover}, {Paredes},
  {Marcote}, {Iwasawa}, {Mold{\'o}n}, {Casares}, {Migliari}, \&
  {Paredes-Fortuny}}]{Ribo2017ApJ_MWC_656}
{Rib{\'o}}, M., {Munar-Adrover}, P., {Paredes}, J.~M., {et~al.} 2017,
  \bibinfo{title}{{The First Simultaneous X-Ray/Radio Detection of the First
  Be/BH System MWC 656},} \apjl, 835, L33, \dodoi{10.3847/2041-8213/835/2/L33}

\bibitem[{R.~D. {Robinson} {et~al.}(2002){Robinson}, {Smith}, \&
  {Henry}}]{Robinson_2002ApJ_Be_Star_Possible_Magnetic_Dynamo}
{Robinson}, R.~D., {Smith}, M.~A., \& {Henry}, G.~W. 2002,
  \bibinfo{title}{{X-Ray and Optical Variations in the Classical Be Star
  {\ensuremath{\gamma}} Cassiopeia: The Discovery of a Possible Magnetic
  Dynamo},} \apj, 575, 435, \dodoi{10.1086/341141}

\bibitem[{A. {Rouco Escorial} {et~al.}(2017){Rouco Escorial}, {Bak Nielsen},
  {Wijnands}, {Cavecchi}, {Degenaar}, \& {Patruno}}]{Rouco2017MNRAS_4U0115p63}
{Rouco Escorial}, A., {Bak Nielsen}, A.~S., {Wijnands}, R., {et~al.} 2017,
  \bibinfo{title}{{The low-luminosity behaviour of the 4U 0115+63 Be/X-ray
  transient},} \mnras, 472, 1802, \dodoi{10.1093/mnras/stx2111}

\bibitem[{A. {Rouco Escorial} {et~al.}(2018){Rouco Escorial}, {van den
  Eijnden}, \& {Wijnands}}]{Rouco_2018AnA_colddisc_quasi_stable_low_acc}
{Rouco Escorial}, A., {van den Eijnden}, J., \& {Wijnands}, R. 2018,
  \bibinfo{title}{{Discovery of accretion-driven pulsations in the prolonged
  low X-ray luminosity state of the Be/X-ray transient GX 304-1},} \aap, 620,
  L13, \dodoi{10.1051/0004-6361/201834572}

\bibitem[{A. {Rouco Escorial} {et~al.}(2019){Rouco Escorial}, {Wijnands},
  {Ootes}, {Degenaar}, {Snelders}, {Kaper}, {Cackett}, \&
  {Homan}}]{Rouco_2019AnA_GROJ1750m27_lowlevel_acc}
{Rouco Escorial}, A., {Wijnands}, R., {Ootes}, L.~S., {et~al.} 2019,
  \bibinfo{title}{{Quiescent X-ray variability in the neutron star Be/X-ray
  transient GRO J1750-27},} \aap, 630, A105,
  \dodoi{10.1051/0004-6361/201834327}

\bibitem[{A. {Rouco Escorial} {et~al.}(2020){Rouco Escorial}, {Wijnands}, {van
  den Eijnden}, {Patruno}, {Degenaar}, {Parikh}, \&
  {Ootes}}]{Rouco_2020AnA_4U_0115p63_crust_cooling_low_level_accretion}
{Rouco Escorial}, A., {Wijnands}, R., {van den Eijnden}, J., {et~al.} 2020,
  \bibinfo{title}{{Recurrent low-level luminosity behaviour after a giant
  outburst in the Be/X-ray transient 4U 0115+63},} \aap, 638, A152,
  \dodoi{10.1051/0004-6361/201936287}

\bibitem[{J. {Roy} {et~al.}(2020){Roy}, {Agrawal}, {Singari}, \&
  {Misra}}]{Roy2020RAA_3A_0726m260}
{Roy}, J., {Agrawal}, P.~C., {Singari}, B., \& {Misra}, R. 2020,
  \bibinfo{title}{{AstroSat observation of the Be/X-ray binary Pulsar 3A
  0726-260 (4U 0728-25)},} Research in Astronomy and Astrophysics, 20, 155,
  \dodoi{10.1088/1674-4527/20/9/155}

\bibitem[{K. {Roy} {et~al.}(2024){Roy}, {Sharma}, {Manikantan}, \&
  {Paul}}]{Roy2024aanda_IGR_J06074p2205}
{Roy}, K., {Sharma}, R., {Manikantan}, H., \& {Paul}, B. 2024,
  \bibinfo{title}{{Detection of a cyclotron line in the Be X-ray pulsar IGR
  J06074+2205},} \aap, 686, A145, \dodoi{10.1051/0004-6361/202348998}

\bibitem[{A. {Salganik} {et~al.}(2025){Salganik}, {Tsygankov}, {Chernyakova},
  {Malyshev}, \& {Poutanen}}]{Salganik2025AnA}
{Salganik}, A., {Tsygankov}, S.~S., {Chernyakova}, M., {Malyshev}, D., \&
  {Poutanen}, J. 2025, \bibinfo{title}{{Discovery of a bimodal luminosity
  distribution in the persistent Be/X-ray pulsar 2RXP
  J130159.6{\textendash}635806},} \aap, 698, A71,
  \dodoi{10.1051/0004-6361/202453378}

\bibitem[{D.~A. {Schwartz} {et~al.}(1980){Schwartz}, {Griffiths},
  {Thorstensen}, {Charles}, \& {Bowyer}}]{Schwartz_1980AJ_HEAO}
{Schwartz}, D.~A., {Griffiths}, R.~E., {Thorstensen}, J.~R., {Charles}, P.~A.,
  \& {Bowyer}, S. 1980, \bibinfo{title}{{Optical identification of 4U 1907+09
  using the HEAO-1 Scanning modulation collimator position.},} \aj, 85, 549,
  \dodoi{10.1086/112710}

\bibitem[{M. {Shidatsu} {et~al.}(2022){Shidatsu}, {Pike}, {Mihara}, {Sugizaki},
  {Negoro}, {Nakajima}, \& {Murata}}]{Shidatsu2022ATel15495_MAXI_J0655m013}
{Shidatsu}, M., {Pike}, S., {Mihara}, T., {et~al.} 2022,
  \bibinfo{title}{{NuSTAR follow-up observation of MAXI J0655-013 reveals 1129s
  pulsations},} The Astronomer's Telegram, 15495, 1

\bibitem[{A. {Slettebak}(1979){Slettebak}}]{Slettebak_1979SSR_Be_stars}
{Slettebak}, A. 1979, \bibinfo{title}{{The Be Stars},} \ssr, 23, 541,
  \dodoi{10.1007/BF00212356}

\bibitem[{R. {Staubert} {et~al.}(2011){Staubert}, {Pottschmidt}, {Doroshenko},
  {Wilms}, {Suchy}, {Rothschild}, \& {Santangelo}}]{Staubert2011aanda}
{Staubert}, R., {Pottschmidt}, K., {Doroshenko}, V., {et~al.} 2011,
  \bibinfo{title}{{Finding a 24-day orbital period for the X-ray binary 1A
  1118-616},} \aap, 527, A7, \dodoi{10.1051/0004-6361/201015737}

\bibitem[{R. {Staubert} {et~al.}(2019){Staubert}, {Tr{\"u}mper}, {Kendziorra},
  {Klochkov}, {Postnov}, {Kretschmar}, {Pottschmidt}, {Haberl}, {Rothschild},
  {Santangelo}, {Wilms}, {Kreykenbohm}, \& {F{\"u}rst}}]{Staubert_2019AnA}
{Staubert}, R., {Tr{\"u}mper}, J., {Kendziorra}, E., {et~al.} 2019,
  \bibinfo{title}{{Cyclotron lines in highly magnetized neutron stars},} \aap,
  622, A61, \dodoi{10.1051/0004-6361/201834479}

\bibitem[{O. {Struve}(1931){Struve}}]{Struve_1931ApJ_Be_stars}
{Struve}, O. 1931, \bibinfo{title}{{On the Origin of Bright Lines in Spectra of
  Stars of Class B},} \apj, 73, 94, \dodoi{10.1086/143298}

\bibitem[{M. {Sugizaki} {et~al.}(2015){Sugizaki}, {Yamamoto}, {Mihara},
  {Nakajima}, \& {Makishima}}]{Sugizaki2015PASJ}
{Sugizaki}, M., {Yamamoto}, T., {Mihara}, T., {Nakajima}, M., \& {Makishima},
  K. 2015, \bibinfo{title}{{Luminosity and spin-period evolution of GX 304-1
  during outbursts from 2009 to 2013 observed with the MAXI/GSC, RXTE/PCA, and
  Fermi/GBM},} \pasj, 67, 73, \dodoi{10.1093/pasj/psv039}

\bibitem[{R. {Sunyaev} {et~al.}(2021){Sunyaev}, {Arefiev}, {Babyshkin},
  {Bogomolov}, {Borisov}, {Buntov}, {Brunner}, {Burenin}, {Churazov},
  {Coutinho}, {Eder}, {Eismont}, {Freyberg}, {Gilfanov}, {Gureyev}, {Hasinger},
  {Khabibullin}, {Kolmykov}, {Komovkin}, {Krivonos}, {Lapshov}, {Levin},
  {Lomakin}, {Lutovinov}, {Medvedev}, {Merloni}, {Mernik}, {Mikhailov},
  {Molodtsov}, {Mzhelsky}, {M{\"u}ller}, {Nandra}, {Nazarov}, {Pavlinsky},
  {Poghodin}, {Predehl}, {Robrade}, {Sazonov}, {Scheuerle}, {Shirshakov},
  {Tkachenko}, \& {Voron}}]{Sunyaev2021AnA}
{Sunyaev}, R., {Arefiev}, V., {Babyshkin}, V., {et~al.} 2021,
  \bibinfo{title}{{SRG X-ray orbital observatory. Its telescopes and first
  scientific results},} \aap, 656, A132, \dodoi{10.1051/0004-6361/202141179}

\bibitem[{B.~G. {Taylor} {et~al.}(1981){Taylor}, {Andresen}, {Peacock}, \&
  {Zobl}}]{Taylor_1981SSRv_EXOSAT}
{Taylor}, B.~G., {Andresen}, R.~D., {Peacock}, A., \& {Zobl}, R. 1981,
  \bibinfo{title}{{The EXOSAT Mission},} \ssr, 30, 479,
  \dodoi{10.1007/BF01246069}

\bibitem[{Y. {Terada} {et~al.}(2006){Terada}, {Mihara}, {Nakajima}, {Suzuki},
  {Isobe}, {Makishima}, {Takahashi}, {Enoto}, {Kokubun}, {Kitaguchi}, {Naik},
  {Dotani}, {Nagase}, {Tanaka}, {Watanabe}, {Kitamoto}, {Sudoh}, {Yoshida},
  {Nakagawa}, {Sugita}, {Kohmura}, {Kotani}, {Yonetoku}, {Angelini}, {Cottam},
  {Mukai}, {Kelley}, {Soong}, {Bautz}, {Kissel}, \&
  {Doty}}]{Terada2006ApJ_3a0535p262}
{Terada}, Y., {Mihara}, T., {Nakajima}, M., {et~al.} 2006,
  \bibinfo{title}{{Cyclotron Resonance Energies at a Low X-Ray Luminosity:
  A0535+262 Observed with Suzaku},} \apjl, 648, L139, \dodoi{10.1086/508018}

\bibitem[{J.~A. {Tomsick} {et~al.}(2008){Tomsick}, {Chaty}, {Rodriguez},
  {Walter}, \& {Kaaret}}]{Tomsick2008ApJ_Chandra}
{Tomsick}, J.~A., {Chaty}, S., {Rodriguez}, J., {Walter}, R., \& {Kaaret}, P.
  2008, \bibinfo{title}{{Chandra Localizations and Spectra of INTEGRAL Sources
  in the Galactic Plane},} \apj, 685, 1143, \dodoi{10.1086/591040}

\bibitem[{J.~A. {Tomsick} {et~al.}(2011){Tomsick}, {Heinke}, {Halpern},
  {Kaaret}, {Chaty}, {Rodriguez}, \&
  {Bodaghee}}]{Tomsick2011ApJ_IGR_J01363p6610}
{Tomsick}, J.~A., {Heinke}, C., {Halpern}, J., {et~al.} 2011,
  \bibinfo{title}{{Confirmation of IGR J01363+6610 as a Be X-ray Binary with
  Very Low Quiescent X-ray Luminosity},} \apj, 728, 86,
  \dodoi{10.1088/0004-637X/728/2/86}

\bibitem[{J.~M. {Torrej{\'o}n} \& A. {Orr}(2001){Torrej{\'o}n} \&
  {Orr}}]{Torrej2001AnA}
{Torrej{\'o}n}, J.~M., \& {Orr}, A. 2001, \bibinfo{title}{{BeppoSAX survey of
  Be/X-ray binary candidates},} \aap, 377, 148,
  \dodoi{10.1051/0004-6361:20011070}

\bibitem[{J.~M. {Torrej{\'o}n} {et~al.}(2013){Torrej{\'o}n}, {Schulz}, {Nowak},
  {Testa}, \& {Rodes}}]{Torrej_2013ApJ_HD_119682}
{Torrej{\'o}n}, J.~M., {Schulz}, N.~S., {Nowak}, M.~A., {Testa}, P., \&
  {Rodes}, J.~J. 2013, \bibinfo{title}{{Hot Thermal X-Ray Emission from the Be
  Star HD 119682},} \apj, 765, 13, \dodoi{10.1088/0004-637X/765/1/13}

\bibitem[{J. {Truemper} {et~al.}(1978){Truemper}, {Pietsch}, {Reppin}, {Voges},
  {Staubert}, \& {Kendziorra}}]{Truemper_1978ApJ_herxm1_cyclo}
{Truemper}, J., {Pietsch}, W., {Reppin}, C., {et~al.} 1978,
  \bibinfo{title}{{Evidence for strong cyclotron line emission in the hard
  X-ray spectrum of Hercules X-1.},} \apjl, 219, L105, \dodoi{10.1086/182617}

\bibitem[{S.~S. {Tsygankov} {et~al.}(2016{\natexlab{a}}){Tsygankov},
  {Lutovinov}, {Doroshenko}, {Mushtukov}, {Suleimanov}, \&
  {Poutanen}}]{Tsygankov_2016AnA_prop_eff_4U_0115p63_N_V_0332p53}
{Tsygankov}, S.~S., {Lutovinov}, A.~A., {Doroshenko}, V., {et~al.}
  2016{\natexlab{a}}, \bibinfo{title}{{Propeller effect in two brightest
  transient X-ray pulsars: 4U 0115+63 and V 0332+53},} \aap, 593, A16,
  \dodoi{10.1051/0004-6361/201628236}

\bibitem[{S.~S. {Tsygankov} {et~al.}(2016{\natexlab{b}}){Tsygankov},
  {Lutovinov}, {Krivonos}, {Molkov}, {Jenke}, {Finger}, \&
  {Poutanen}}]{Tsygankov2016MNRAS}
{Tsygankov}, S.~S., {Lutovinov}, A.~A., {Krivonos}, R.~A., {et~al.}
  2016{\natexlab{b}}, \bibinfo{title}{{NuSTAR discovery of a cyclotron
  absorption line in the transient X-ray pulsar 2S 1553-542},} \mnras, 457,
  258, \dodoi{10.1093/mnras/stv2849}

\bibitem[{S.~S. {Tsygankov} {et~al.}(2017{\natexlab{a}}){Tsygankov},
  {Mushtukov}, {Suleimanov}, {Doroshenko}, {Abolmasov}, {Lutovinov}, \&
  {Poutanen}}]{Tsygankov2017AnA_cold_disc}
{Tsygankov}, S.~S., {Mushtukov}, A.~A., {Suleimanov}, V.~F., {et~al.}
  2017{\natexlab{a}}, \bibinfo{title}{{Stable accretion from a cold disc in
  highly magnetized neutron stars},} \aap, 608, A17,
  \dodoi{10.1051/0004-6361/201630248}

\bibitem[{S.~S. {Tsygankov} {et~al.}(2017{\natexlab{b}}){Tsygankov},
  {Wijnands}, {Lutovinov}, {Degenaar}, \&
  {Poutanen}}]{Tsygankov2017MNRAS_pulsars_in_quiescence}
{Tsygankov}, S.~S., {Wijnands}, R., {Lutovinov}, A.~A., {Degenaar}, N., \&
  {Poutanen}, J. 2017{\natexlab{b}}, \bibinfo{title}{{The X-ray properties of
  Be/X-ray pulsars in quiescence},} \mnras, 470, 126,
  \dodoi{10.1093/mnras/stx1255}

\bibitem[{S.~S. {Tsygankov} {et~al.}(2026){Tsygankov}, {Lipunova},
  {Suleimanov}, {Salganik}, {Mushtukov}, {Forsblom}, {Tavleev}, {Kuzin}, \&
  {Poutanen}}]{Tsygankov_2026_real_time_observations_of_the_transition}
{Tsygankov}, S.~S., {Lipunova}, G., {Suleimanov}, V.~F., {et~al.} 2026,
  \bibinfo{title}{{Real-time observations of the transition to the quiescent
  state in an accreting magnetised neutron star: No propeller required?},}
  arXiv e-prints, arXiv:2608.10733, \dodoi{10.48550/arXiv.2608.10733}

\bibitem[{D. {Tub{\'\i}n-Arenas} {et~al.}(2024){Tub{\'\i}n-Arenas}, {Krumpe},
  {Lamer}, {Haase}, {Sanders}, {Brunner}, {Homan}, {Schwope}, {Georgakakis},
  {Poppenhaeger}, {Traulsen}, {K{\"o}nig}, {Merloni}, {Gueguen}, {Strong}, \&
  {Liu}}]{Tubin_2024AnA_erosita_upperlimit}
{Tub{\'\i}n-Arenas}, D., {Krumpe}, M., {Lamer}, G., {et~al.} 2024,
  \bibinfo{title}{{The eROSITA upper limits. Description and access to the
  data},} \aap, 682, A35, \dodoi{10.1051/0004-6361/202346773}

\bibitem[{D.~A. {Verner} {et~al.}(1996){Verner}, {Ferland}, {Korista}, \&
  {Yakovlev}}]{Verner_1996ApJ_crosssection}
{Verner}, D.~A., {Ferland}, G.~J., {Korista}, K.~T., \& {Yakovlev}, D.~G. 1996,
  \bibinfo{title}{{Atomic Data for Astrophysics. II. New Analytic Fits for
  Photoionization Cross Sections of Atoms and Ions},} \apj, 465, 487,
  \dodoi{10.1086/177435}

\bibitem[{L.~B.~F.~M. {Waters} {et~al.}(1989){Waters}, {Pols}, {Hogeveen},
  {Cote}, \& {van den Heuvel}}]{Waters_1989AnA_Be_p_WD}
{Waters}, L.~B.~F.~M., {Pols}, O.~R., {Hogeveen}, S.~J., {Cote}, J., \& {van
  den Heuvel}, E.~P.~J. 1989, \bibinfo{title}{{The formation and detectability
  of Be + white dwarf systems.},} \aap, 220, L1

\bibitem[{M.~G. {Watson} {et~al.}(1981){Watson}, {Warwick}, \&
  {Ricketts}}]{Watson1981MNRAS}
{Watson}, M.~G., {Warwick}, R.~S., \& {Ricketts}, M.~J. 1981,
  \bibinfo{title}{{2S 1145-619: an X-ray pulsar in an eccentric binary
  system?},} \mnras, 195, 197, \dodoi{10.1093/mnras/195.2.197}

\bibitem[{M.~C. {Weisskopf} {et~al.}(2000){Weisskopf}, {Tananbaum}, {Van
  Speybroeck}, \& {O'Dell}}]{Weisskopf_2000SPIE_chandrapaper}
{Weisskopf}, M.~C., {Tananbaum}, H.~D., {Van Speybroeck}, L.~P., \& {O'Dell},
  S.~L. 2000, \bibinfo{title}{{Chandra X-ray Observatory (CXO): overview},} in
  Society of Photo-Optical Instrumentation Engineers (SPIE) Conference Series,
  Vol. 4012, X-Ray Optics, Instruments, and Missions III, ed. J.~E. {Truemper}
  \& B.~{Aschenbach}, 2--16, \dodoi{10.1117/12.391545}

\bibitem[{M. {Wenger} {et~al.}(2000){Wenger}, {Ochsenbein}, {Egret}, {Dubois},
  {Bonnarel}, {Borde}, {Genova}, {Jasniewicz}, {Lalo{\"e}}, {Lesteven}, \&
  {Monier}}]{Simbad_Wenger_2000AnA}
{Wenger}, M., {Ochsenbein}, F., {Egret}, D., {et~al.} 2000,
  \bibinfo{title}{{The SIMBAD astronomical database. The CDS reference database
  for astronomical objects},} \aaps, 143, 9, \dodoi{10.1051/aas:2000332}

\bibitem[{W.~A. {Wheaton} {et~al.}(1979){Wheaton}, {Doty}, {Primini}, {Cooke},
  {Dobson}, {Goldman}, {Hecht}, {Howe}, {Hoffman}, {Scheepmaker}, {Tsiang},
  {Lewin}, {Matteson}, {Gruber}, {Baity}, {Rotschild}, {Knight}, {Nolang}, \&
  {Peterson}}]{Wheaton_1979_4U0115p63_cyclo}
{Wheaton}, W.~A., {Doty}, J.~P., {Primini}, F.~A., {et~al.} 1979,
  \bibinfo{title}{{An absorption feature in the spectrum of the pulsed hard
  X-ray flux from 4U0115 + 63},} \nat, 282, 240, \dodoi{10.1038/282240a0}

\bibitem[{R. {Wijnands} \& N. {Degenaar}(2016){Wijnands} \&
  {Degenaar}}]{Wijnands2016MNRAS_V0332p53_n_4U0115p63}
{Wijnands}, R., \& {Degenaar}, N. 2016, \bibinfo{title}{{Meta-stable low-level
  accretion rate states or neutron star crust cooling in the Be/X-ray
  transients V0332+53 and 4U 0115+63},} \mnras, 463, L46,
  \dodoi{10.1093/mnrasl/slw096}

\bibitem[{J. {Wilms} {et~al.}(2000){Wilms}, {Allen}, \&
  {McCray}}]{Wilms_2000ApJ_abund}
{Wilms}, J., {Allen}, A., \& {McCray}, R. 2000, \bibinfo{title}{{On the
  Absorption of X-Rays in the Interstellar Medium},} \apj, 542, 914,
  \dodoi{10.1086/317016}

\bibitem[{C.~A. {Wilson} {et~al.}(1997){Wilson}, {Finger}, {Harmon}, {Scott},
  {Wilson}, {Bildsten}, {Chakrabarty}, \&
  {Prince}}]{Wilson_1997ApJ_GS_0834m430}
{Wilson}, C.~A., {Finger}, M.~H., {Harmon}, B.~A., {et~al.} 1997,
  \bibinfo{title}{{A Sequence of Outbursts from the Transient X-Ray Pulsar GS
  0834-430},} \apj, 479, 388, \dodoi{10.1086/303841}

\bibitem[{R.~B. {Wilson} {et~al.}(1994){Wilson}, {Harmon}, {Fishman}, {Finger},
  {Stollberg}, {Pendleton}, {Briggs}, {Prince}, {Bildsten}, {Chakrabarty},
  {Rubin}, \& {Zhang}}]{Wilson1994AIPC_GRO_J1008m57}
{Wilson}, R.~B., {Harmon}, B.~A., {Fishman}, G.~J., {et~al.} 1994,
  \bibinfo{title}{{Discovery of the Hard X-Ray Pulsar GRO J1008-57 by BATSE},}
  in American Institute of Physics Conference Series, Vol. 308, The Evolution
  of X-ray Binariese, ed. S.~{Holt} \& C.~S. {Day} (AIP), 451,
  \dodoi{10.1063/1.45985}

\bibitem[{K.~S. {Wood} {et~al.}(1984){Wood}, {Meekins}, {Yentis}, {Smathers},
  {McNutt}, {Bleach}, {Byram}, {Chupp}, {Friedman}, \&
  {Meidav}}]{Wood_1984ApJS_HEAO1}
{Wood}, K.~S., {Meekins}, J.~F., {Yentis}, D.~J., {et~al.} 1984,
  \bibinfo{title}{{The HEAO A-1 X-ray source catalog.},} \apjs, 56, 507,
  \dodoi{10.1086/190992}

\bibitem[{T. {Yamamoto} {et~al.}(2011){Yamamoto}, {Sugizaki}, {Mihara},
  {Nakajima}, {Yamaoka}, {Matsuoka}, {Morii}, \&
  {Makishima}}]{Yamamoto2011PASJ_gx_304m1}
{Yamamoto}, T., {Sugizaki}, M., {Mihara}, T., {et~al.} 2011,
  \bibinfo{title}{{Discovery of a Cyclotron Resonance Feature in the X-Ray
  Spectrum of GX 304-1 with RXTE and Suzaku during Outbursts Detected by MAXI
  in 2010},} \pasj, 63, S751, \dodoi{10.1093/pasj/63.sp3.S751}

\bibitem[{J. {Yan} {et~al.}(2012){Yan}, {Zurita Heras}, {Chaty}, {Li}, \&
  {Liu}}]{Yan2012ApJ}
{Yan}, J., {Zurita Heras}, J.~A., {Chaty}, S., {Li}, H., \& {Liu}, Q. 2012,
  \bibinfo{title}{{Multi-wavelength Study of the Be/X-Ray Binary MXB
  0656-072},} \apj, 753, 73, \dodoi{10.1088/0004-637X/753/1/73}

\bibitem[{Y. {Zhao} {et~al.}(2019){Zhao}, {Heinke}, {Tsygankov}, {Ho},
  {Potekhin}, \& {Shaw}}]{Zhao2019MNRAS}
{Zhao}, Y., {Heinke}, C.~O., {Tsygankov}, S.~S., {et~al.} 2019,
  \bibinfo{title}{{Soft excess in the quiescent Be/X-ray pulsar RX
  J0812.4-3114},} \mnras, 488, 4427, \dodoi{10.1093/mnras/stz1946}

\end{thebibliography}

\end{document}